\documentclass[manyauthors,nocleardouble,COMPASS]{cernphprep}

\usepackage{bm}
\usepackage{color}

\usepackage{amssymb}
\usepackage{amsmath}
\usepackage{amsbsy}
\usepackage{times}
\usepackage{epsfig}
\usepackage{colordvi}
\usepackage{graphicx}
\usepackage{wrapfig,rotating}
\usepackage[numbers, square, comma, sort&compress]{natbib}
\usepackage{hyperref} 
\usepackage{multicol}
\usepackage{booktabs}
\usepackage{multirow}
\usepackage{lineno}
\RequirePackage[T1]{fontenc}
\newcommand\EPJSTYLE{no}

\newsavebox\myboxA
\newsavebox\myboxB
\newlength\mylenA

\newcommand{\ttGeV}{(\textrm{GeV}\!/c)^2}
\newcommand{\tmGeV}{(\textrm{GeV}\!/c)^{-2}}
\newcommand{\tpGeV}{t'}
\newcommand{\mmGeV}{\textrm{GeV}\!/c^2}

\newcommand{\mtpGeV}{m_{3\pi}}
\newcommand{\bpGeV}{b_{\textrm{prim}}}
\newcommand{\bdGeV}{b_{\textrm{diff}}}

\makeatletter

\DeclareSymbolFont{letters}     {OML}{cmm}{m}{it}
\DeclareSymbolFont{symbols}     {OMS}{cmsy}{m}{n}
\DeclareSymbolFont{largesymbols}{OMX}{cmex}{m}{n}
\begin{document}
\begin{titlepage}
\PHnumber{2014--041}
\PHdate{10 March 2014}

\title{Measurement of radiative widths of $a_2(1320)$ and $\pi_2(1670)$}

\Collaboration{The COMPASS Collaboration}
\ShortAuthor{The COMPASS Collaboration}

\begin{abstract}
The COMPASS Collaboration at CERN has investigated the reaction $\pi^- \gamma
\rightarrow \pi^-\pi^-\pi^+$ embedded in the Primakoff reaction of
$190~\textrm{GeV}$ pions scattering in the Coulomb field of a lead target,
$\pi^- \text{Pb} \rightarrow \pi^-\pi^-\pi^+ \text{Pb}$.  Exchange of quasi-real
photons is selected by isolating the sharp Coulomb peak observed at momentum
transfer below $0.001~\ttGeV$.  Using a partial-wave analysis the
amplitudes and relative phases of the $a_2(1320)$ and $\pi_2(1670)$ mesons have
been extracted, and the Coulomb and the diffractive contributions have been
disentangled.  Measuring absolute production cross sections we have determined
the radiative width of the $a_2(1320)$ to be $\Gamma_0(a_2(1320) \rightarrow
\pi\gamma) = (358 \pm 6_{\textrm{stat}} \pm 42_{\textrm{syst}})~\textrm{keV}$.
As the first measurement, $\Gamma_0(\pi_2(1670) \rightarrow \pi\gamma) = (181 \pm
11_{\textrm{stat}} \pm 27_{\textrm{syst}})~\textrm{keV} \cdot
(\textrm{BR}^{\textrm{PDG}}_{f_2 \pi}/\textrm{BR}_{f_2 \pi})$ is obtained for
the radiative width of the $\pi_2(1670)$, where in this analysis the branching
ratio $\textrm{BR}^{\textrm{PDG}}_{f_2 \pi}=0.56$ has been used.  We compare
these values to previous measurements and theoretical predictions.
\end{abstract}
\vfill
\Submitted{(to be submitted to EPJA)}
\end{titlepage}

{\pagestyle{empty}
%
%

\section*{The COMPASS Collaboration}
\label{app:collab}
\renewcommand\labelenumi{\textsuperscript{\theenumi}~}
\renewcommand\theenumi{\arabic{enumi}}
\begin{flushleft}
C.~Adolph\Irefn{erlangen},
R.~Akhunzyanov\Irefn{dubna}, 
M.G.~Alekseev\Irefn{triest_i},
G.D.~Alexeev\Irefn{dubna}, 
A.~Amoroso\Irefnn{turin_u}{turin_i},
V.~Andrieux\Irefn{saclay},
V.~Anosov\Irefn{dubna}, 
A.~Austregesilo\Irefnn{cern}{munichtu},
B.~Bade{\l}ek\Irefn{warsawu},
F.~Balestra\Irefnn{turin_u}{turin_i}
% ...
\end{flushleft}

%
%

\begin{Authlist}
\item \Idef{bielefeld}{Universit\"at Bielefeld, Fakult\"at f\"ur Physik, 33501 Bielefeld, Germany\Arefs{f}}
\item \Idef{bochum}{Universit\"at Bochum, Institut f\"ur Experimentalphysik, 44780 Bochum, Germany\Arefs{f}\Arefs{ll}}
\item \Idef{bonniskp}{Universit\"at Bonn, Helmholtz-Institut f\"ur  Strahlen- und Kernphysik, 53115 Bonn, Germany\Arefs{f}}
\item \Idef{bonnpi}{Universit\"at Bonn, Physikalisches Institut, 53115 Bonn, Germany\Arefs{f}}
\item \Idef{brno}{Institute of Scientific Instruments, AS CR, 61264 Brno, Czech Republic\Arefs{g}}
\item \Idef{calcutta}{Matrivani Institute of Experimental Research \& Education, Calcutta-700 030, India\Arefs{h}}
\item \Idef{dubna}{Joint Institute for Nuclear Research, 141980 Dubna, Moscow region, Russia\Arefs{i}}
\item \Idef{erlangen}{Universit\"at Erlangen--N\"urnberg, Physikalisches Institut, 91054 Erlangen, Germany\Arefs{f}}
\item \Idef{freiburg}{Universit\"at Freiburg, Physikalisches Institut, 79104 Freiburg, Germany\Arefs{f}\Arefs{ll}}
\item \Idef{cern}{CERN, 1211 Geneva 23, Switzerland}
\item \Idef{liberec}{Technical University in Liberec, 46117 Liberec, Czech Republic\Arefs{g}}
\item \Idef{lisbon}{LIP, 1000-149 Lisbon, Portugal\Arefs{j}}
\item \Idef{mainz}{Universit\"at Mainz, Institut f\"ur Kernphysik, 55099 Mainz, Germany\Arefs{f}}
\item \Idef{miyazaki}{University of Miyazaki, Miyazaki 889-2192, Japan\Arefs{k}}
\item \Idef{moscowlpi}{Lebedev Physical Institute, 119991 Moscow, Russia}
\item \Idef{munichlmu}{Ludwig-Maximilians-Universit\"at M\"unchen, Department f\"ur Physik, 80799 Munich, Germany\Arefs{f}\Arefs{l}}
\item \Idef{munichtu}{Technische Universit\"at M\"unchen, Physik Department, 85748 Garching, Germany\Arefs{f}\Arefs{l}}
\item \Idef{nagoya}{Nagoya University, 464 Nagoya, Japan\Arefs{k}}
\item \Idef{praguecu}{Charles University in Prague, Faculty of Mathematics and Physics, 18000 Prague, Czech Republic\Arefs{g}}
\item \Idef{praguectu}{Czech Technical University in Prague, 16636 Prague, Czech Republic\Arefs{g}}
\item \Idef{protvino}{State Scientific Center Institute for High Energy Physics of National Research Center `Kurchatov Institute', 142281 Protvino, Russia}
\item \Idef{saclay}{CEA IRFU/SPhN Saclay, 91191 Gif-sur-Yvette, France\Arefs{ll}}
\item \Idef{telaviv}{Tel Aviv University, School of Physics and Astronomy, 69978 Tel Aviv, Israel\Arefs{m}}
\item \Idef{triest_i}{Trieste Section of INFN, 34127 Trieste, Italy}
\item \Idef{triest_u}{University of Trieste, Department of Physics, 34127 Trieste, Italy}
\item \Idef{triest_ictp}{Abdus Salam ICTP, 34151 Trieste, Italy}
\item \Idef{turin_u}{University of Turin, Department of Physics, 10125 Turin, Italy}
\item \Idef{turin_i}{Torino Section of INFN, 10125 Turin, Italy}
\item \Idef{turin_p}{University of Eastern Piedmont, 15100 Alessandria, Italy}
\item \Idef{warsaw}{National Centre for Nuclear Research, 00-681 Warsaw, Poland\Arefs{n} }
\item \Idef{warsawu}{University of Warsaw, Faculty of Physics, 00-681 Warsaw, Poland\Arefs{n} }
\item \Idef{warsawtu}{Warsaw University of Technology, Institute of Radioelectronics, 00-665 Warsaw, Poland\Arefs{n} }
\item \Idef{yamagata}{Yamagata University, Yamagata, 992-8510 Japan\Arefs{k} }
\end{Authlist}
%
%
\vspace*{-\baselineskip}\renewcommand\theenumi{\alph{enumi}}
\begin{Authlist}
\item \Adef{a}{Also at Instituto Superior T\'ecnico, Universidade de Lisboa, Lisbon, Portugal}
\item \Adef{aa}{Also at Department of Physics, Pusan National University, Busan 609-735, Republic of Korea and at Physics Department, Brookhaven National Laboratory, Upton, NY 11973, U.S.A. }
\item \Adef{bb}{Supported by the DFG Research Training Group Programme 1102  ``Physics at Hadron Accelerators''}
\item \Adef{b}{Also at Chubu University, Kasugai, Aichi, 487-8501 Japan\Arefs{k}}
\item \Adef{c}{Also at KEK, 1-1 Oho, Tsukuba, Ibaraki, 305-0801 Japan}
\item \Adef{cc}{Also at Moscow Institute of Physics and Technology, Moscow Region, 141700, Russia}
\item \Adef{y}{present address: National Science Foundation, 4201 Wilson Boulevard, Arlington, VA 22230, United States}
\item \Adef{x}{present address: RWTH Aachen University, III. Physikalisches Institut, 52056 Aachen, Germany}
\item \Adef{e}{Also at GSI mbH, Planckstr.\ 1, D-64291 Darmstadt, Germany}
\item \Adef{f}{Supported by the German Bundesministerium f\"ur Bildung und Forschung}
\item \Adef{g}{Supported by Czech Republic MEYS Grants ME492 and LA242}
\item \Adef{h}{Supported by SAIL (CSR), Govt.\ of India}
\item \Adef{i}{Supported by CERN-RFBR Grants 08-02-91009 and 12-02-91500}
\item \Adef{j}{\raggedright Supported by the Portuguese FCT - Funda\c{c}\~{a}o para a Ci\^{e}ncia e Tecnologia, COMPETE and QREN, Grants CERN/FP/109323/2009, CERN/FP/116376/2010 and CERN/FP/123600/2011}
\item \Adef{k}{Supported by the MEXT and the JSPS under the Grants No.18002006, No.20540299 and No.18540281; Daiko Foundation and Yamada Foundation}
\item \Adef{l}{Supported by the DFG cluster of excellence `Origin and Structure of the Universe' (www.universe-cluster.de)}
\item \Adef{ll}{Supported by EU FP7 (HadronPhysics3, Grant Agreement number 283286)}
\item \Adef{m}{Supported by the Israel Science Foundation, founded by the Israel Academy of Sciences and Humanities}
\item \Adef{n}{Supported by the Polish NCN Grant DEC-2011/01/M/ST2/02350}
\end{Authlist}

\clearpage}
\section{Introduction}
\label{sec:intro}

Radiative decays of mesons are an important tool for the investigation of their
internal structure as the electromagnetic transition operators are well known
and probe the difference between the initial and final state mesons in terms of
their electric charge or magnetic current distributions.  The established
$a_2(1320)\rightarrow \pi\gamma$ decay constitutes a magnetic quadrupole
transition. The $\pi_2(1670)\rightarrow\pi\gamma$ decay was not measured
before. It represents an electric quadrupole transition, which is expected to
probe the charge distribution of the orbitally excited meson out to large
distances.  Radiative transitions can be calculated using the meson wave
function obtained in various quark models.  In addition, the vector meson
dominance model is used to relate $\rho\pi$ and $\gamma\pi$ decays via the
$\rho-\gamma$ equivalence.  Various calculations do exist for the radiative
width of the $a_{2}(1320)$.  Applying vector meson dominance, a width of $375
\pm 50~\textrm{keV}$ was calculated by ref.~\cite{rosner_radwidths}.  Using a
relativistic quark model for the meson wave function, a value of
$324~\textrm{keV}$ was extracted \cite{aznauryan_oganesyan}, and
$235~\textrm{keV}$ was derived from a covariant oscillator quark model
\cite{ishida_et_al}.  Newer calculations in this model framework yield
$237~\textrm{keV}$ \cite{maeda_cov_osc_qm}.  The covariant oscillator quark
model was also used for a prediction of the radiative width of the
$\pi_2(1670)$; for two different model versions values of $335~\textrm{keV}$ and
$521~\textrm{keV}$ are given in ref.~\cite{maeda_cov_osc_qm}.

The direct measurement of electromagnetic couplings using radiative decays of
mesons is difficult, as the corresponding branching ratios are small, and
background from processes containing $\pi^0\rightarrow \gamma\gamma$ or
$\eta\rightarrow \gamma\gamma$ with one or more of the photons lost may be
significant. An alternative access to the radiative transition amplitudes is
given by $\pi\gamma$ scattering as provided by Primakoff production of the
resonances under investigation, where an ultra-relativistic ({\it
  i.e.\ }quasi-stable) pion beam scatters off the quasi-real photons of the
electromagnetic field of a heavy nucleus.  The respective flux is given by the
Weizs\"{a}cker-Williams equivalent-photon approximation \cite{equiv_ph_approx},
which relates the experimentally observed cross section $\sigma_{\pi A}$ to the
cross section of real photon scattering $\sigma_{\pi\gamma}$ as

\begin{align}
  \frac{\textrm{d}\sigma_{\pi A}}{\textrm{d}s\, \textrm{d}t'\,\textrm{d}\Phi}\ =\ 
  \frac{\alpha\cdot Z^2}{\pi (s-m_\pi^2)}\cdot F_{\mbox{\tiny eff}}^2(t') \cdot
  \frac{t'}{(t'+t_{\min})^2} \cdot
  \frac{\textrm{d}\sigma_{\pi\gamma}(s)}{\textrm{d}\Phi} \ . 
\label{eq:intro:WW_xsec}  
\end{align} 

The positive quantity $t' = \vert t \vert - t_{\textrm{min}}$ contains the
four-momentum transfer squared $t = (p_{\pi}-p_{X})^2$ and
$t_{\textrm{min}}=[(s-m_\pi^2)/2E_{\mbox{\tiny beam}}]^2$ with $ m = m_{X} =
\sqrt{s}$ being the invariant mass of a final state $X$ given by $s = (p_{\pi} +
p_{\gamma})^2$.  The symbol $\textrm{d}\Phi$ denotes the phase space element as
given in ref.~\cite{pdg2012}, eq.~(43.11), and $Z$ is the charge of the nucleus
with mass number $A$.  

We approximate the form factor $F_{\mbox{\tiny eff}}^2(t')$ by means of the
sharp-radius approach of refs.~\cite{faeldt_2009_2013cor} and
\cite{faeldt_2010}, thus taking into account the distortion of the pionic wave
functions in the Coulomb field.  We use $|F^u_C(t', t_{\textrm{min}})|^2$ given
in eq.~(27) of ref.~\cite{faeldt_2009_2013cor}, which also includes the
Weizs\"{a}cker-Williams term $t'/(t'+t_{\textrm{min}})^2$.  For the extended
charge distribution of the lead nucleus, we take a sharp radius of $r_u =
6.52~\textrm{fm}$.

The cross section for the production and decay of a broad resonance $X$ with
spin $J$ and nominal mass $m_0$, averaged over its spin projections, is
parameterised by a relativistic Breit-Wigner function. Modified for the case of
pion-induced Primakoff production, it reads

\begin{align}
\ifthenelse{\equal{\EPJSTYLE}{yes}}
{
\frac{\textrm{d}\sigma}{\textrm{d}m\textrm{d}t'} =& 16 \alpha Z^2 (2J+1) 
\left( \frac{m}{m^2 - m_{\pi}^2} \right)^3 \nonumber \\
&\cdot  
\frac{m_0^2 \, \Gamma_{\pi\gamma}(m) \, \Gamma_{\textrm{final}} (m)}{(m^2 - m_0^2)^2 + m_0^2 \Gamma^2_{\textrm{total}}(m)} \nonumber \\ &\cdot
\ \frac{t'}{(t' + t_{\textrm{min}})^2} \ F^2_{\textrm{eff}}(t')\ . 
}{
\frac{\textrm{d}\sigma}{\textrm{d}m\textrm{d}t'} = 16 \alpha Z^2 (2J+1) 
\left( \frac{m}{m^2 - m_{\pi}^2} \right)^3
\frac{m_0^2 \, \Gamma_{\pi\gamma}(m) \, \Gamma_{\textrm{final}} (m)}{(m^2 - m_0^2)^2 + m_0^2 \Gamma^2_{\textrm{total}}(m)} \ \cdot
\ \frac{t'}{(t' + t_{\textrm{min}})^2} \  F^2_{\textrm{eff}}(t')\ .
}
\label{eq:intro:xsec_primakoff_resonance}
\end{align}

Here, $\Gamma_{\pi\gamma}(m) = f_{\pi\gamma}^{\textrm{dyn}}(m)\cdot \Gamma_0(X
\rightarrow \pi\gamma)$ is the mass-dependent radiative width with
$f^{\textrm{dyn}}_{\pi\gamma}$ the kinematic factor discussed in
sect.~\ref{sec:extract:ingred} multiplied by the nominal radiative width
$\Gamma_0(X \rightarrow \pi\gamma)$ that is the subject of this paper.  The
symbol $\Gamma_{\textrm{total}}(m)$ denotes the total mass-dependent width of
the resonance $X$ (see eq.~\eqref{eq:app:BW_massdep_width} below), and
$\Gamma_{\textrm{final}}(m)$ its mass-dependent partial width for the decay into
the investigated final state

\begin{align}
\Gamma_{\textrm{final}}(m)
= f_{{\text{final}}}^{\textrm{dyn}}(m)\cdot \Gamma_0(m_0) \cdot \textrm{CG} \cdot \textrm{BR}\ ,
\label{eq:intro:gamma_final}
\end{align}
where $\textrm{CG}$ is the relevant squared isospin Clebsch-Gordan coefficient
of the resonance decay into the investigated final state with branching ratio
$\textrm{BR}$ and $\Gamma_0(m_0)$ is the nominal width of the resonance at its
nominal mass.

Integrating eq.~\eqref{eq:intro:xsec_primakoff_resonance} over the relevant
ranges in $m$ and $t'$, the radiative width is found to be related to the
absolute cross section $\sigma_{\textrm{Primakoff}, X}$ via a constant $C_X$
that is calculated according to eq.~\eqref{eq:extract:C}:

\begin{align}
\ifthenelse{\equal{\EPJSTYLE}{yes}}
{
\sigma_{\textrm{Primakoff}, X} &= \int_{m_1}^{m_2} \int_{0}^{t'_{\textrm{max}}}  \frac{\textrm{d}\sigma}{\textrm{d}m\textrm{d}t'}  \,\textrm{d}t' \, \textrm{d}m \nonumber \\
	&=  \Gamma_0(X \rightarrow \pi\gamma) \cdot  C_X \ . 
}{
\sigma_{\textrm{Primakoff}, X} = \int_{m_1}^{m_2} \int_{0}^{t'_{\textrm{max}}}  \frac{\textrm{d}\sigma}{\textrm{d}m\textrm{d}t'}  \,\textrm{d}t' \, \textrm{d}m =  \Gamma_0(X \rightarrow \pi\gamma) \cdot  C_X \ . 
}
\label{eq:intro:xsec_radwidth}
\end{align}
Thus the radiative width $\Gamma_0(X \rightarrow \pi\gamma)$ can be determined
from the number of events $N_{X, \textrm{prim}}$ experimentally observed from
Primakoff production,
\begin{equation}
 \Gamma_0(X \rightarrow \pi\gamma) = \frac{ N_{X, \textrm{prim}}/\epsilon_{X}}{ C_X \cdot L \cdot \textrm{CG} \cdot \textrm{BR} \cdot \epsilon_{\textrm{resol}}} \ ,
\label{eq:intro:radwidth_from_data}
\end{equation}
with $\epsilon_{X}$ being the acceptance of the experimental apparatus and the
event selection procedure, and $L$ the integrated luminosity corresponding to
the analysed data set. Effects due to the finite resolution in $t'$ are absorbed
by $\epsilon_{\textrm{resol}}$, which reflects the migration of events from the
sharp peak near $t'\approx 0$ to higher values outside of our selected $t'$
region.

\section{Primakoff production of resonances in the $\pi^-\pi^-\pi^+$ final state}
\label{sec:prim_resonances}

\ifthenelse{\equal{\EPJSTYLE}{yes}}
{
\begin{figure*}
}{
\begin{figure}
}
\includegraphics[width=0.99\textwidth]{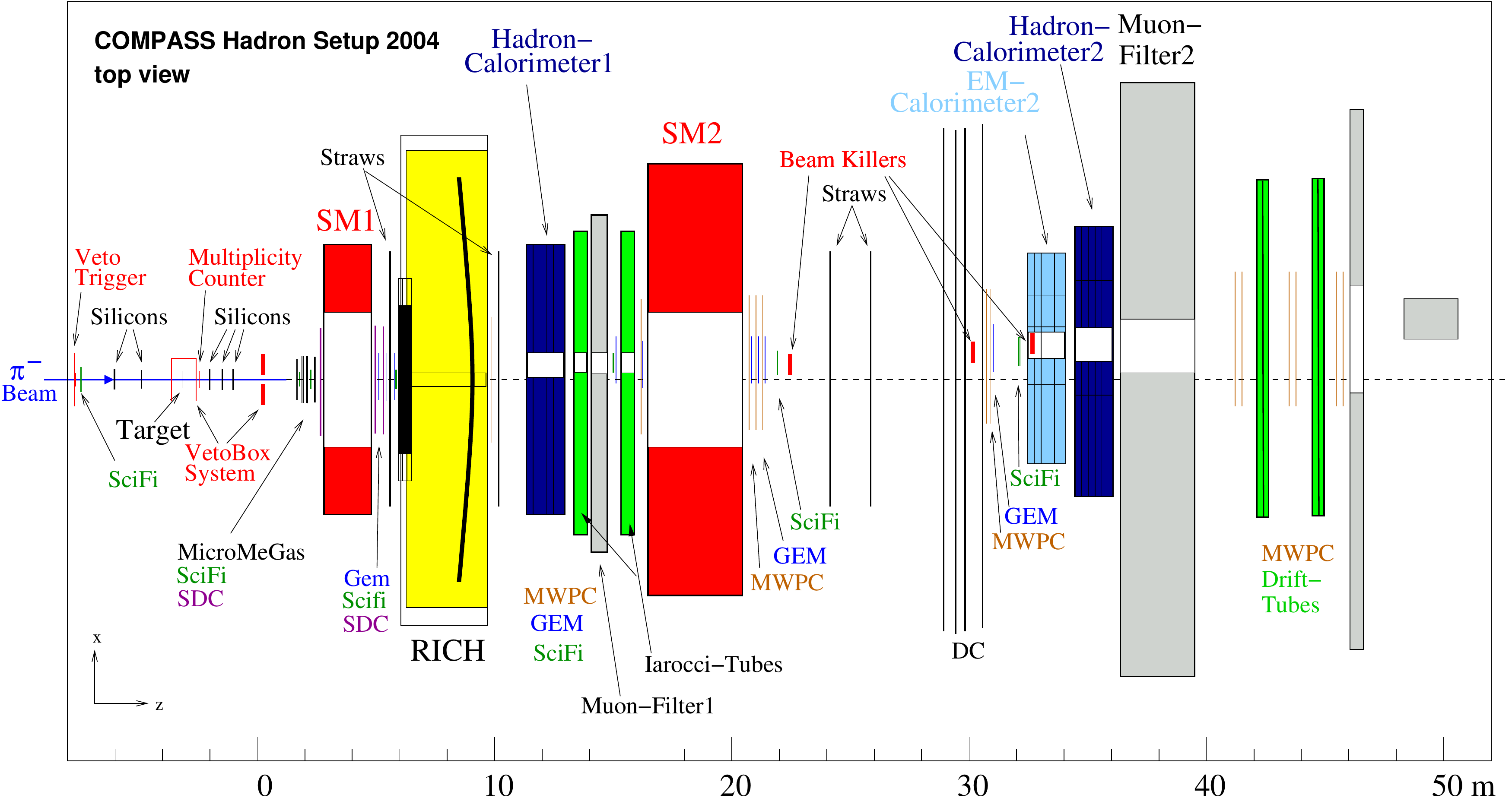}
\caption{Sketch of the experimental setup of the Hadron Run in the year 2004.}
\label{fig:setup2004}
\ifthenelse{\equal{\EPJSTYLE}{yes}}
{
\end{figure*}
}{
\end{figure}
}

The COMPASS experiment located at the CERN Super Proton Synchrotron features a
large-acceptance and high-precision spectrometer \cite{compass_spectrometer}. It
offers very good conditions to study reactions of high-energy beam particles
impinging on fixed targets at low to intermediate momentum transfers.  Its
acceptance covers mostly the phase space for final state particles emerging in
forward direction.  The data presented in the following were recorded in 2004
using a $190~\textrm{GeV}$ negative hadron beam, which consists of
$96.8\%~\pi^-$, $2.4\%~K^-$, and $0.8\%~\bar{p}$ at the COMPASS target.  The
experimental setup is sketched in fig.~\ref{fig:setup2004}.  The target is
surrounded by a veto system designed to reject non-exclusive forward or
large-angle reactions.  High-precision silicon micro-strip detectors with a
spatial resolution of $8-11~\mu\textrm{m}$ make vertex reconstruction possible
for smallest scattering angles.  The two spectrometer stages, which are arranged
around the magnets SM1 and SM2, are both equipped with a variety of detectors
for tracking, calorimetry and particle identification.  The target consisted of
$3~\textrm{mm}$ of lead disks, which were deployed either as one continuous disk
or as two disks with $2~\textrm{mm}$ and $1~\textrm{mm}$ thickness,
respectively. The latter two target disks, which were separated by
$10~\textrm{cm}$ along the beam, allowed additional systematic studies.  The
present analysis uses events recorded with the so-called multiplicity trigger
that selects at least two charged outgoing particles at scattering angles
smaller than $50~\textrm{mrad}$.  For this purpose, a scintillator disk of
$5~\textrm{mm}$ thickness and a diameter of $5~\textrm{cm}$ was placed about
$62~\textrm{cm}$ downstream of the target. The hits in this scintillator had to
be in coincidence with the beam trigger and an energy deposit of several
$\textrm{GeV}$ in the hadronic calorimeter HCAL2.

\begin{figure}
\begin{center}
\ifthenelse{\equal{\EPJSTYLE}{yes}}
{\includegraphics[width=0.99\columnwidth]{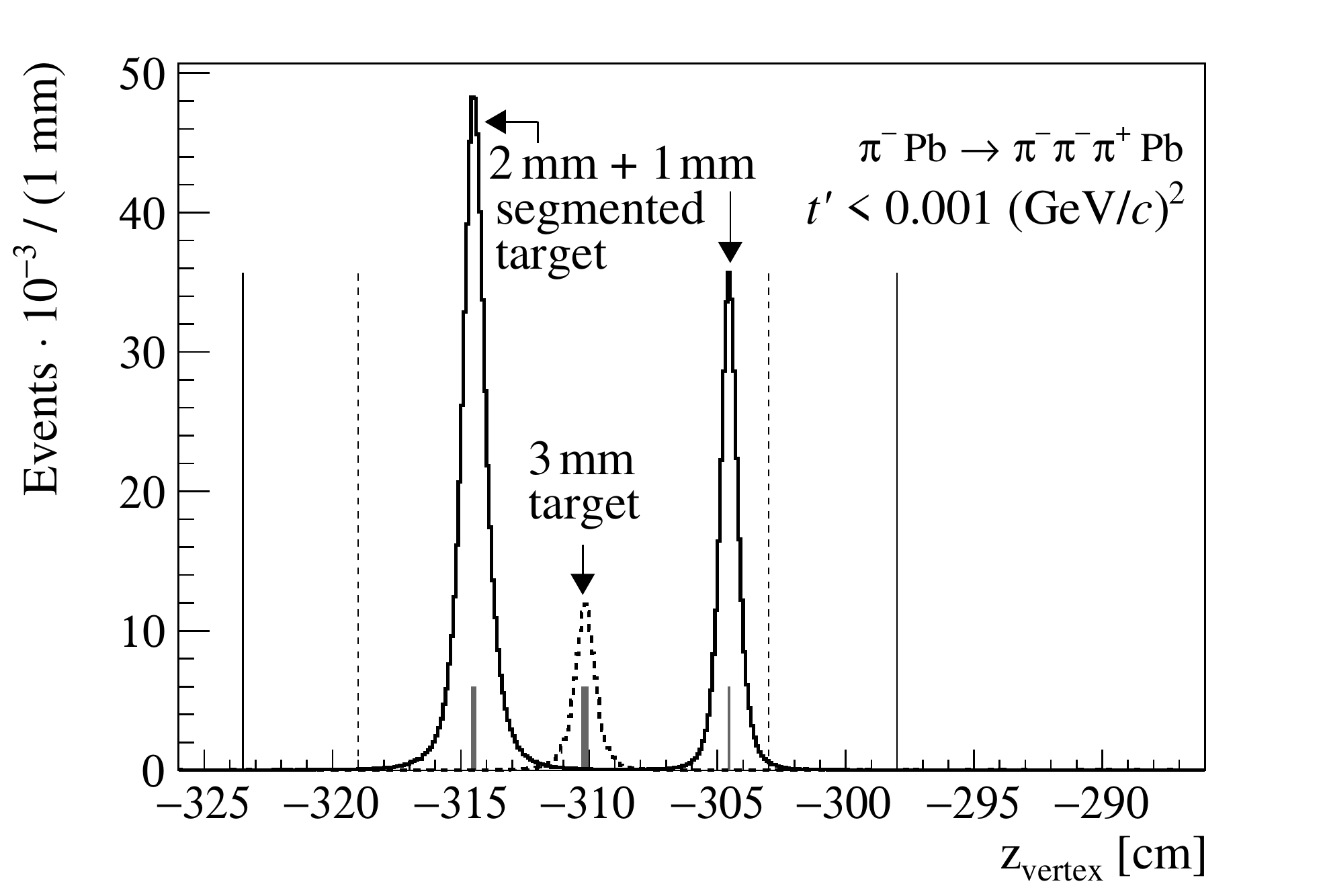}}
{\includegraphics[width=0.49\columnwidth]{fig2.pdf}}
\caption{Distribution of the vertex positions along the beam axis for the $3\pi$
  final-state events with $\tpGeV<0.001~\ttGeV$.  The distributions of the
  reconstructed vertices for the two target setups (solid and dashed lines) are
  complemented with vertical thin lines indicating the cuts applied to the
  respective data sets. The grey boxes represent the nominal thicknesses and
  positions of the target disks. The numbers of entries in the two histogram
  sets reflect the different measurement times with the two target setups.}
\label{fig:vertices_prim}
\end{center}
\end{figure}

\begin{figure}
\begin{center}
\ifthenelse{\equal{\EPJSTYLE}{yes}}
{\includegraphics[width=0.99\columnwidth]{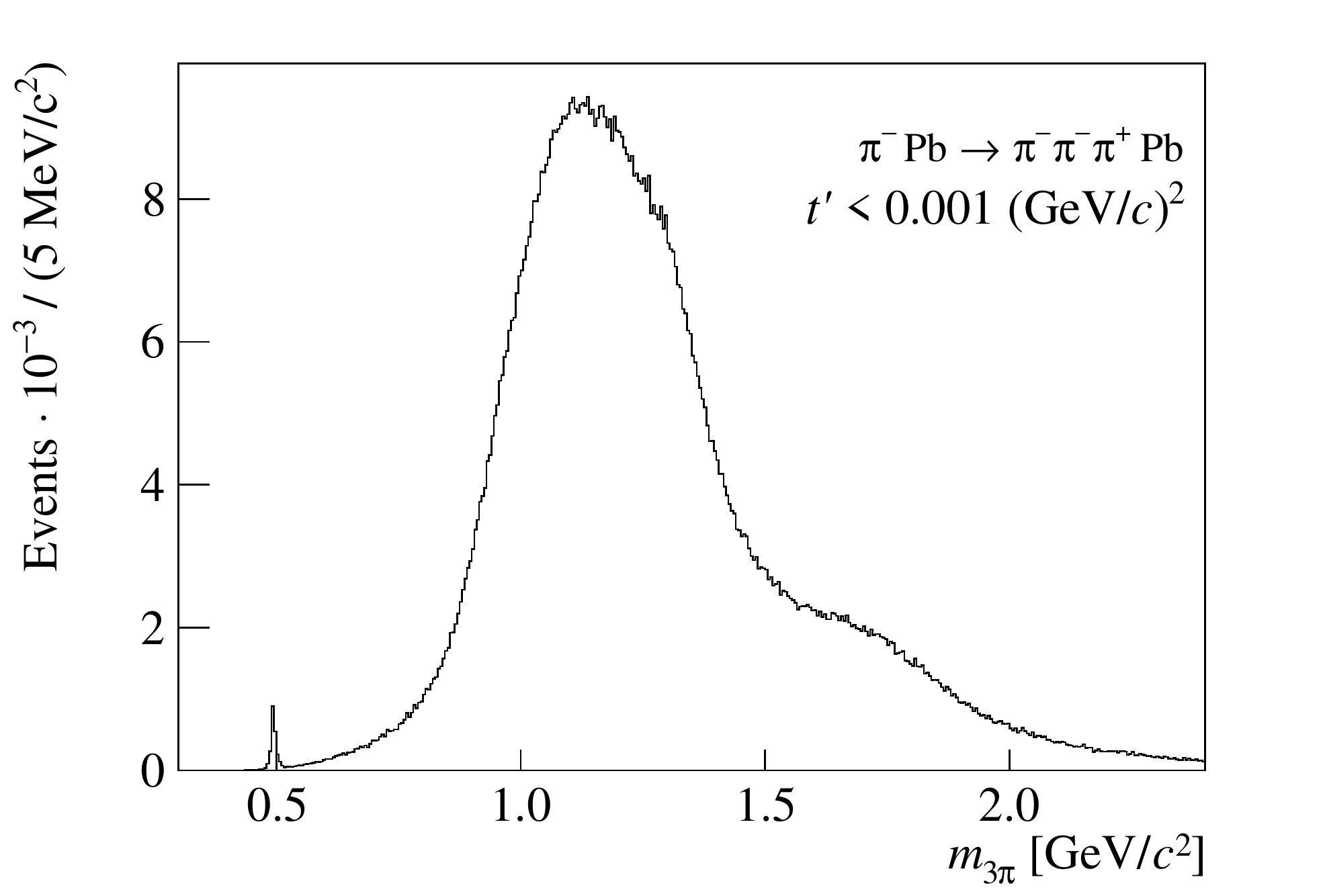}}
{\includegraphics[width=0.49\columnwidth]{fig3.pdf}}
\caption{Invariant mass spectrum of the $3\pi$ final-state events with
  $\tpGeV<0.001~\ttGeV$.  The sharp peak at $\mtpGeV\approx 0.493~\mmGeV$ originates from
  in-flight decays of beam kaons into the investigated final state. }
\label{fig:mass_prim}
\end{center}
\end{figure}

For the present data analysis, events are required to have exactly three charged
outgoing particles with charge signature $(- - +)$. These are assumed to be
pions. A common vertex fit between these particles and the incoming beam
particle must be consistent with an interaction in the lead target as indicated
in fig.~\ref{fig:vertices_prim}.  The summed energy of the three outgoing pions
$E_{3\pi}$ needs to match the mean beam energy within $\pm 4~\textrm{GeV}$ to
assure an exclusive $\pi^- \textrm{Pb} \rightarrow \pi^- \pi^- \pi^+
{\textrm{Pb}}$ reaction.  About 1~million events were recorded with
$\tpGeV<0.001~\ttGeV$, {\it i.e.\ }in the Primakoff $t'$ region.  Their invariant
$3\pi$ mass spectrum is shown in fig.~\ref{fig:mass_prim}, where the main
contributions from diffractive production of the $a_1(1260)$ and $\pi_2(1670)$
resonances are clearly visible.  The low-mass region $\mtpGeV< 0.72~\mmGeV$ has been
the focus of the measurement of chiral dynamics using the same data set
\cite{compass_3pichpt}.  The small peak at $\mtpGeV\approx 0.493~\mmGeV$ originates
from the in-flight decays of beam kaons into $\pi^-\pi^-\pi^+$.

\begin{figure}
\ifthenelse{\equal{\EPJSTYLE}{yes}}
{\includegraphics[width=0.99\columnwidth]{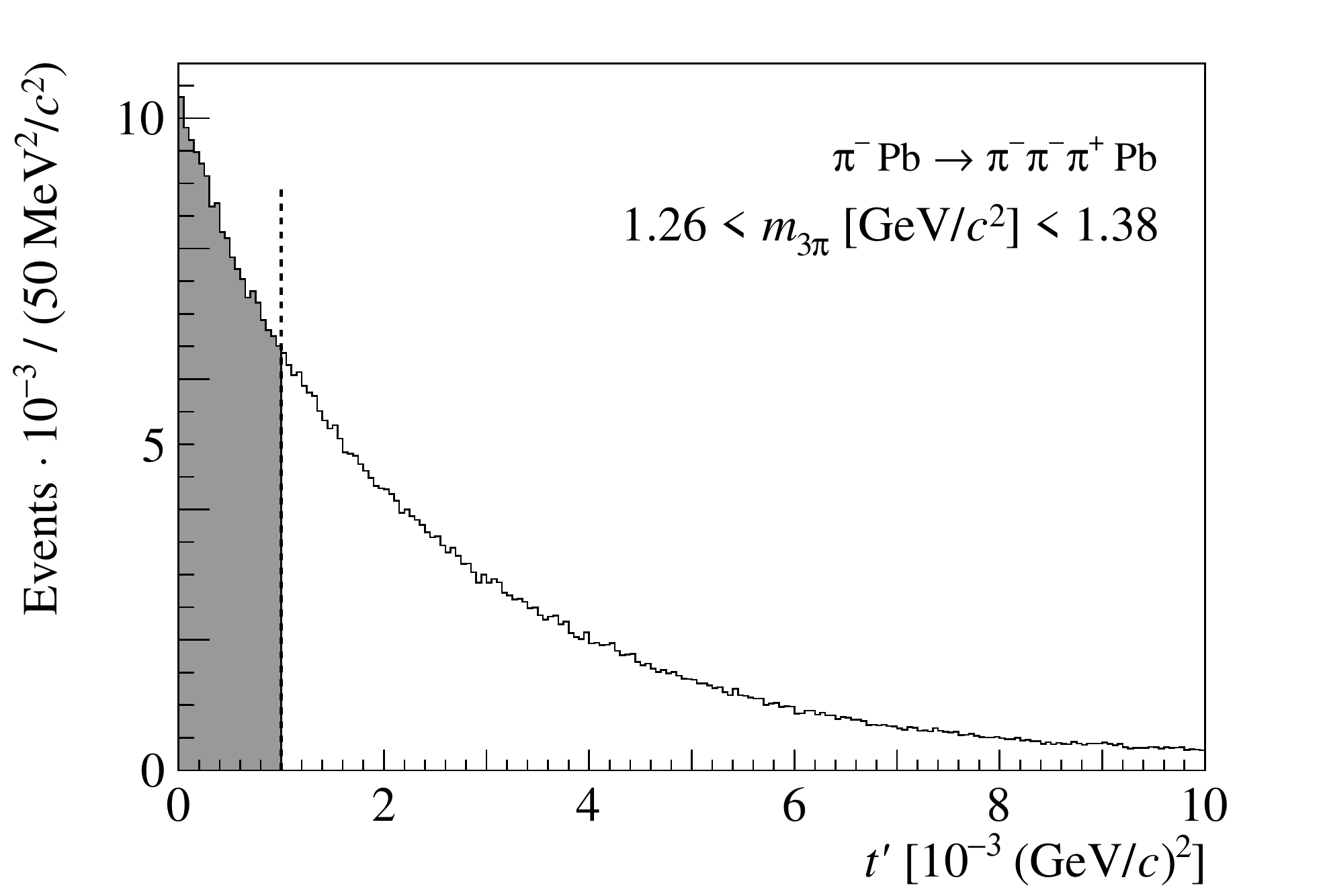}\\
 {\centerline{(a)}}\\
\includegraphics[width=0.99\columnwidth]{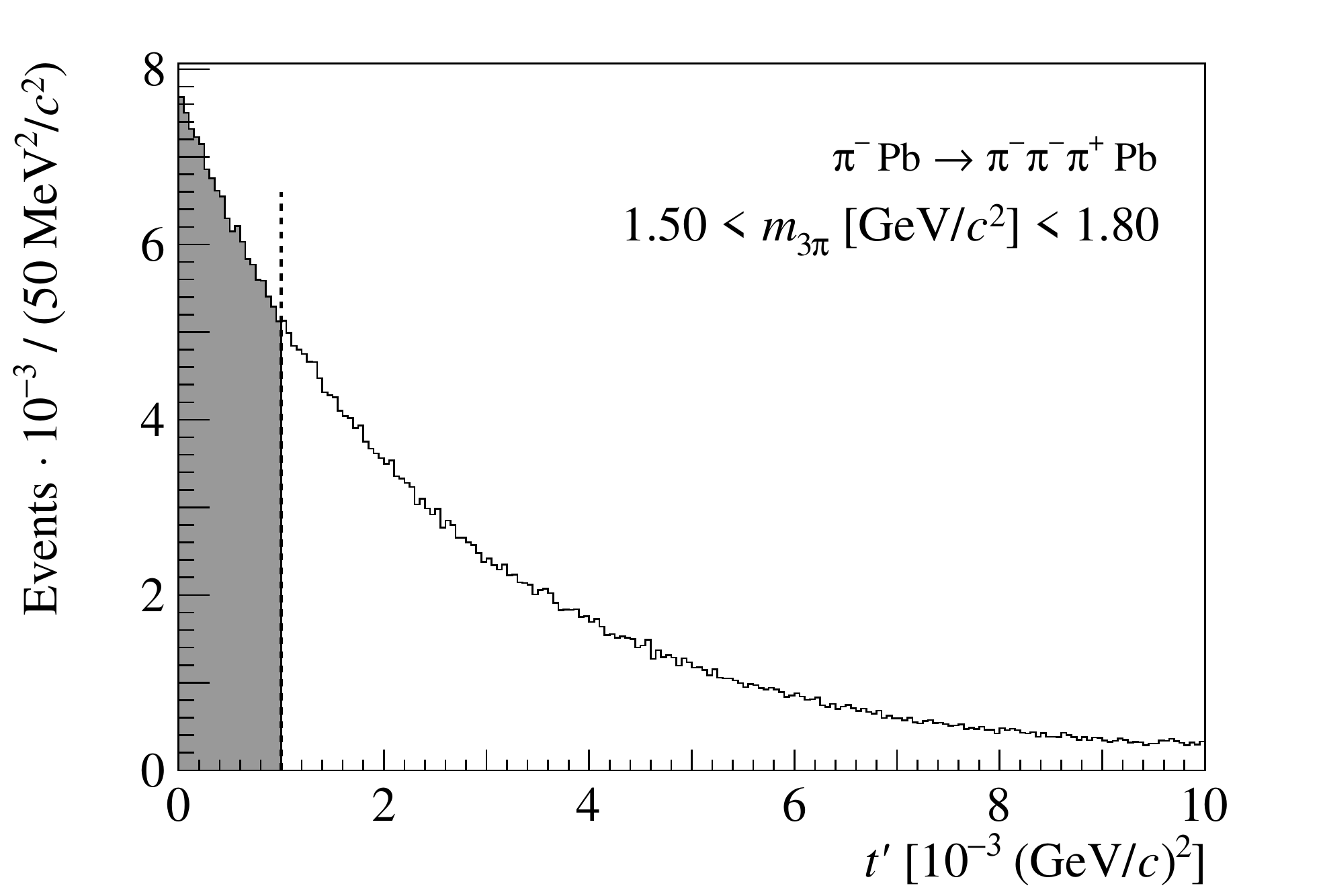} \\
 {\centerline{(b)}} }
{ \includegraphics[width=0.49\columnwidth]{fig4a.pdf} 
\includegraphics[width=0.49\columnwidth]{fig4b.pdf} \\
 {\centerline{(a) \hspace{.47\columnwidth} (b)}} }
\caption{Momentum transfer distributions in the investigated mass regions
  containing (a) the $a_2(1320)$ or (b) the $\pi_2(1670)$. The Primakoff region,
  $\tpGeV<0.001~\ttGeV$, is highlighted. }
\label{fig:tprime_a2reg_pi2reg}
\end{figure}

For the extraction of the resonant components contained in this mass spectrum, a
partial-wave analysis (PWA) is carried out as summarised in
sect.~\ref{sec:prim_res:pwa_lowt}.  The specific features of a PWA at very low
$t'$ are summarised in sect.~\ref{sec:prim_res:resol}, followed by the
presentation of the Primakoff production of $a_2(1320)$ and $\pi_2(1670)$ in
sect.~\ref{sec:prim_res:a2_pi2}.  The momentum transfer distributions for the
investigated mass regions, {\it i.e.} $1.26\ \mmGeV < \mtpGeV < 1.38\ \mmGeV$ containing the
$a_2(1320)$ and $1.50\ \mmGeV < \mtpGeV < 1.80\ \mmGeV$ containing the $\pi_2(1670)$, are
presented in fig.~\ref{fig:tprime_a2reg_pi2reg}. A sharp increase is observed
with $t' \rightarrow 0$, where the Primakoff process contributes in addition to
the dominant diffractive production. These figures demonstrate the necessity of
special methods to extract the Primakoff process.

\subsection{Partial-wave analysis of the $\pi^-\pi^-\pi^+$ system at very low $t'$}
\label{sec:prim_res:pwa_lowt}
In the first step of the partial-wave analysis, the data are divided into bins
of the invariant three-pion mass that in the following is denoted by $m$.  The
experimentally observed cross section $\Delta\sigma_{m}(\tau, t')$, in terms of
acceptance-corrected intensity, is parameterised by

\begin{align}
\ifthenelse{\equal{\EPJSTYLE}{yes}}
{
& \Delta\sigma_{m}(\tau, t') = \frac{1}{L\cdot  \epsilon_X\cdot \epsilon_{\textrm{resol}}} \cdot\nonumber \\
& \sum\limits_{\epsilon = \pm 1}\,\sum\limits_{r=1}^{N_{r}} \left\vert \sum\limits_{i}\;T_{ir}^{\epsilon}(m)\;\overline{f}_{i}^{\epsilon}(t', m) \,\overline{\psi}_{i}^{\epsilon}(\tau, m)
\right\vert ^{2} \ .
}{
 \Delta\sigma_{m}(\tau, t')
=\frac{1}{L\cdot  \epsilon_X\cdot \epsilon_{\textrm{resol}}} \cdot
\sum\limits_{\epsilon = \pm 1}\,\sum\limits_{r=1}^{N_{r}} \left\vert \sum\limits_{i}\;T_{ir}^{\epsilon}(m)\;\overline{f}_{i}^{\epsilon}(t', m) \,\overline{\psi}_{i}^{\epsilon}(\tau, m)
\right\vert ^{2} \ .
}
\label{eq:pwa:xsec_mindep}
\end{align}
The symbol $\overline{\psi}_{i}^{\epsilon}(\tau, m)$ denotes the normalised
decay amplitude of a particular partial wave $i$, depending only on $\tau$
within the mass bin. Here, $\tau$ is the vector of the independent phase-space
variables that parameterise the final-state kinematics, {\it i.e.} 5-dimensional
for a three-body final state. The normalisation of the decay amplitude is chosen
such that the integral of the amplitude squared over the full phase space is
equal to 1.  Each decay amplitude is multiplied by its corresponding $t'$
dependence $\overline{f}_{i}^{\epsilon}(t', m)$:

\begin{align}
\ifthenelse{\equal{\EPJSTYLE}{yes}}
{
\overline{\psi}_{i}^{\epsilon}(\tau, m) = \frac{\psi_{i}^{\epsilon}(\tau, m)} {\sqrt{\int \left\vert \psi_{i}^{\epsilon}(\tau, m) \right\vert^{2} \text{d} \Phi(\tau)}} \nonumber \\
\textrm{and} \quad
\overline{f}_{i}^{\epsilon}(t', m) = \frac{f_{i}^{\epsilon}(t', m)} {\sqrt{\int \left\vert f_{i}^{\epsilon}(t', m) \right\vert^{2} \text{d} t'}} \ .
}{
\overline{\psi}_{i}^{\epsilon}(\tau, m) = \frac{\psi_{i}^{\epsilon}(\tau, m)} {\sqrt{\int \left\vert \psi_{i}^{\epsilon}(\tau, m) \right\vert^{2} \text{d} \Phi(\tau)}} \quad \textrm{and} \quad
\overline{f}_{i}^{\epsilon}(t', m) = \frac{f_{i}^{\epsilon}(t', m)} {\sqrt{\int \left\vert f_{i}^{\epsilon}(t', m) \right\vert^{2} \text{d} t'}} \ .
}
\label{eq:pwa:norm_amp}
\end{align}
The $t'$ dependences are either following the experimental data or are obtained
from a dedicated Monte Carlo study as explained later. At this stage, also
resolution effects of the spectrometer are taken into account.  The
complex-valued numbers $T_{ir}^{\epsilon}(m)$ in eq.~\eqref{eq:pwa:xsec_mindep}
are the transition amplitudes that represent the strengths of the individual
amplitudes $i$ and their phases.  They are assumed to be constant within each
mass bin so that $\Delta \sigma_m$ depends only on the phase-space parameter
vector $\tau$.  The parameterisation of the cross section is optimised using
$T_{ir}^{\epsilon}$ as fitting parameters in an extended maximum-likelihood fit,
taking into account the geometrical acceptance of the spectrometer obtained from
a dedicated Monte Carlo simulation as described in
appendix~\ref{sec:app:likelihood}.  Since the PWA is performed in bins of mass
$m$ or momentum transfer $t'$, respectively, resolution effects in these
variables are not unfolded by the employed acceptance correction.

The decay amplitudes $\psi_{i}^{\epsilon}(\tau, m)$ of the three-pion final
states are parameterised in the isobar model by subsequent two-particle decays,
{\it i.e.} the three-pion resonance decays first into a single $\pi^-$ and a
di-pion resonance, referred to as the isobar in the following, which decays
further into a $\pi^+\pi^-$ pair. The amplitudes are given in the
Gottfried-Jackson reference system \cite{gottfried_jackson,hansen_3mesons} and
denoted as $J^{PC}M^{\epsilon}\textrm{\{isobar\}}[L] \pi$, giving the quantum
numbers of the three-pion resonance $J^{PC}$, its spin projection $M$ onto the
beam axis, its reflectivity $\epsilon$, the isobar, and the angular momentum $L$
between the isobar and the unpaired $\pi^-$.  The amplitudes are
Bose-symmetrised in the two $\pi^{-}$.

The reflectivity $\epsilon=\pm 1$ describes the symmetry or antisymmetry of the
decay amplitude under a reflection through the production plane.  In the
so-called reflectivity basis the amplitudes have the quantum numbers $\epsilon$
and $M \geq 0$ \cite{chung_technique2}.  They are combinations of the two
amplitudes with the customary quantum numbers $+M$ and $-M$. Parity conservation
demands that the two contributions $\epsilon=\pm 1$ are added incoherently.
Natural parity of the exchange particle holds for the photon (with total spin
and parity $J^P = 1^-$) and the pomeron (Regge trajectory with $P = (-1)^{J}$),
and this leads to the expectation of observing only $\epsilon = +1$.  The
assumption of natural parity exchange leads to the appearance of $J^{PC}=2^{++}$
resonances only with $M=1$, while {\it e.g.\ }for $J^{PC}=2^{-+}$ resonances
both $M=0$ and $M=1$ are allowed.
 
The rank $N_{r}$ introduces the number of independent sets of coherent
amplitudes.  Choosing $N_r>1$ allows effectively for incoherence between
contributing partial waves as expected {\it e.g.\ }for different helicity final
states of the unobserved recoil particle. However, in the kinematic range under
investigation we do not expect this to play a role as we expect coherent
scattering on the whole nucleus, and thus set $N_r=1$. Nevertheless, apparent
incoherence effects occur due to resolution. These are taken into account by the
partial-coherence concept that is explained in appendix~\ref{sec:app:part_coh}.

The physical parameters are extracted from the spin-density matrix
\begin{equation}
  \rho_{ij}^{\epsilon}=\sum\limits_{r}T_{ir}^{\epsilon}T_{jr}^{\epsilon *} \; .
  \label{eq:pwa:spin_density}
\end{equation}
In particular, its diagonal elements determine the intensities $\mathcal{I}_i$
of the specific amplitudes $i$, and the relative phases $\varphi_{ij}$ between
two amplitudes $i$ and $j$ are contained in the non-diagonal elements ($i \neq
j$):
\begin{equation}
  \mathcal{I}^{\epsilon}_{i} = \rho_{ii}^{\epsilon}
  \quad \textrm{and} \quad
  \rho_{ij}^{\epsilon} = \vert \rho_{ij}^{\epsilon} \vert \,\textrm{e}^{i\,\varphi_{ij}^{\epsilon}} \; .
  \label{eq:pwa:int_phase}
\end{equation}
A partial-wave analysis of data covering only the very low momentum transfer
$t'$, as carried out here for the extraction of Primakoff contributions, has two
particular features in addition to the resolution effects that are discussed
later.

First, there are two production mechanisms contributing at $t' \approx 0$,
diffractive and Primakoff production.  They can be distinguished by the spin-projection $M$ of
the produced system.  The $t'$ dependence of the cross section for diffractively
produced states with spin-projection $M$ is given \cite{perl} by:
\begin{equation}
\textrm{d}\sigma /\textrm{d}t' \propto t'^{M} \textrm{e}^{-b(m)t'} \quad (\textrm{with} \ M\geq0), 
\label{eq:spec:diff_xsec_tM}
\end{equation}
where $b(m)$ is the slope that depends on the mass $m$ of the produced system as
well as on the size of the target nucleus. Thus for events at lowest momentum
transfer $\tpGeV<0.001~\ttGeV$, only intermediate states with $M=0$ are produced
diffractively, while diffractive production with $M=1$ is expected to be
negligible.  Primakoff production populates intermediate states with $M=1$ as
the helicities of quasi-real photons are $\lambda_{\gamma} = \pm 1$.  The spin
projection $M=0$ is suppressed for quasi-real photons of very small virtuality.
Following the assumptions listed above, the $t'$ dependences
$f_{i}^{\epsilon}(t', m)$ will follow a pure diffractive behaviour for $M=0$
amplitudes and the pure Primakoff shape folded with the experimental resolution
in case of $M=1$ (see sect.~\ref{sec:prim_res:resol}).

Secondly, in addition to the (isobaric) decays of resonances there are
non-resonant scattering processes populating the same final state.  In the case
of quasi-real photon exchange and for the low-mass region, these processes can
be calculated in Chiral Perturbation Theory (ChPT) \cite{kaiser, kaiser2}. This
was implemented as special amplitudes to the PWA, and the tree-level
calculations were probed successfully up to $\mtpGeV\leq 0.72~\mmGeV$, see
ref.\cite{compass_3pichpt}.  Higher-order ChPT calculations include loops and
$\rho$ contributions. They are expected to describe further non-resonant
contributions at higher masses, and thus are used for the present analysis in
addition to the chiral amplitude used in the low-mass analysis
\cite{compass_3pichpt} (see appendices~\ref{sec:appendix:chpt},
\ref{sec:appendix:waveset}, and table~\ref{tab:waveset}).

\subsection{Features of $t'$ spectra at values of very low $t'$ and resolution effects}
\label{sec:prim_res:resol}
\begin{figure}
\begin{center}
\ifthenelse{\equal{\EPJSTYLE}{yes}}
{\includegraphics[width=0.99\columnwidth]{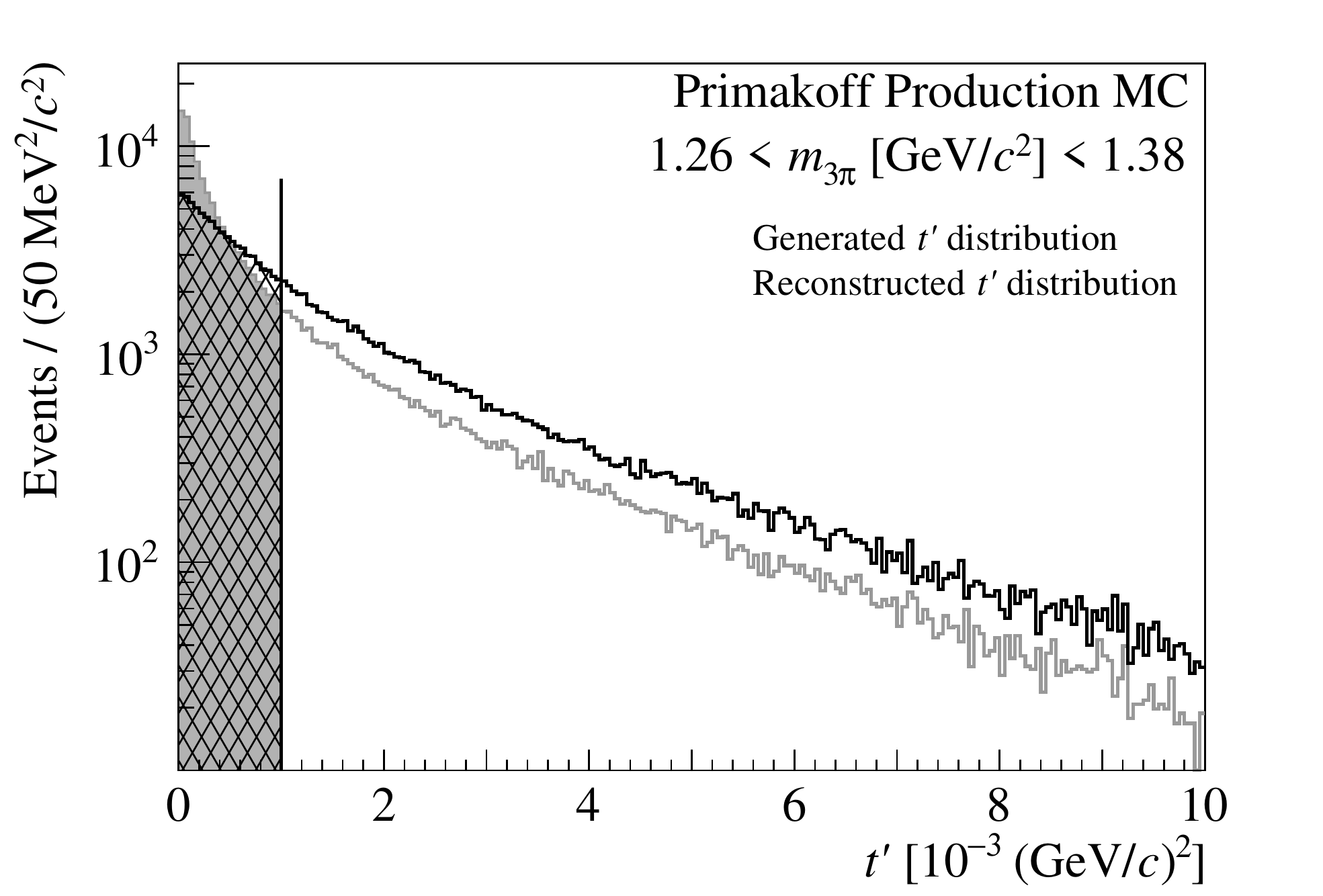}}
{\includegraphics[width=0.49\columnwidth]{fig5.pdf}}
\caption{Illustration of the migration of events due to the limited resolution
  of the spectrometer in $t'$. The simulated events follow the cross section
  given by eq.~\eqref{eq:intro:WW_xsec} with the form factor introduced in
  sect.~\ref{sec:intro}, as given by refs.~\cite{faeldt_2009_2013cor} and
  \cite{faeldt_2010}. For details see text.} 
\label{fig:tprime_mig_a2}
\end{center}
\end{figure}
Due to the high energy of the incoming beam, the outgoing particles are strongly
boosted in the forward direction, and the opening angles between the decay
particles of the $\pi^-\pi^-\pi^+$ final state are small.  At very low momentum
transfer $t' \approx 0$, the scattering angle between the incoming pion and the
produced resonance is extremely small, making the measurement sensitive to
resolution effects.

The impact of the finite $t'$ resolution was studied using a dedicated Monte
Carlo (MC) simulation.  Events generated with a $t'$ dependence according to
eq.~\eqref{eq:intro:WW_xsec} have been processed using the full chain of the
simulation and reconstruction software.  The distribution of both the generated
and the reconstructed values of $t'$ of those events that pass all analysis cuts
are depicted in fig.~\ref{fig:tprime_mig_a2} for the mass window around the mass
of the $a_2(1320)$ as an example.  The original shape of the cross section with
a sharp peak at $t_{\textrm{min}}$ is modified significantly.  For
$\tpGeV<0.001~\ttGeV$, an approximately exponential behaviour $\textrm{d}
\sigma_{\textrm{rec, Primakoff}}/\textrm{d}t' \propto \exp
(-b_{\textrm{prim}}(m) t')$ is observed for
the reconstructed Primakoff MC events.  The slope parameter
$\bpGeV(m)$ was found to change from $\bpGeV(0.5~\mmGeV) \approx 1500\ \tmGeV$ to
$\bpGeV(2.5~\mmGeV) \approx 700\ \tmGeV$.  This experimentally expected $t'$ dependence is
imposed on the Primakoff amplitudes in the PWA by the factor
$\overline{f}_{i}^{\epsilon}(t', m)$ in eq.~\eqref{eq:pwa:xsec_mindep} with
$f_{i}^{\epsilon}(t', m) = \exp(- \frac{1}{2}b_{\textrm{prim}}(m) t')$ for all
amplitudes with $M=1$. In addition, the rescaling factor
$\epsilon_{\textrm{resol}}$ in eq.~\eqref{eq:intro:radwidth_from_data}, which
takes into account the migration of events above or below the upper $t'$ limit,
is estimated from this study. This rescaling factor complements the calculation
of $C_X$ used in eq.~\eqref{eq:intro:radwidth_from_data} when integrating
eq.~\eqref{eq:intro:xsec_primakoff_resonance} over $t'$ as shown in
eq.~\eqref{eq:intro:xsec_radwidth}.  Figure~\ref{fig:tprime_mig_a2} depicts the
distribution following the cross section not containing resolution effects in
grey and the experimentally expected distribution marked in black. Both
histograms are based on the properties of fully reconstructed events only, as
the detection and reconstruction efficiencies are expected to be taken care of
by the acceptance correction of the PWA, which is denoted by $\epsilon_X$ in
eq.~\eqref{eq:intro:radwidth_from_data}. The rescaling factor
$\epsilon_{\textrm{resol}}$ is given by the ratio of the integrals of the grey
and the black-marked histograms in the indicated range $\tpGeV<0.001~\ttGeV$. It
results in $\epsilon_{\textrm{resol}} \approx 0.74$.

The experimental $t'$ dependence for the observed diffractive production was
determined by statistical subtraction. For this method, the diffractive
contribution was modelled by $\textrm{d}\sigma_{\textrm{diff}}/\textrm{d}t'
\propto \exp (-b_{\textrm{diff}} (m) t')$, due to the overall predominat
diffractive $M=0$ contribution in the data. The $t'$ distributions were modelled
as described above. The full data set was divided into mass bins and fitted by
the sum of these two contributions, with $\bdGeV(m)$ as a fit
parameter. The resulting trend from $\bdGeV(0.5~\mmGeV) \approx 420\ \tmGeV$ to
$\bdGeV(2.5~\mmGeV) 
\approx 320\ \tmGeV$ is used for the $t'$ dependence of the diffractive $M=0$ amplitudes
in the PWA, {\it i.e.\ }the $\overline{f}_{i}^{\epsilon}(t', m)$ in
eq.~\eqref{eq:pwa:xsec_mindep}.

An additional effect of the finite resolution at very low $t'$ stems from the
presence of two coherent production processes with very different $t'$
dependence. The finite resolution leads to a statistical mixing of events with
different $t'$ and thus to a partial loss of the coherence between the different
production amplitudes. This can be taken into account by setting the rank $N_r >
1$ in eq.~\eqref{eq:pwa:xsec_mindep}.  However, in the present analysis,
amplitudes with $M=0$ are observed to be coherent with respect to one another,
as are those with $M=1$. Thus $N_r = 1$ is actually chosen, while the reduced
coherence between these two sets of amplitudes due to resolution is taken into
account by the concept of partial coherence (see appendix~\ref{sec:app:part_coh}).

Furthermore, at very small $t'$ the production plane defined by the incoming
pion and the outgoing system $X$ is known with low precision at small scattering
angles.  In this case, the process is almost collinear so that the production
plane cannot be defined reliably and the contributions from $\epsilon = +1$ and
$\epsilon = -1$ are poorly distinguishable.  Thus at the limit of the extremely
small $t'$ observed for the photon peak, the full intensity of the physical
$\epsilon=+1$ amplitude is reconstructed with approximately equal amounts of
$\epsilon=+1$ and $\epsilon=-1$ contributions for each amplitude with $M=1$.
This introduces an artificial factor $\sqrt{2}$ in the amplitudes, which,
however, is not considered separately in the following.  The total intensity
observed is thus contained and conserved in the incoherent sum of these two
contributions as stated in eq.~\eqref{eq:pwa:xsec_mindep}.  This effect has been
reproduced in a dedicated Monte Carlo simulation, with data being generated with
amplitudes containing only positive reflectivity. Passing these data through the
standard simulation and reconstruction chain, the same amount of negative
reflectivity contributions appeared as in the experimental data.

\subsection{Primakoff production of $a_2(1320)$ and $\pi_2(1670)$}
\label{sec:prim_res:a2_pi2}
\ifthenelse{\equal{\EPJSTYLE}{yes}}
{
\begin{figure*}
\includegraphics[width=0.49\textwidth]{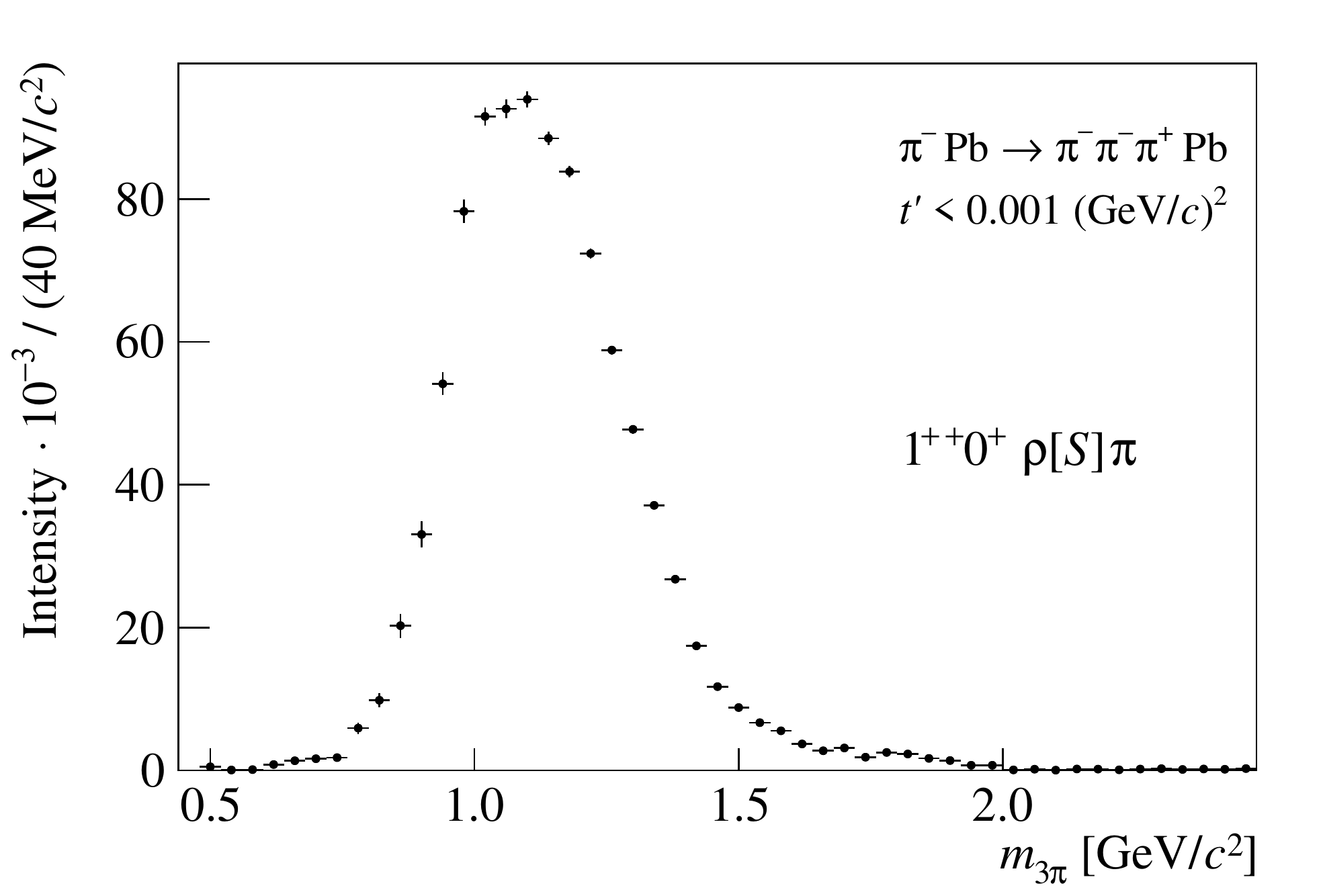}
\includegraphics[width=0.49\textwidth]{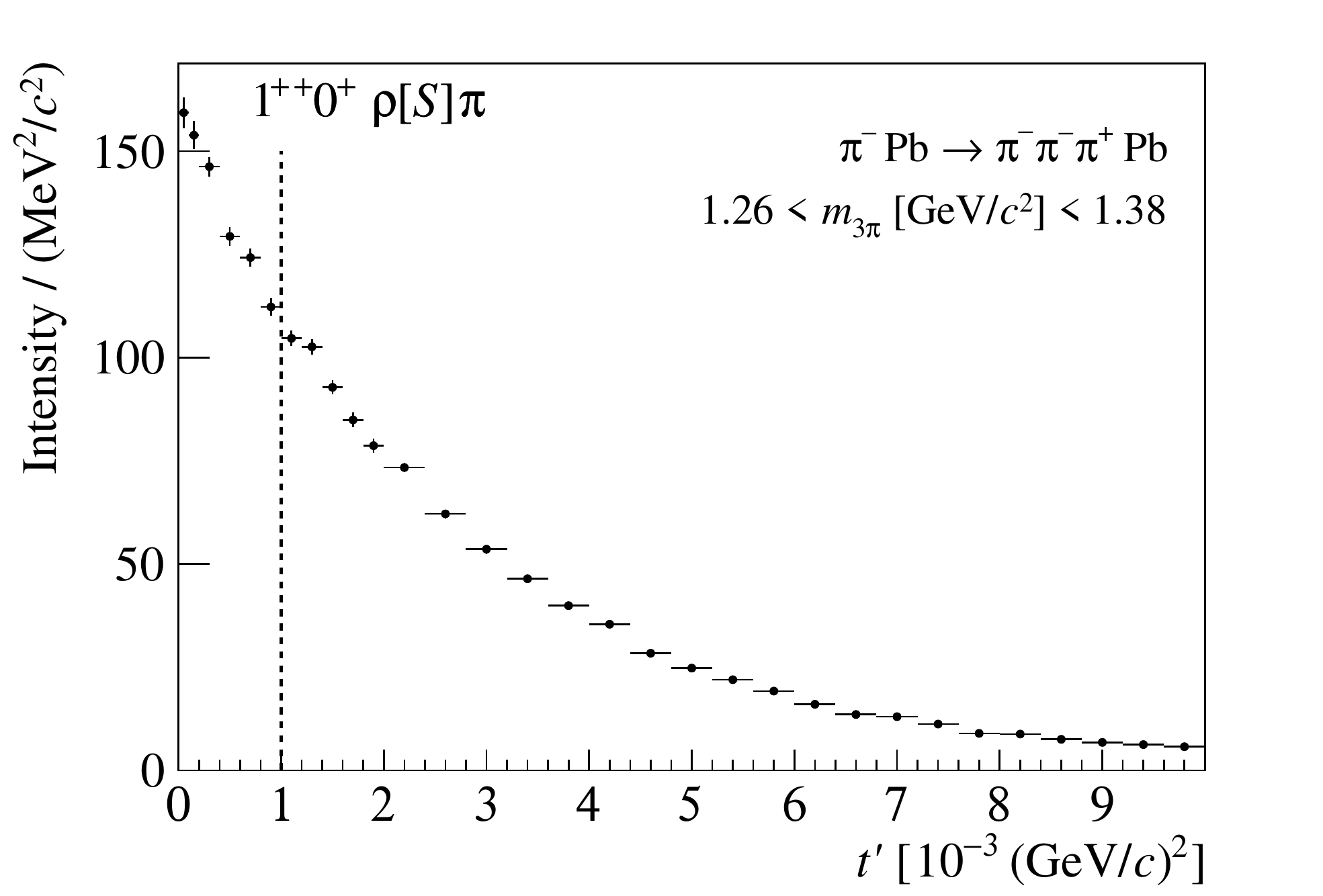}\\
 {\centerline{(a) \hspace{.47\textwidth} (d)}}\\
\includegraphics[width=0.49\textwidth]{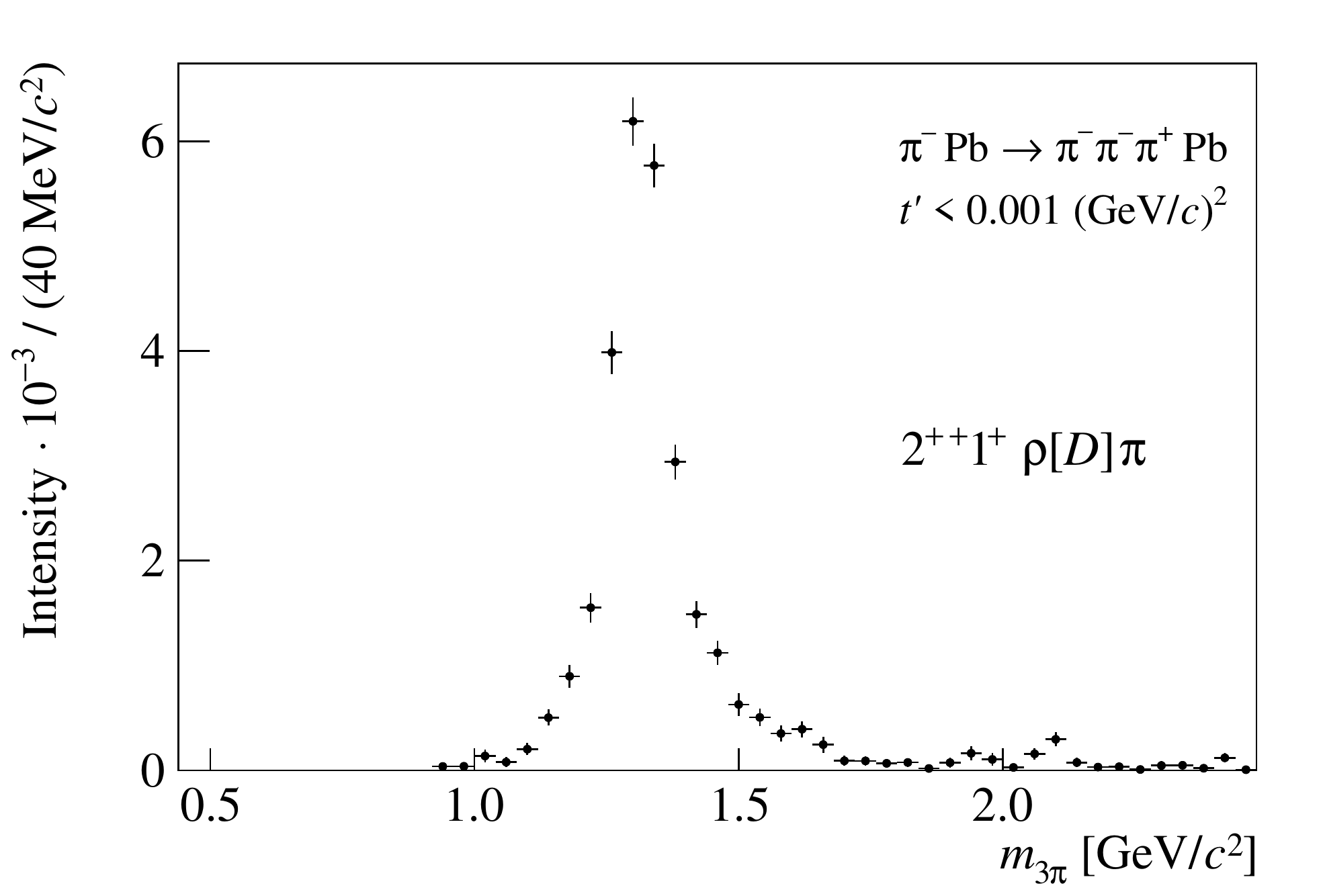}
\includegraphics[width=0.49\textwidth]{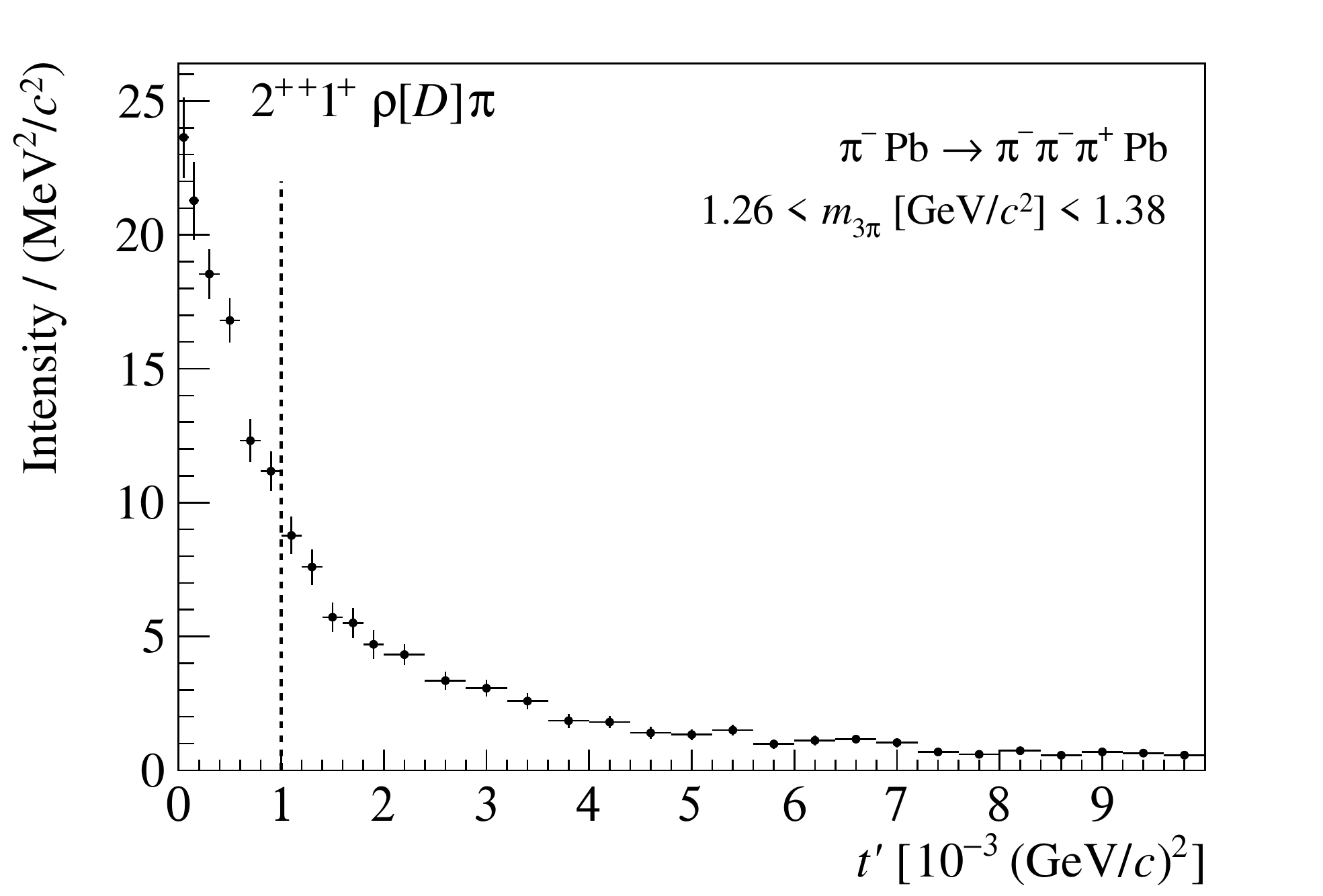}\\
 {\centerline{(b) \hspace{.47\textwidth} (e)}} \\
\includegraphics[width=0.49\textwidth]{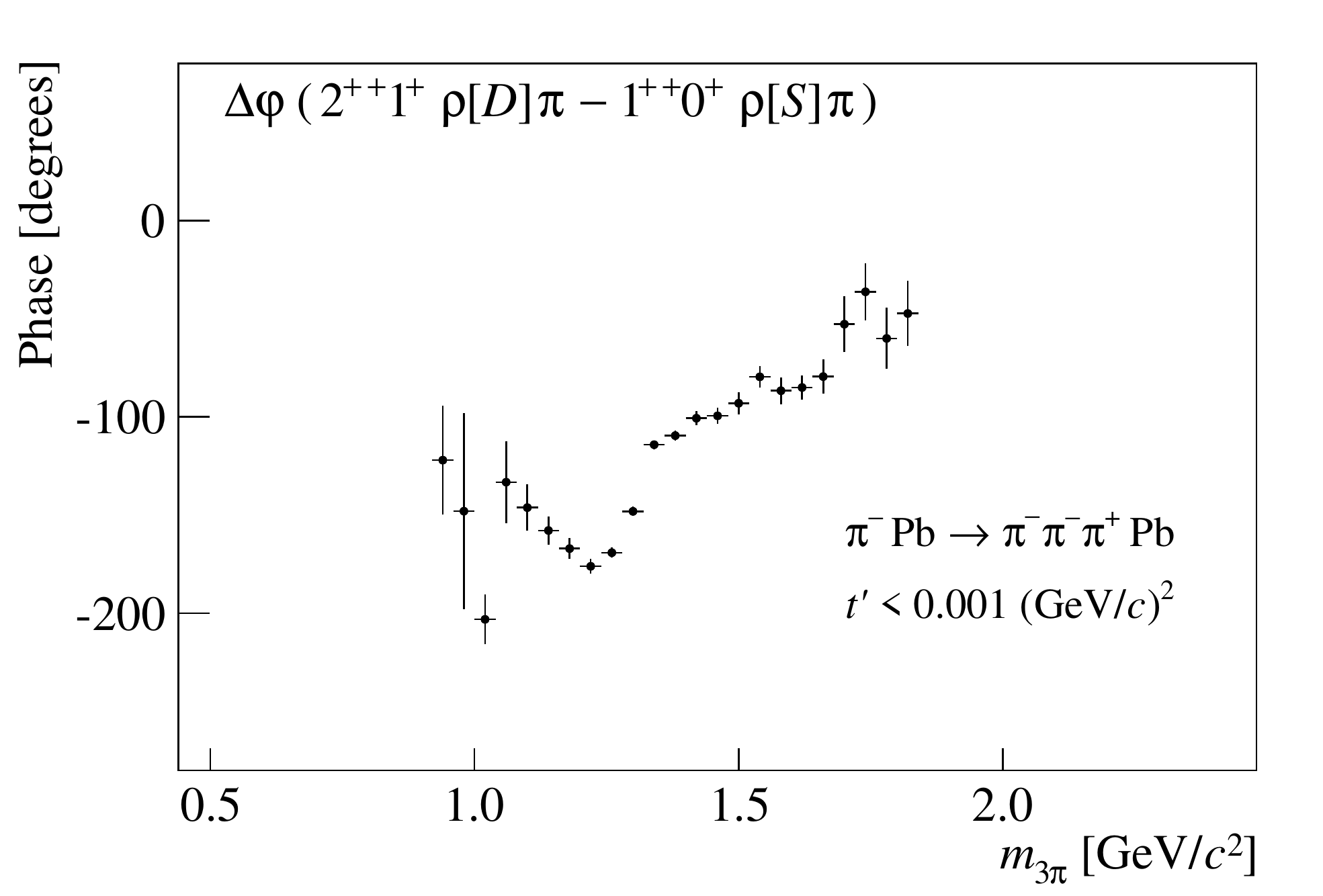}
\includegraphics[width=0.49\textwidth]{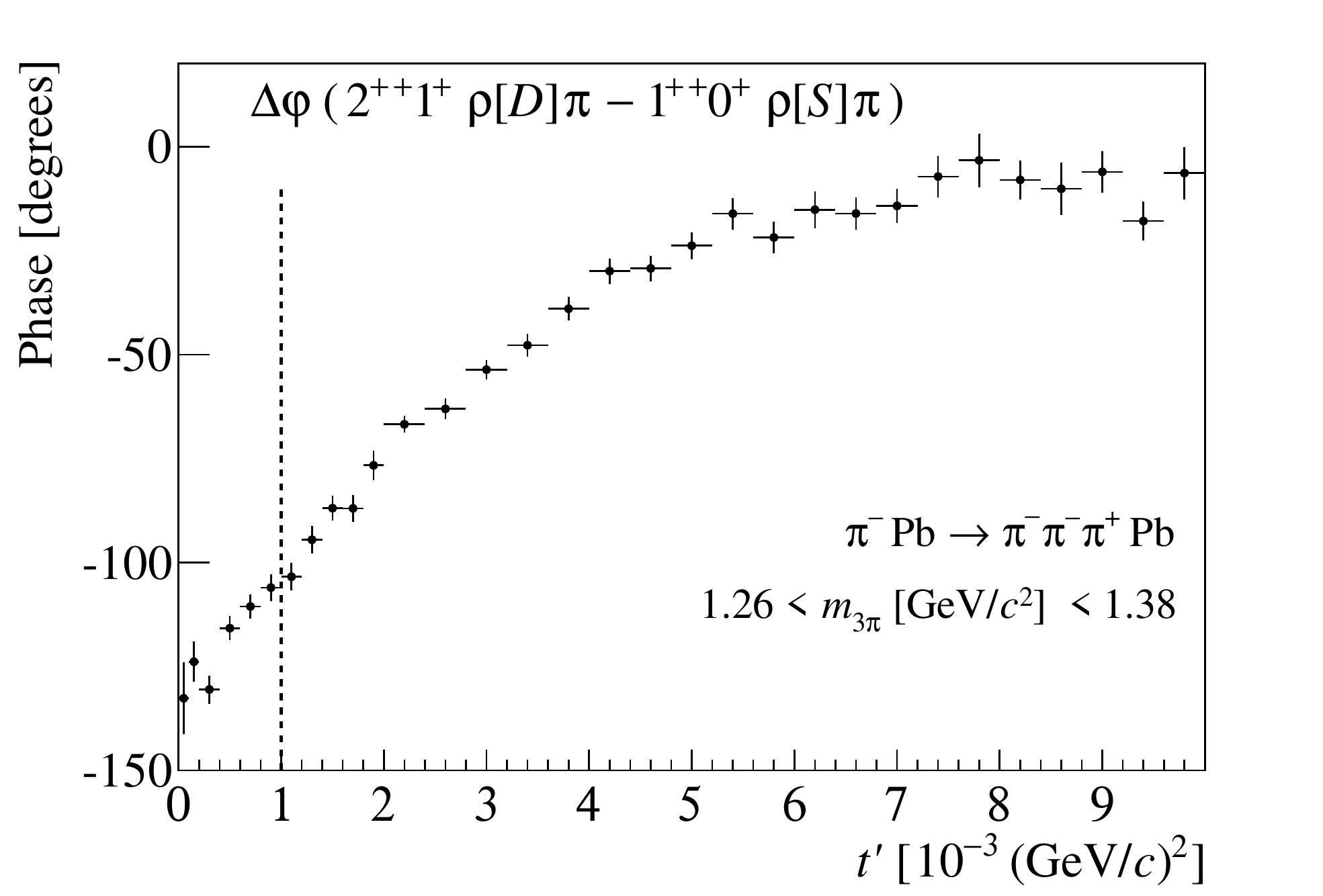} \\
 {\centerline{(c) \hspace{.47\textwidth} (f)}}
 }
 {
 \begin{figure}
\includegraphics[width=0.49\columnwidth]{fig6a.pdf}
\includegraphics[width=0.49\columnwidth]{fig6b.pdf}\\
 {\centerline{(a) \hspace{.47\columnwidth} (d)}}\\
\includegraphics[width=0.49\columnwidth]{fig6c.pdf}
\includegraphics[width=0.49\columnwidth]{fig6d.pdf}\\
 {\centerline{(b) \hspace{.47\columnwidth} (e)}} \\
\includegraphics[width=0.49\columnwidth]{fig6e.pdf}
\includegraphics[width=0.49\columnwidth]{fig6f.pdf} \\
 {\centerline{(c) \hspace{.47\columnwidth} (f)}}
 }
\caption{Intensities of $a_1(1260)$ (top), $a_2(1320)$ (middle) and their
  relative phase (bottom) in bins of three-pion mass (left) and $t'$ (right),
  depicting the Primakoff production of the $a_2(1320)$. For details see text.}
\label{fig:prim_a2}
\ifthenelse{\equal{\EPJSTYLE}{yes}}
{
\end{figure*}
}{
\end{figure}
}
\ifthenelse{\equal{\EPJSTYLE}{yes}}
{
\begin{figure*}
}{
\begin{figure}
}
\includegraphics[width=0.49\textwidth]{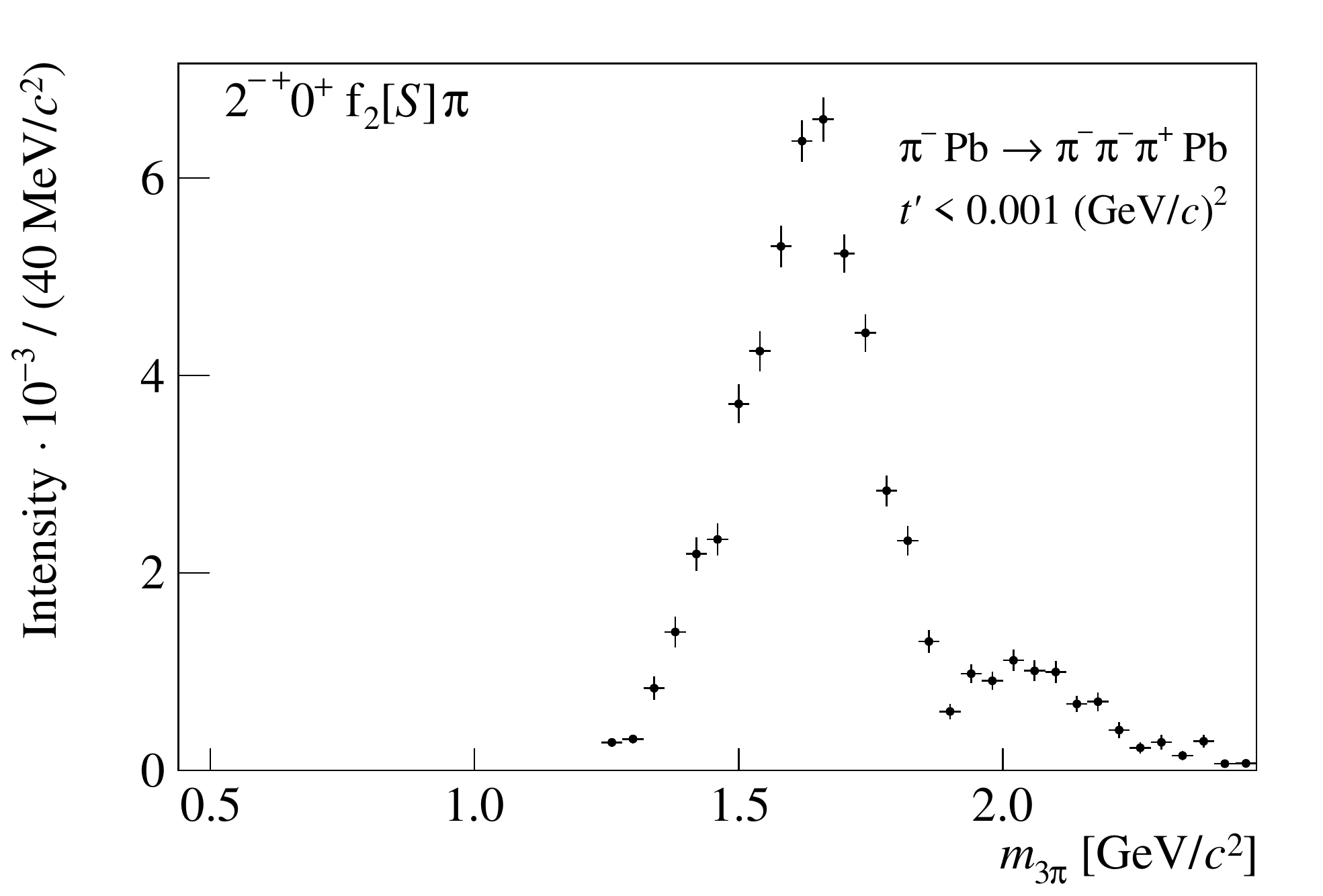}
\includegraphics[width=0.49\textwidth]{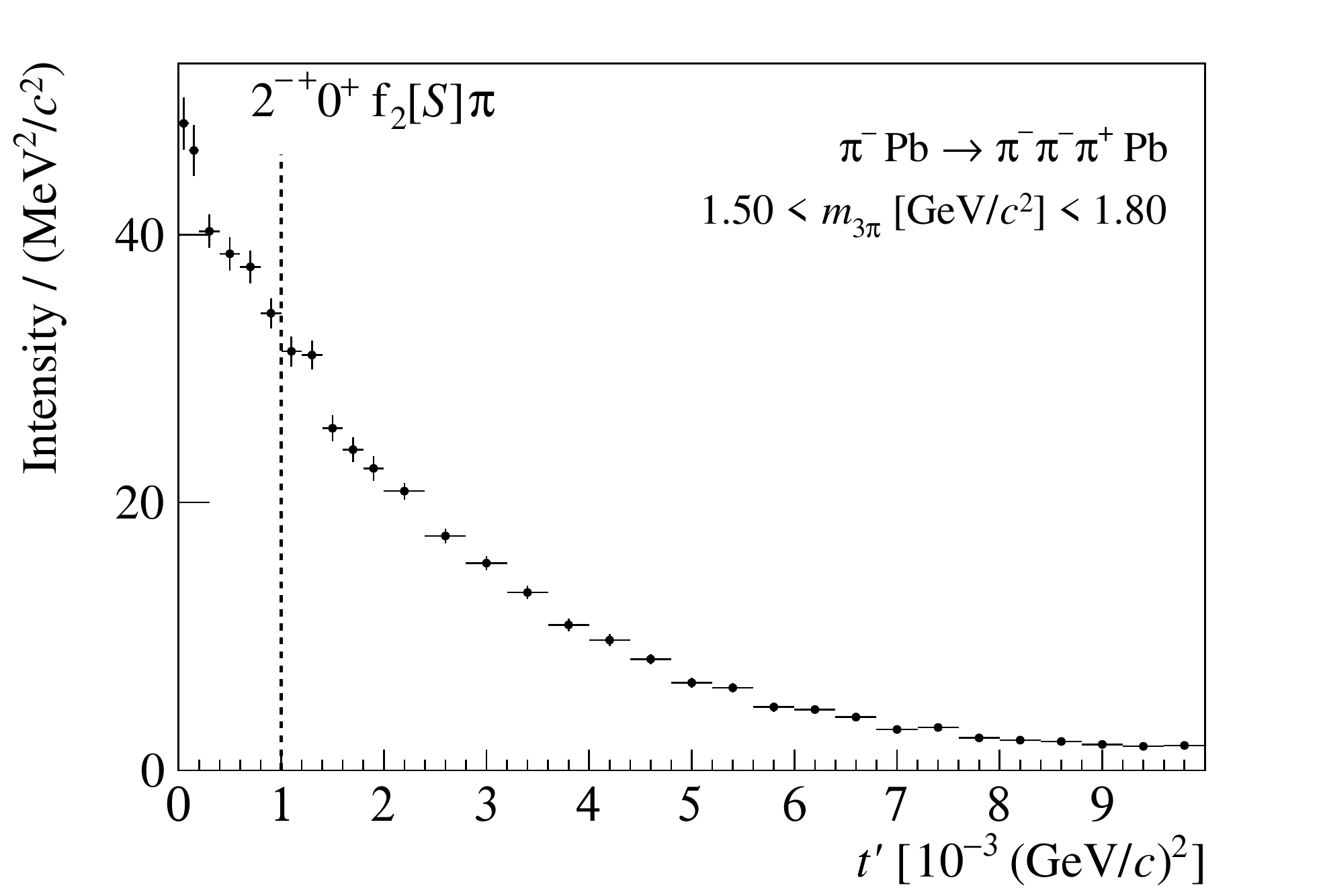}\\
 {\centerline{(a) \hspace{.47\textwidth} (d)}} \\
\includegraphics[width=0.49\textwidth]{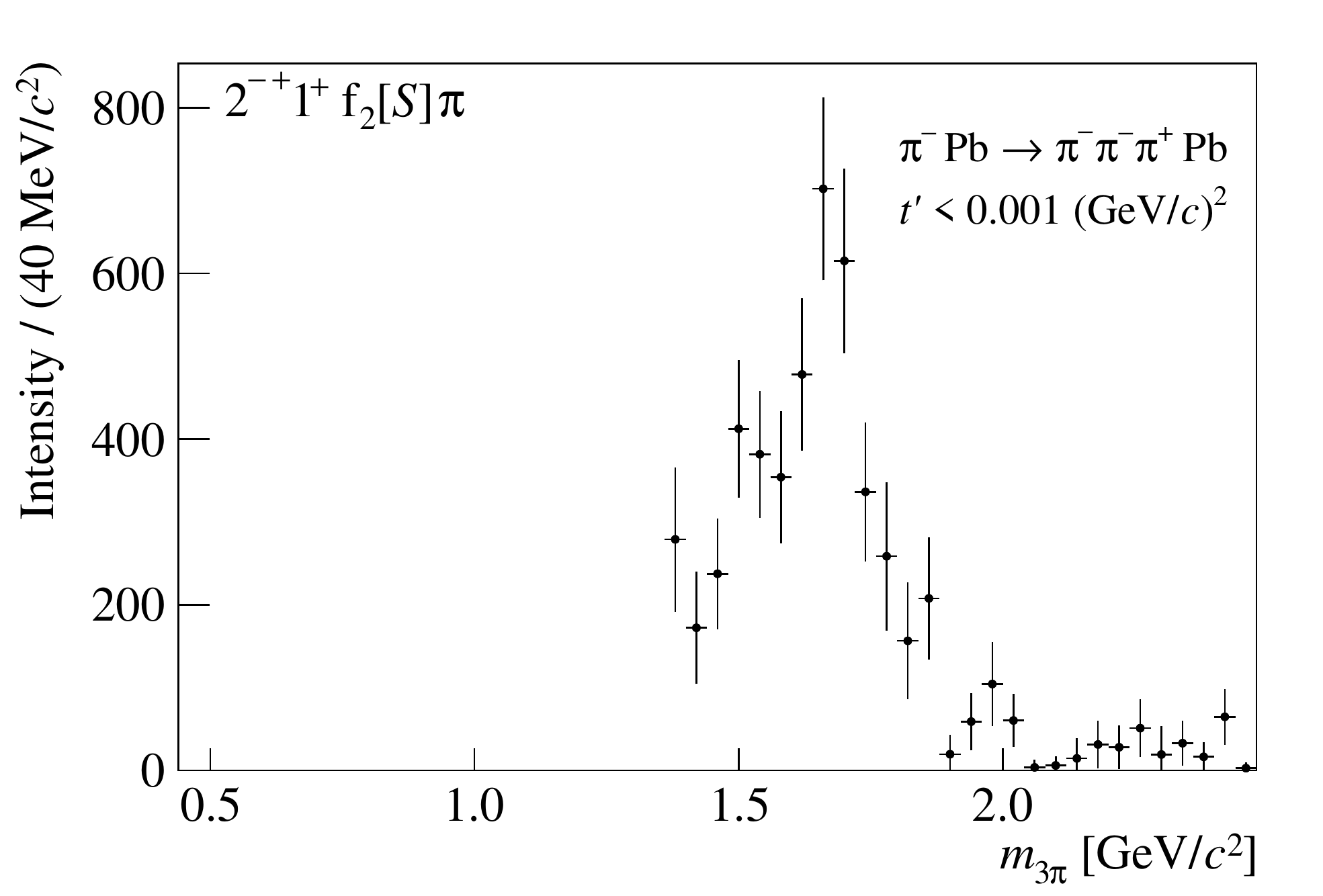}
\includegraphics[width=0.49\textwidth]{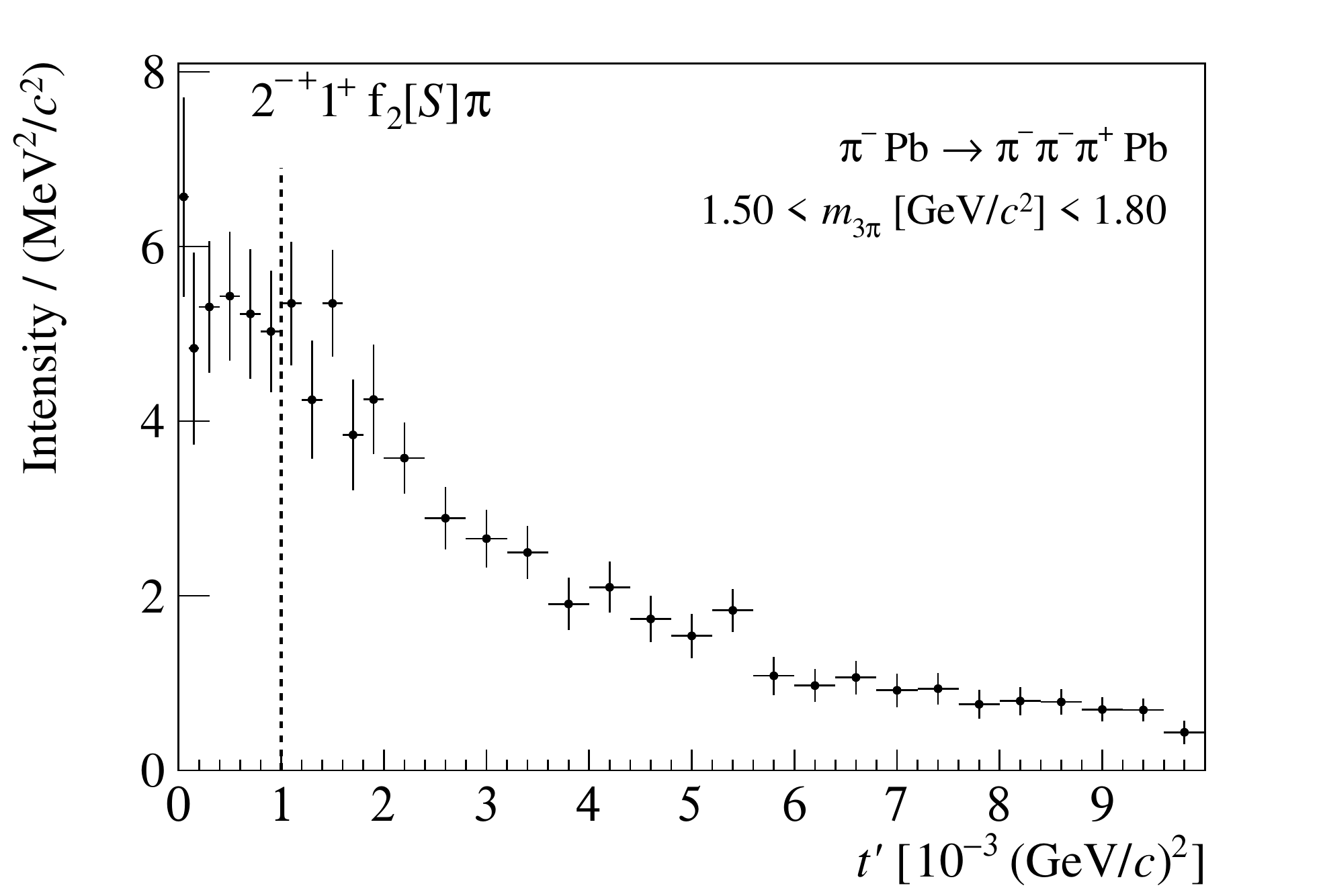}\\
 {\centerline{(b) \hspace{.47\textwidth} (e)}} \\
\includegraphics[width=0.49\textwidth]{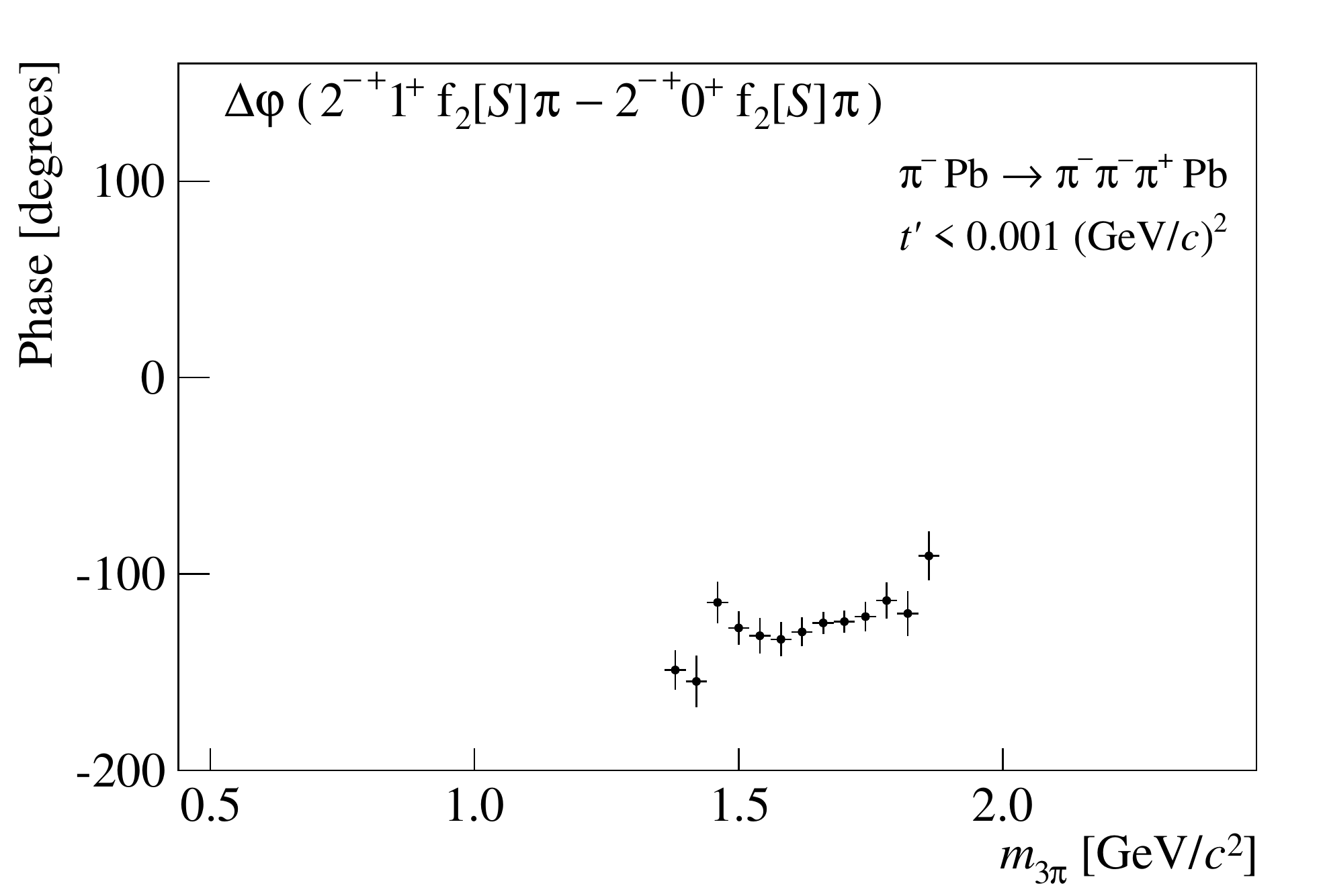}
\includegraphics[width=0.49\textwidth]{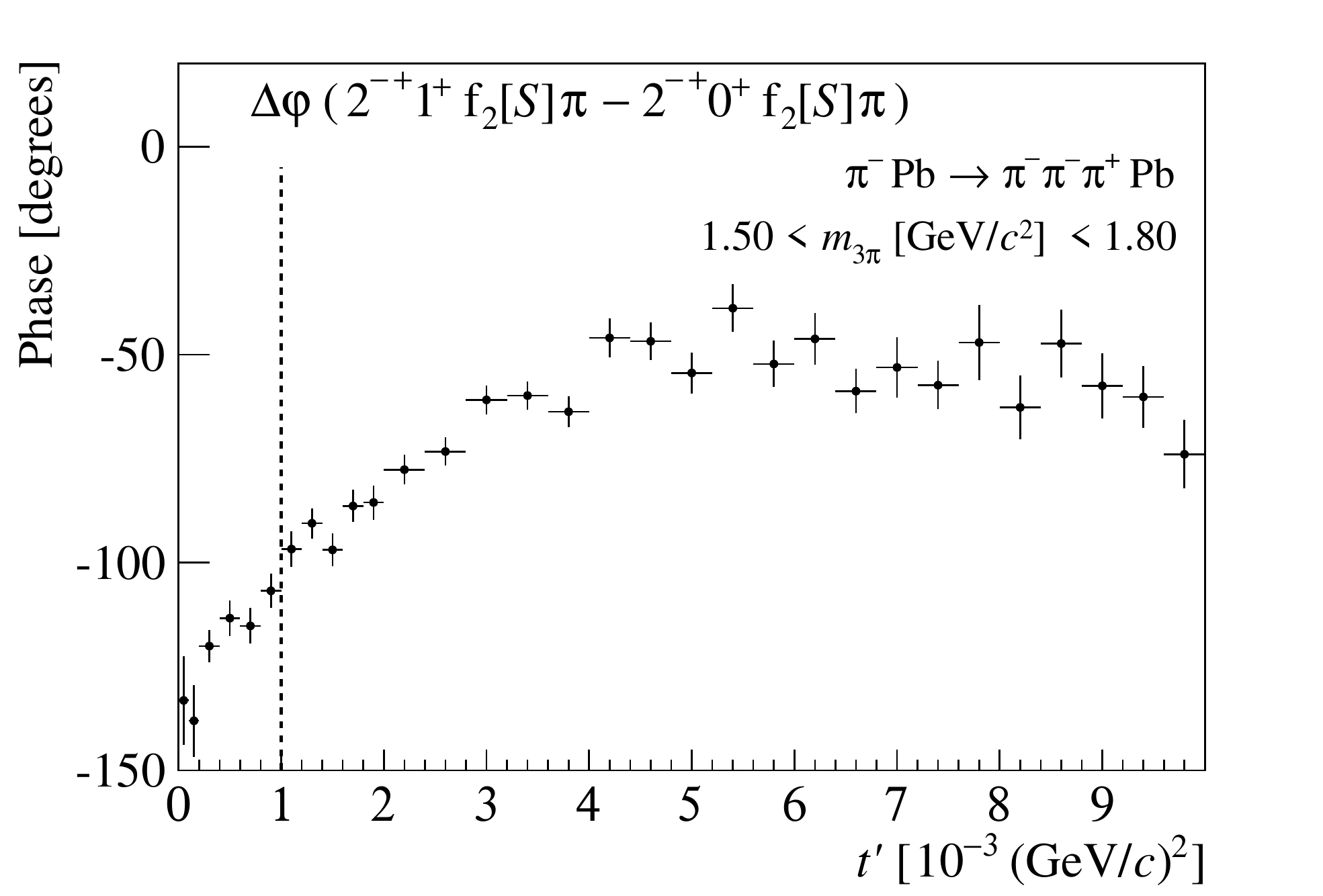} \\
 {\centerline{(c) \hspace{.47\textwidth} (f)}}
\caption{Intensities of the diffractively produced $\pi_2(1670)$ with $M=0$
  (top), the $\pi_2(1670)$ with $M=1$ (middle) and their relative phase (bottom)
  in bins of three-pion mass (left) and $t'$ (right), showing the Primakoff
  production of the $\pi_2(1670)$ with $M=1$. For details see text.}
\label{fig:prim_pi2}
\ifthenelse{\equal{\EPJSTYLE}{yes}}
{
\end{figure*}
}{
\end{figure}
}
The PWA results related to the Primakoff production of the $a_2(1320)$ are shown
in fig.~\ref{fig:prim_a2}.  The extracted intensity of the $1^{++}0^{+}
\rho[S]\pi$ decay amplitude, which is known to contain the diffractively
produced $a_{1}(1260)$, is shown in fig.~\ref{fig:prim_a2}(a). The intensity of
the $2^{++}1^{+} \rho[D]\pi$ decay amplitude, where the $a_2(1320)$ is expected,
is shown in fig.~\ref{fig:prim_a2}(b). The relative phase between these two
amplitudes is shown in fig.~\ref{fig:prim_a2}(c).  Here, the PWA was performed
in $40~\textrm{MeV}\!/c^2$ mass bins and covered $\tpGeV<0.001~\ttGeV$.  The phase
variation with respect to the $3\pi$ mass shows a clear rise at $\mtpGeV\approx
1.32~\mmGeV$, {\it i.e.\ }at the nominal mass of the $a_{2}(1320)$, indicating its
resonant behaviour.  The change of this phase with the momentum transfer $t'$ is
extracted performing a separate PWA in bins of $t'$ using only one mass bin that
contains the major part of the $a_2(1320)$.  This mass bin covers the range
$1.26\ \mmGeV < \mtpGeV < 1.38\ \mmGeV$, {\it i.e.\ }it is chosen significantly broader than the
usual $40~\textrm{MeV}\!/c^2$. The mass dependence is introduced by the
respective Breit-Wigner functions as factors in the decay amplitudes of
$a_{1}(1260)$ and $a_{2}(1320)$, while $t'$ dependences are not applied.  In
fig.~\ref{fig:prim_a2}(d) and (e) the resulting intensities of the same
amplitudes containing the $a_{1}(1260)$ and $a_{2}(1320)$, respectively, are
shown, this time in bins of $t'$.  In fig.~\ref{fig:prim_a2}(f), the relative
phase between these two decay amplitudes in bins of $t'$ shows the transition of
the production process from Primakoff production to diffractive dissociation of
the pion into the $a_{2}(1320)$ in the depicted range of $t'$.  The latter is
characterised by an approximately constant phase at $\tpGeV>0.006$.  In the
region of interest $\tpGeV<0.001~\ttGeV$, the relative phase $\Delta\varphi$ of the two
production amplitudes covers the range between $110^{\circ}$ and $130^{\circ}$.
This indicates that interference of diffractive and Primakoff production of the
$a_2(1320)$ in this $t'$ range is small.

Figure~\ref{fig:prim_pi2}(a) shows the intensity of the $2^{-+}0^{+} f_2[S]\pi$
partial wave with $M=0$, which contains the diffractively produced
$\pi_{2}(1670)$, and fig.~\ref{fig:prim_pi2}(b) the intensity of the
$2^{-+}1^{+} f_2[S]\pi$ amplitude with $M=1$.  Their relative phase as obtained
from the PWA is shown in fig.~\ref{fig:prim_pi2}(c) as a function of the
three-pion mass.  The phase shows a constant behaviour around the nominal mass
of the $\pi_{2}(1670)$.  This phase locking indicates the presence of the same
resonance $\pi_2(1670)$ in both spin projections $M=0$ and $M=1$, which are
allowed for $J^{PC}=2^{-+}$ amplitudes for natural parity exchange as explained
before.  Again, a separate PWA was performed in bins of momentum transfer $t'$
while using a broad three-pion mass interval covering the main part of the width
of the $\pi_2(1670)$, {\it i.e.\ }$1.50\ \mmGeV < \mtpGeV < 1.80\ \mmGeV$.  For this PWA fit the
decay amplitudes of the significant partial waves are multiplied by
mass-dependent functions containing sums of the relevant Breit-Wigner functions
and additional background as given in appendix~\ref{sec:appendix:tbins_pi2}.
The relative phase between the $M=0$ and $M=1$ components of the $\pi_2(1670)$
(fig.~\ref{fig:prim_pi2} (d, e, f)) demonstrates the transition from Primakoff
to diffractive production of the $\pi_{2}(1670)$ with $M=1$ in the depicted
range of $t'$.  Again, we observe the relative phase being approximately
$90^{\circ}$ in the region of interest $\tpGeV<0.001~\ttGeV$, which limits interference
effects between diffractive and Primakoff production in this $t'$ range.  This
allows the separation of the two production processes by a fit with a sum of two
non-interfering contributions.

At this point we do not make any statement about resonances in $J^{PC}=1^{++}$
or $J^{PC}=1^{-+}$ amplitudes with $M=1$.  Such amplitudes are present in the
fit (see table~\ref{tab:waveset}) and collect non-negligible intensities, but
their interpretation in terms of resonances is not obvious. The clarification of
their nature is beyond the scope of the present paper.

\ifthenelse{\equal{\EPJSTYLE}{yes}}
{
\begin{figure*}
}{
\begin{figure}
}
\includegraphics[width=0.49\textwidth]{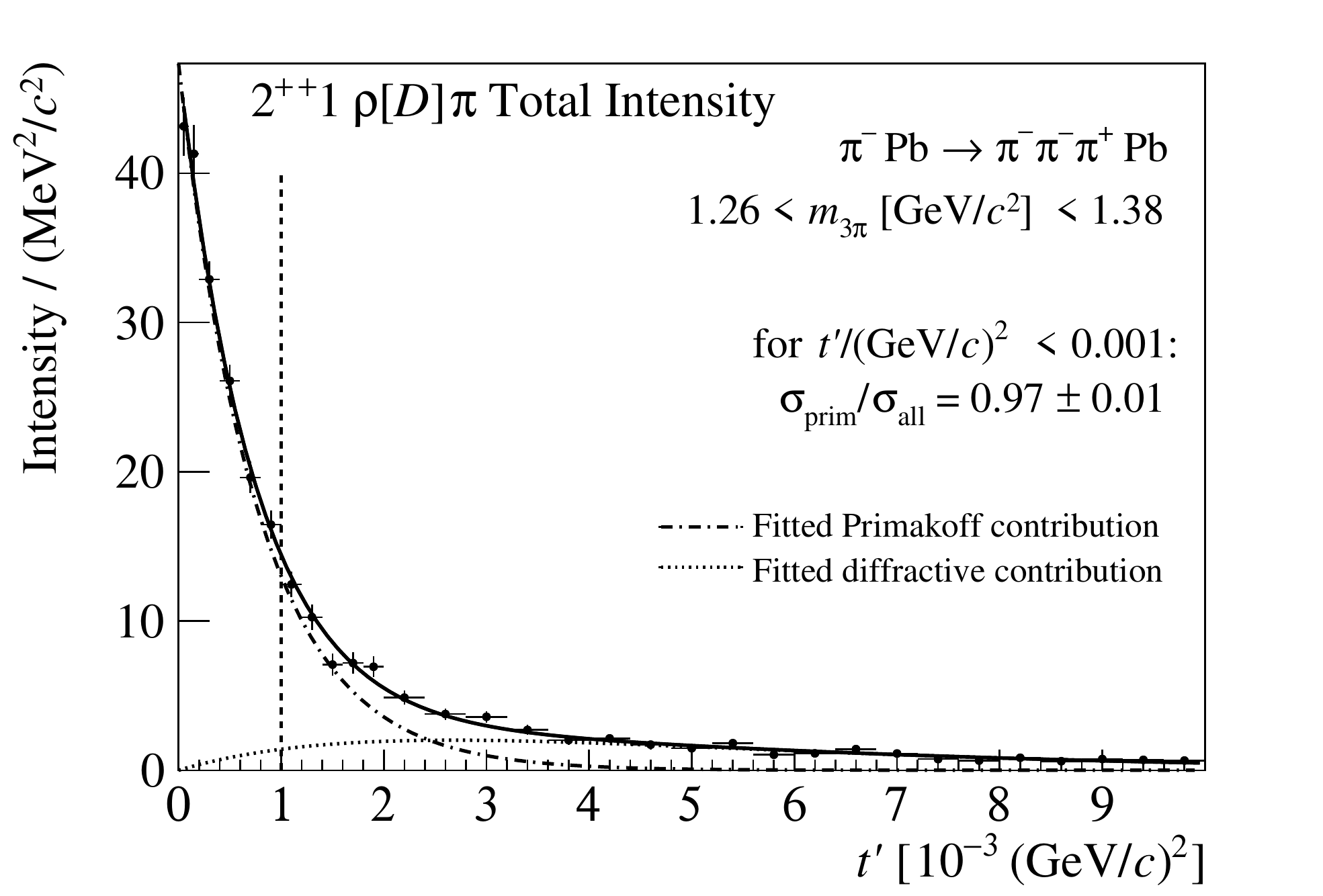}
\includegraphics[width=0.49\textwidth]{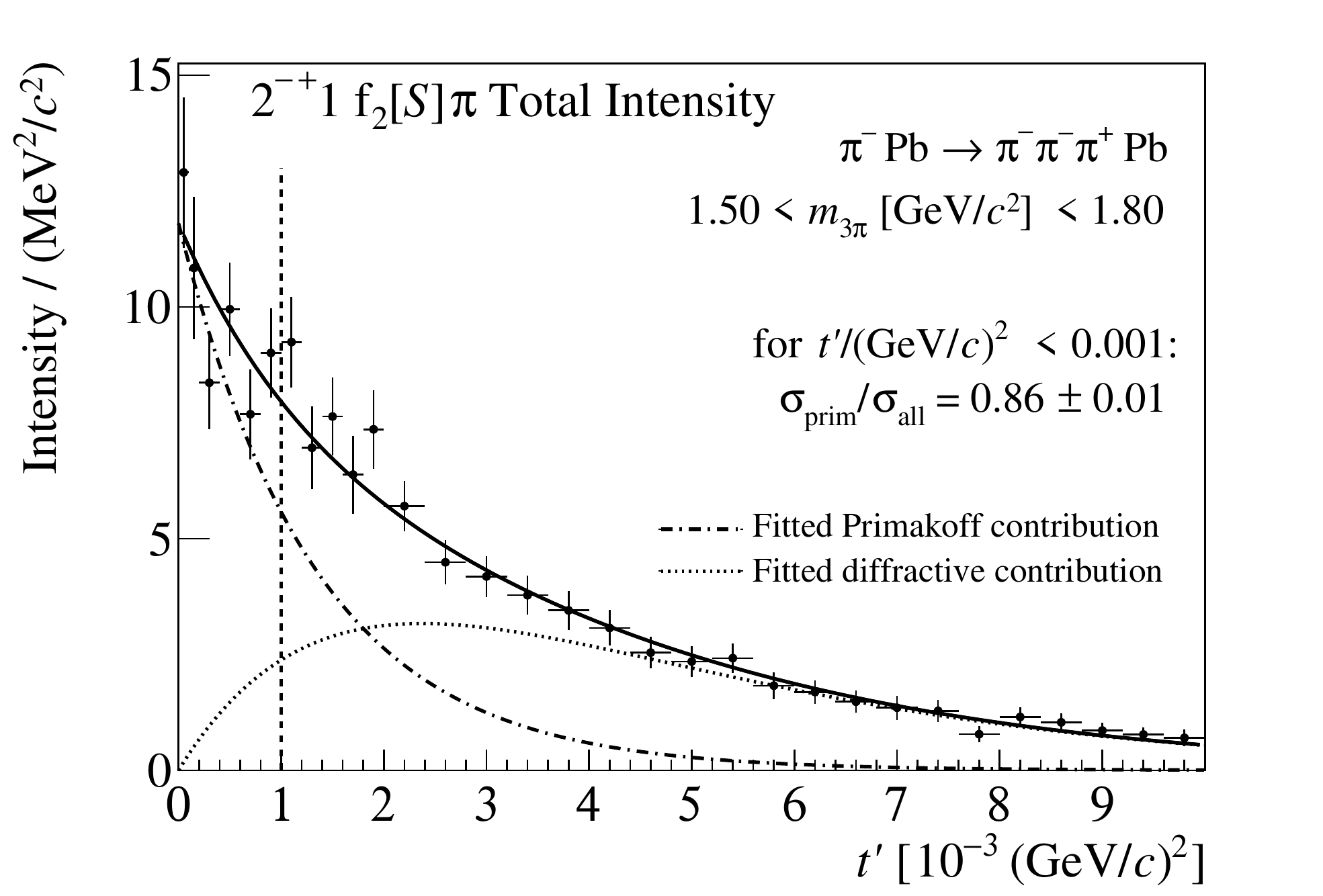}\\
 {\centerline{(a) \hspace{.47\textwidth} (b)}} 
\caption{Total intensities of (a) the $J^{PC}M=2^{++}1$ and (b) the
  $J^{PC}M=2^{-+}1$ states decaying into $f_2[S]\pi$ in bins of $t'$. Both are
  fitted by a sum of two exponentials to extract the fraction of Primakoff
  production of the $a_2(1320)$ and the $\pi_2(1670)$. The full line refers in
  both cases to the sum of the two contributions. For details see text.}
\label{fig:prim_totals}
\ifthenelse{\equal{\EPJSTYLE}{yes}}
{
\end{figure*}
}{
\end{figure}
}

The total intensities of the amplitudes with $J^{PC}M=2^{++}1$ in the
$\rho[D]\pi$ and $J^{PC}M=2^{-+}1$ in the $f_2 [S]\pi$ decay channel are
depicted in fig.~\ref{fig:prim_totals}.  They are fitted by an incoherent sum of
Primakoff and diffractive productions to obtain the relative strengths of both
contributions to be evaluated in the indicated range $\tpGeV<0.001~\ttGeV$. This
procedure is justified by the expected relative phase of $\approx90^{\circ}$
between the photon and pomeron amplitudes, which is caused by the photon being
real, while the pomeron corresponds to imaginary potential due to the absorptive
nature of the strong interaction.  For pomeron exchange, no phase difference
between $M=0$ and $M=1$ amplitudes is expected.  For this fit the Primakoff
production is parameterised by $\textrm{d}\sigma_{\textrm{prim}}/\textrm{d}t'
\propto \exp(-b_{\textrm{prim}}(m) t')$ instead of the extremely sharp form
given by eq.~\eqref{eq:intro:WW_xsec}, as discussed in
sect.~\ref{sec:prim_res:resol}.  The diffractive contribution, in turn, is
parameterised by $\sigma_{\textrm{diff}} \propto t' \exp (-b_{\textrm{diff}}(m)
t')$, as for $M=1$ in eq.~\eqref{eq:spec:diff_xsec_tM}.

The resulting slopes amount to $\bpGeV(m_{a_2})=(1292\pm 53)\ \tmGeV$ and
$\bdGeV(m_{a_2})=(374\pm 25)\ \tmGeV$ for the $a_2(1320)$
(fig.~\ref{fig:prim_totals}(a)). The quoted fit uncertainties take into account
the error estimates from the PWA as they are indicated on the data points in
fig.~\ref{fig:prim_totals}. These slope parameters are in fair agreement with
the expectations from the simulation of Primakoff production, namely
$b_{\textrm{prim,MC}}\approx 980\ \tmGeV$ and
$b_{\textrm{diff,MC}}\approx 370\ \tmGeV$.  These expected values are
obtained following the procedure explained in sect.~\ref{sec:prim_res:resol}.

For the $2^{-+} 1 f_2 [S]\pi$ total intensity in the mass region of the
$\pi_2(1670)$ (fig.~\ref{fig:prim_totals} (b)), the separation of the two
contributions is difficult as the respective parameters are highly
correlated. Therefore, in our analysis we imposed the condition that the
fractions of Primakoff and diffractive contributions are equal at that value of
$t'$, where the relative phase between the $M=0$ and $M=1$ amplitudes
(fig.~\ref{fig:prim_pi2} (f)) is $45^{\circ}$ below the phase at $\tpGeV\approx
0.01\ \ttGeV$.  This is approximately achieved limiting the parameter range to
$\bpGeV\geq 750\ \tmGeV$. It results in $\bdGeV=(421\pm 20)\ \tmGeV$ and
$\bpGeV=750 \ \tmGeV$ at its lower
limit, in rather good agreement with the expected values of
$b_{\textrm{diff,MC}}\approx 350\ \tmGeV$ and
$b_{\textrm{prim,MC}}\approx 760\ \tmGeV$ from the procedure described in
sect.~\ref{sec:prim_res:resol}.

The ratio $\sigma_{\textrm{prim}}/\sigma_{\textrm{all}}$ is obtained by
integrating the contributions in the range $\tpGeV<0.001~\ttGeV$, yielding $0.97\pm
0.01$ for the $a_2(1320)$ and $0.86 \pm 0.07$ for the $\pi_2(1670)$. The
uncertainties quoted are obtained from varying the upper and lower limits of
$b_{\textrm{prim}}$ and $b_{\textrm{diff}}$. For the $a_2(1320)$, fits are
performed with limiting $\bpGeV\leq 980\ \tmGeV$ and requiring $\bdGeV\geq 320\ \tmGeV$.  For
the $\pi_2(1670)$, fits with $\bpGeV$ down to $500\ \tmGeV$ are taken into account.
This is done to account for the neglect of interference between the Primakoff
and the diffractive contributions.

\section{Extraction of the radiative widths}
\label{sec:rad_widths}
For the extraction of the radiative width of a resonance, we integrate
eq.~\eqref{eq:intro:xsec_primakoff_resonance} over the range $0\ \ttGeV <\tpGeV
<0.001\ \ttGeV$
and over the relevant mass ranges containing the $a_2(1320)$ and $\pi_2(1670)$.

\subsection{Parameterisation of mass-dependent widths}
\label{sec:extract:ingred}
An important ingredient for the determination of the radiative widths is the
accurate mathematical description of the mass spectra observed.  Apart from the
damping of higher masses introduced by $t_{\textrm{min}}$ appearing in
eq.~\eqref{eq:intro:xsec_primakoff_resonance}, this concerns in particular the
total and partial mass-dependent decay widths.  They both enter into fits of the
PWA intensities containing Breit-Wigner parameterisations, which are used for
the extraction of $N_X/\epsilon_X$ and for the calculation of the normalisation
constant $C_X$ that is needed for the evaluation of
eq.~\eqref{eq:intro:radwidth_from_data}.  The exact line shape has to describe
properly the tails towards lower and higher masses which are very asymmetric, as
we assume that those are not mocked up by background but belong to the
resonances under investigation.

If a resonance decays only via two-body decays into particles, the width of
which can be neglected, and if the decay channels do not interfere, the
mass-dependent total width of the resonance can be written as
\begin{align}
\ifthenelse{\equal{\EPJSTYLE}{yes}}
{
\Gamma_{\textrm{total}}(m)  = \sum_{n} \Gamma_n(m) & = \sum_{n} \textrm{BR}_n \; \underbrace{\frac{m_{0}}{m} \frac{p_{n}}{p_{0\,n}} \frac{F^{2}_{L}(p_{n})}{F^{2}_{L}(p_{0\,n})}}_{ f^{\textrm{dyn}}_n(m) } \; \Gamma_{0} \nonumber \\
& \quad \text{with} \quad \Gamma_{0} \equiv \Gamma_{\textrm{total}}(m_{0}) \ .
}{
\Gamma_{\textrm{total}}(m) = \sum_{n} \Gamma_n(m) = \sum_{n} \textrm{BR}_n\; \underbrace{\frac{m_{0}}{m} \frac{p_{n}}{p_{0\,n}} \frac{F^{2}_{L}(p_{n})}{F^{2}_{L}(p_{0\,n})}}_{ f^{\textrm{dyn}}_n(m) } \; \Gamma_{0} \quad \text{with} \quad \Gamma_{0} \equiv \Gamma_{\textrm{total}}(m_{0}) \ . 
}
\label{eq:app:BW_massdep_width}
\end{align}
This expression contains a sum over the partial widths $\Gamma_{n}$ of all
possible decay channels $n$ of this resonance (with their corresponding
normalised branching fractions $\textrm{BR}_n$).  The two-body breakup momentum
$p_{n}$ is the momentum of the daughter particles of a particular decay $n$ of a
parent state with mass $m$ in its centre-of-mass frame, and $L$ is the orbital
angular momentum between the two daughter particles.  The symbol $F_{L}$
specifies the angular momentum barrier factors as given by
ref.~\cite{quigg_hippel}.  The additional index ``0'' denotes the values of
width and breakup momentum at the nominal mass $m_0$ of the resonance. In cases
where the branching fractions are unknown, a Breit-Wigner function with constant
width $\Gamma_{0}$ is usually chosen as an approximation.

A more accurate parameterisation of the mass dependence of $\Gamma_n(m)$ is
preferable especially in the case of multi-particle decays with short-lived
decay products, so that the widths of the daughter particles can be taken into
account properly.  Hence, in order to include properly also sub-threshold
contributions, the term $p_n F_{L}^{2}(p_n)$ is replaced by the integral over
the respective decay amplitude $\int \vert \psi_{n} \vert ^2 \text{d}\Phi(\tau)$
\cite{bowler_a1param}. The effect is depicted in fig.~\ref{fig:extract:pspace}.
The description using angular momentum barrier factors (from
eq.~\eqref{eq:app:BW_massdep_width}, dashed lines) starts from the nominal
$\textrm{\{isobar\}}\pi$ thresholds only, which are $\approx
0.9~\textrm{GeV}\!/c^2$ for the $\rho\pi$ decay and $\approx
1.5~\textrm{GeV}\!/c^2$ for the $f_2 \pi$ decay. The description based on $\int
\vert \psi_n \vert^2 \text{d}\Phi(\tau)$ for the considered decay channels
$n=\rho\pi$ and $n=f_2\pi$ starts from the summed mass of the final state
particles ({\it i.e.} $\approx 0.42~\textrm{GeV}\!/c^2$ for three pions), so
that it describes the low-mass tail which is considerable, particularly for the
$\pi_2(1670)$.  In the figure and in the following, the index ``$n$'' is dropped
for $p_n$ and $p_{0n}$, and those are understood to be the appropriate breakup
momenta.

In many cases, the shape of a specific resonance does not support the use of the
term $m_0/m$ in eq.~\eqref{eq:app:BW_massdep_width}, which introduces additional
damping at higher masses.  Reference~\cite{rho_shape} even suggests that the
term $(m_0/m)^\alpha$ may be used with arbitrary $\alpha$ adjusted to the data.
In this analysis, where also the damping behaviour of $m$ resulting from the
integrated $t'$ dependences from eq.~\eqref{eq:intro:xsec_primakoff_resonance}
is taken into account, a better fit to the mass spectrum is obtained when
omitting the term $m_0/m$ in the parameterisation of the mass-dependent widths.
\ifthenelse{\equal{\EPJSTYLE}{yes}}
{
\begin{figure*}
}{
\begin{figure}
}
\includegraphics[width=0.49\textwidth]{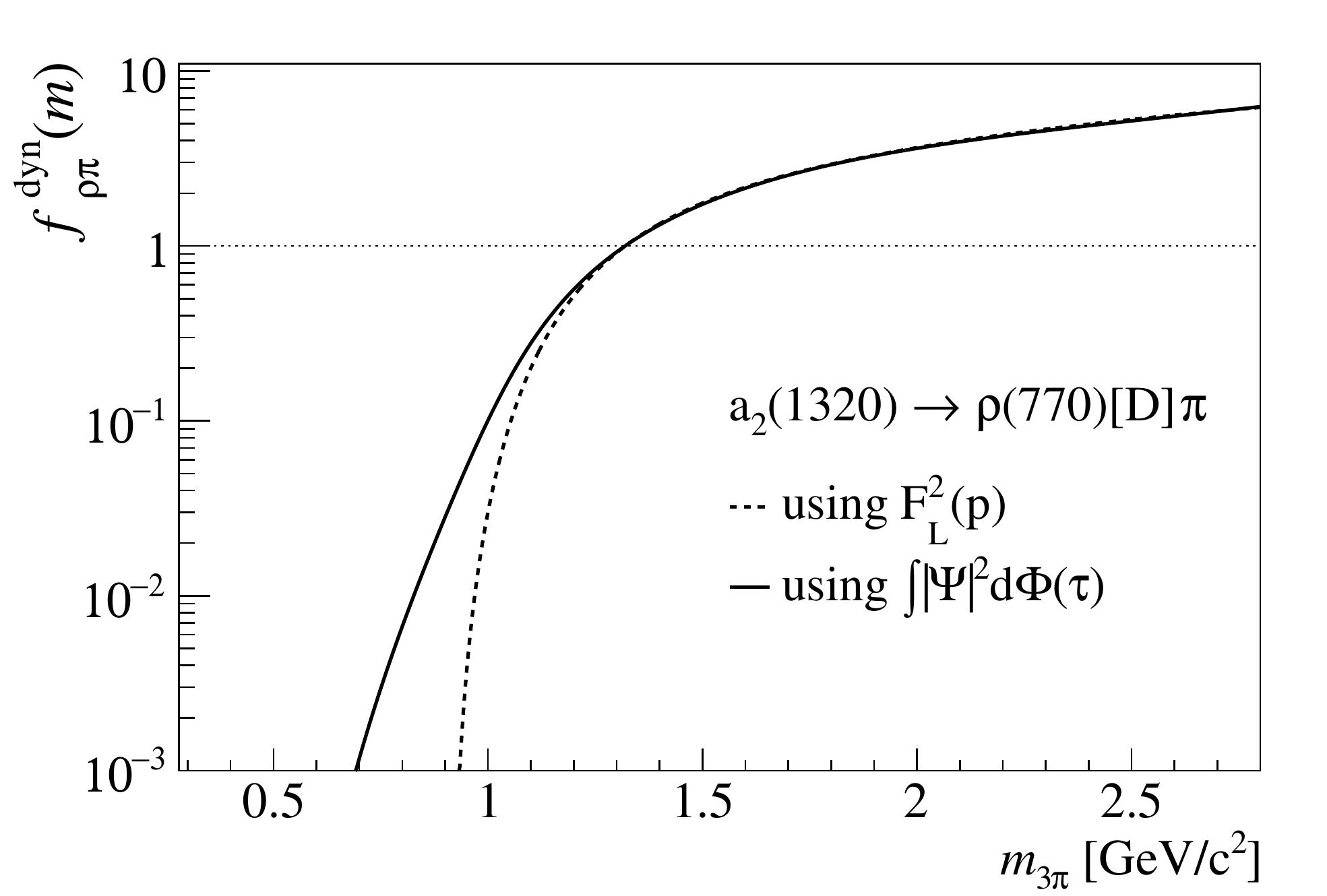}
\includegraphics[width=0.49\textwidth]{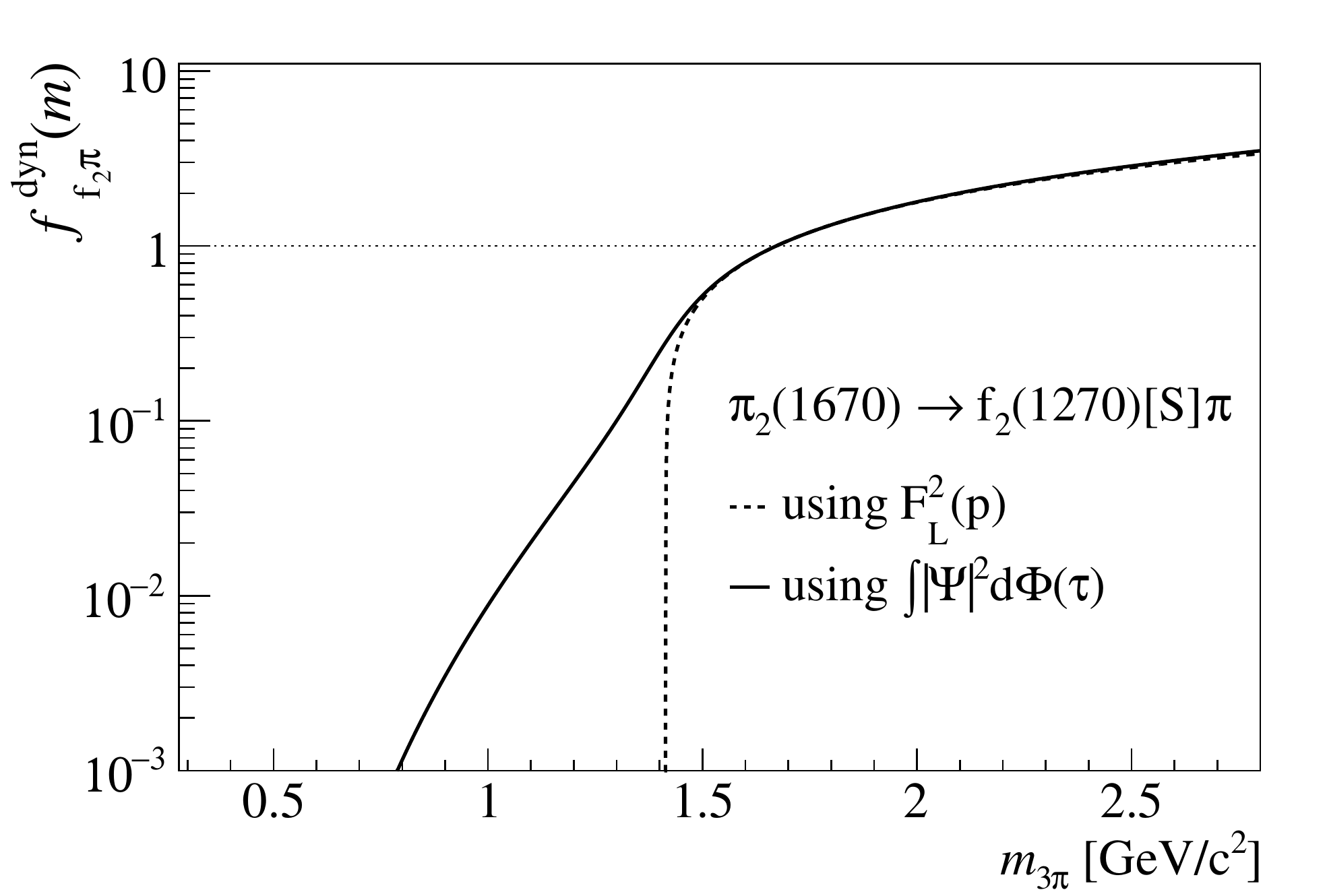}\\
 {\centerline{(a) \hspace{.47\textwidth} (b)}} 
\caption{Comparison of the dynamical factors for (a) the $a_{2}(1320)
  \rightarrow \rho(770)[D]\pi$ and (b) the $\pi_{2}(1670) \rightarrow
  f_{2}(1270)[S]\pi$ decays using angular momentum barrier factors or $\int
  \vert \psi \vert^2 \textrm{d}\Phi(\tau)$ (for details see text). The lines
  extent in the latter case, with non-zero values for $f^{\textrm{dyn}}$, down
  to the three-pion threshold $m_{3\pi} = 3 m_{\pi} \approx
  0.42~\textrm{GeV}\!/c^2$, while in the case of $F_L$ the range is limited to
  $m_{3\pi}>m_\pi+m_{\{\textrm{isobar}\}}$ .}
\label{fig:extract:pspace}
\ifthenelse{\equal{\EPJSTYLE}{yes}}
{
\end{figure*}
}{
\end{figure}
}
\begin{figure}
\begin{center}
\ifthenelse{\equal{\EPJSTYLE}{yes}}
{\includegraphics[width=0.99\columnwidth]{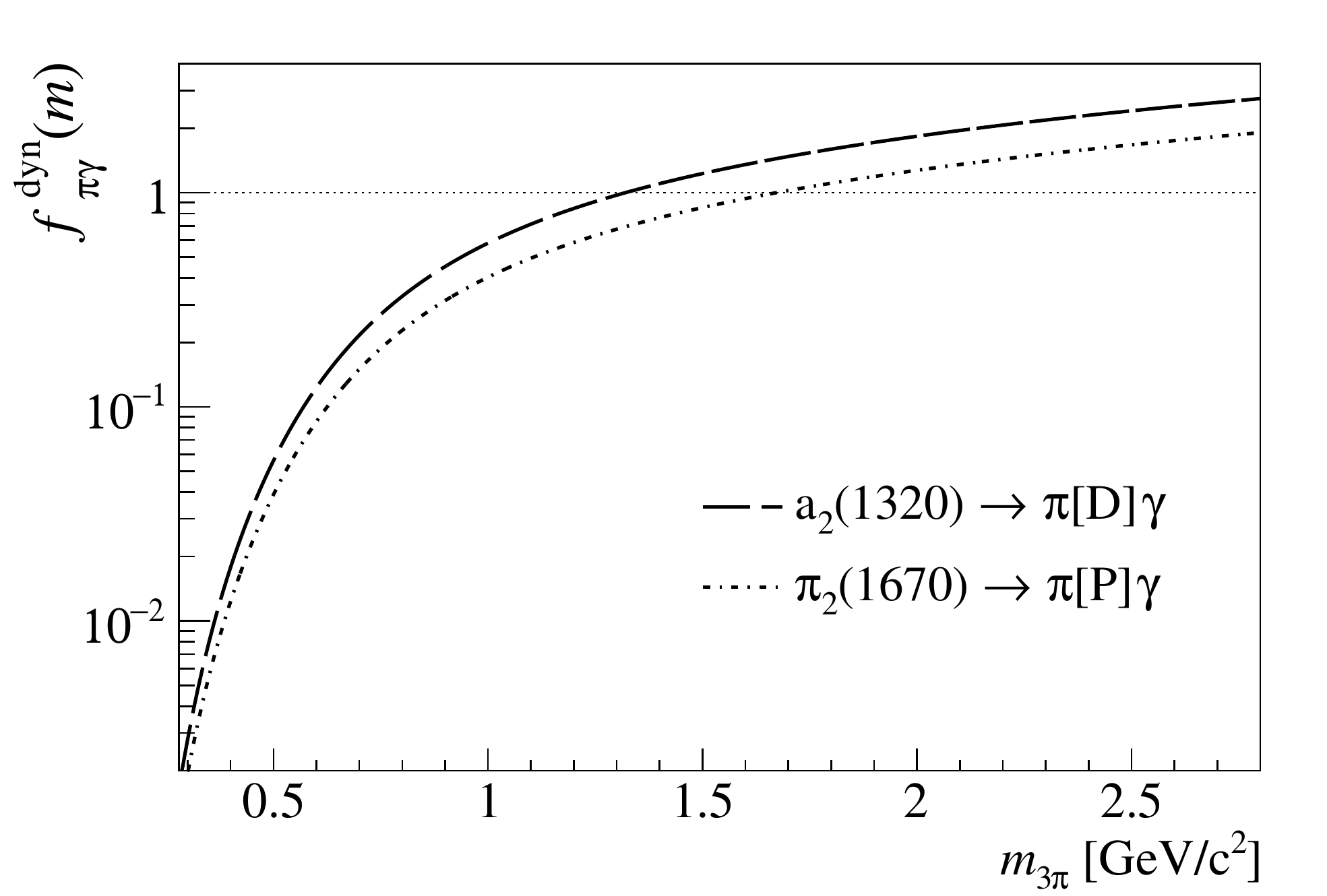}}
{\includegraphics[width=0.49\columnwidth]{fig10.pdf}}
\caption{Shapes of the dynamical factors  $f^{\textrm{dyn}}_{\pi\gamma}(m)$ for the $\pi\gamma$ decays of the $a_2(1320)$ and the $\pi_2(1670)$.}
\label{fig:extract:pspace_pig}
\end{center}
\end{figure}

The mass-dependent partial width of the $X \rightarrow \pi\gamma$ decay, $\Gamma_{\pi\gamma}(m)$, reads  
\begin{equation}
\Gamma_{\pi\gamma}(m) =  \underbrace{ \frac{p}{p_0} \; \frac{F^2_L(p)}{F^2_L(p_0)} }_{ f^{\textrm{dyn}}_{\pi\gamma}(m) } \; \cdot \Gamma_0(X \rightarrow \pi\gamma) \ ,
\label{eq:extract:dynwidth_pigamma}     
\end{equation}
where we use $L=2$ ($D$-wave) and $L=1$ ($P$-wave) for the $\pi\gamma$ decays of
the $a_2(1320)$ and $\pi_2(1670)$ resonance, respectively.  The use of the
$F_L$-dependent factor, which is depicted in fig.~\ref{fig:extract:pspace_pig},
may be disputible. However, it has an effect of only about $1\%$ and $3\%$ on
the final result for the $a_2(1320)$ and the $\pi_2(1670)$, respectively,
compared to using $f^{\textrm{dyn}}_{\pi\gamma}(m)=(p/p_0)^3$ as suggested in
ref.~\cite{rosner_radwidths}.

For the mass-dependent partial decay widths $\Gamma_{\textrm{final}}(m)$ the
widths of the isobars have to be taken into account, {\it i.e.} they are
parametrised as
\begin{equation}
\Gamma_{\textrm{final}}(m) = \underbrace{ \frac{ \int \vert \psi_{\textrm{final}}(m, \tau) \vert ^2 \text{d}\Phi(\tau)}{ \int \vert \psi_{\textrm{final}}(m_0, \tau) \vert ^2 \text{d}\Phi(\tau)} }_{ f^{\textrm{dyn}}_{\pi\textrm{\{isobar\}}}(m) }
\ \cdot\ \Gamma_0(X \rightarrow \pi\textrm{\{isobar\}}) \ . 
\label{eq:extract:dynwidth_final}
\end{equation}
The $a_2(1320)$ is observed in the total intensity of the $J^{PC}M=2^{++}1$
waves decaying into $\rho[D]\pi$, {\it i.e.}\ \{isobar\} = $\rho(770)$ and
$L=2$, so that $\psi_{\textrm{final}}=\psi_{2^{++}1^{+} \rho(770)[D]\pi}$ is
used.

The total width $\Gamma_{\textrm{total}}(m)$ of the $a_2(1320)$ is derived from
eq.~\eqref{eq:app:BW_massdep_width} using the constant $\Gamma_{0} =
\Gamma_{0}(m_0(a_2))$. The total width is taken as the sum of the partial widths
of the two main decay channels $\rho\pi$ and $\eta\pi$.  As the other decay
channels are neglected as described below, we calculate ``renormalised''
branching fractions $\textrm{BR}_n$ from their known branching fractions
$\textrm{BR}^{\textrm{PDG}}_n$ from ref.~\cite{pdg2012}.  Those are given in
table~\ref{tab:a2_branching} together with the used parameterisations of the
phase space.  The $\omega\pi\pi$ decay channel
($\textrm{BR}^{\textrm{PDG}}_{\omega\pi\pi} = 0.105$) and the $K \bar{K}$ decay
channel ($\textrm{BR}^{\textrm{PDG}}_{K \bar{K}} = 0.049$) are not taken into
account, as the treatment of the $\omega\pi\pi$ decay in the framework of
two-particle decays is delicate due to unknown branching fractions into
$b_1(1235)\pi$ and $\omega\rho$, which both are sub-threshold at the nominal
mass of the $a_2(1320)$.  The $K\bar{K}$ channel has an even smaller branching
fraction than the $\omega\pi\pi$ decay channel and its radiative decay width is
not included in $\Gamma_{\textrm{total}}$ here as well due to its smallness,
$\textrm{BR}_{\pi\gamma}<0.01$.
\begin{table}[h]
\caption{Branching fractions $\textrm{BR}^{\textrm{PDG}}_{n}$ and
  $\textrm{BR}_{n}$ and parameterisations of phase space
  $f^{\textrm{dyn}}_{n}(m)$ for the two main decay channels $\rho\pi$ and
  $\eta\pi$, as used for the description of the mass-dependent total width of
  the $a_2(1320)$.}
\begin{center}
\begin{tabular}{l||l|l|l}
$n$ & $\textrm{BR}^{\textrm{PDG}}_{n}$ & $\textrm{BR}_{n}$ & $f^{\textrm{dyn}}_{n}(m)$ \\[2pt]
\hline\hline
\rule{0pt}{15pt}$\rho[D]\pi$ & $0.701$ & $0.82$ &
$\dfrac{\int \vert \psi(m) \vert ^2 \textrm{d}\Phi(\tau)^{\rule{0pt}{0.018\textwidth}} }{ \int \vert \psi(m_0) \vert ^2 \text{d}\Phi(\tau)_{\rule{0pt}{0.024\textwidth}} }$
\\[5pt]
\hline
\rule{0pt}{15pt}$\eta[D]\pi$ & $0.145$ & $0.18$ & 
$\dfrac{p}{p_{0}} \cdot \dfrac{F^2_{2}(p)^{\rule{0pt}{0.018\textwidth}}}{F^2_{2}(p_{0})}$
\\[5pt]
\end{tabular}
\end{center}
\label{tab:a2_branching}
\end{table}

The $\pi_2(1670)$ is observed in the total intensity of the $J^{PC}M=2^{-+}1$
amplitudes decaying into $f_2[S]\pi$, {\it i.e}.\ \{isobar\} = $f_{2}(1270)$ and
$L=0$, so that $\psi_{\textrm{final}}=\psi_{2^{-+}0^{+} f_2(1270) [S]\pi}$ is
used.

The parameterisation of the mass-dependent width of the $\pi_2(1670)$ is more
complicated.  The $\pi_2(1670)$ decays mainly into $3\pi$
($\textrm{BR}^{\textrm{PDG}}_{3\pi}=0.96$), which includes decays into $f_2\pi$
($\textrm{BR} ^{\textrm{PDG}}_{f_2 \pi}=0.56$), $\rho\pi$
($\textrm{BR}^{\textrm{PDG}}_{\rho\pi}=0.31$), $\sigma\pi$
($\textrm{BR}^{\textrm{PDG}}_{\sigma\pi}=0.11$) and $(\pi\pi)_S \pi$
($\textrm{BR}^{\textrm{PDG}}_{(\pi\pi)_S \pi}=0.09$) \cite{pdg2012}. All these
decays are also observed in the COMPASS experiment.  However, the incoherent sum
in eq.~\eqref{eq:app:BW_massdep_width} is questionable for the different $3\pi$
final states, as they interfere significantly. In addition, the branching
fractions $\textrm{BR}^{\textrm{PDG}}$ are quoted for ``$3\pi$'', but do not
distinguish between the charged and the neutral channel, where they are expected
to differ due to the different isospins of $\rho$ and $f_2$.  For this analysis,
we only take into account the decay $\pi_2(1670) \rightarrow f_2(1270) [S] \pi$
with its branching fraction $\textrm{BR}^{\textrm{PDG}}_{f_2 \pi}=0.56$ taken
from ref.~\cite{pdg2012}.  For the mass-dependent width we use
$\Gamma_{\textrm{total}}(m) = f^{\textrm{dyn}}_{\textrm{final}}(m)\Gamma_0(m_0)$
since the exact shape of the Breit-Wigner function does not matter for the
signal strength of the $\pi_2(1670)$ at the current level of accuracy.

\subsection{Acceptance-corrected PWA intensities}
\label{sec:extract:pwa_int}

\ifthenelse{\equal{\EPJSTYLE}{yes}}
{
\begin{figure*}
}{
\begin{figure}
}
\includegraphics[width=0.49\textwidth]{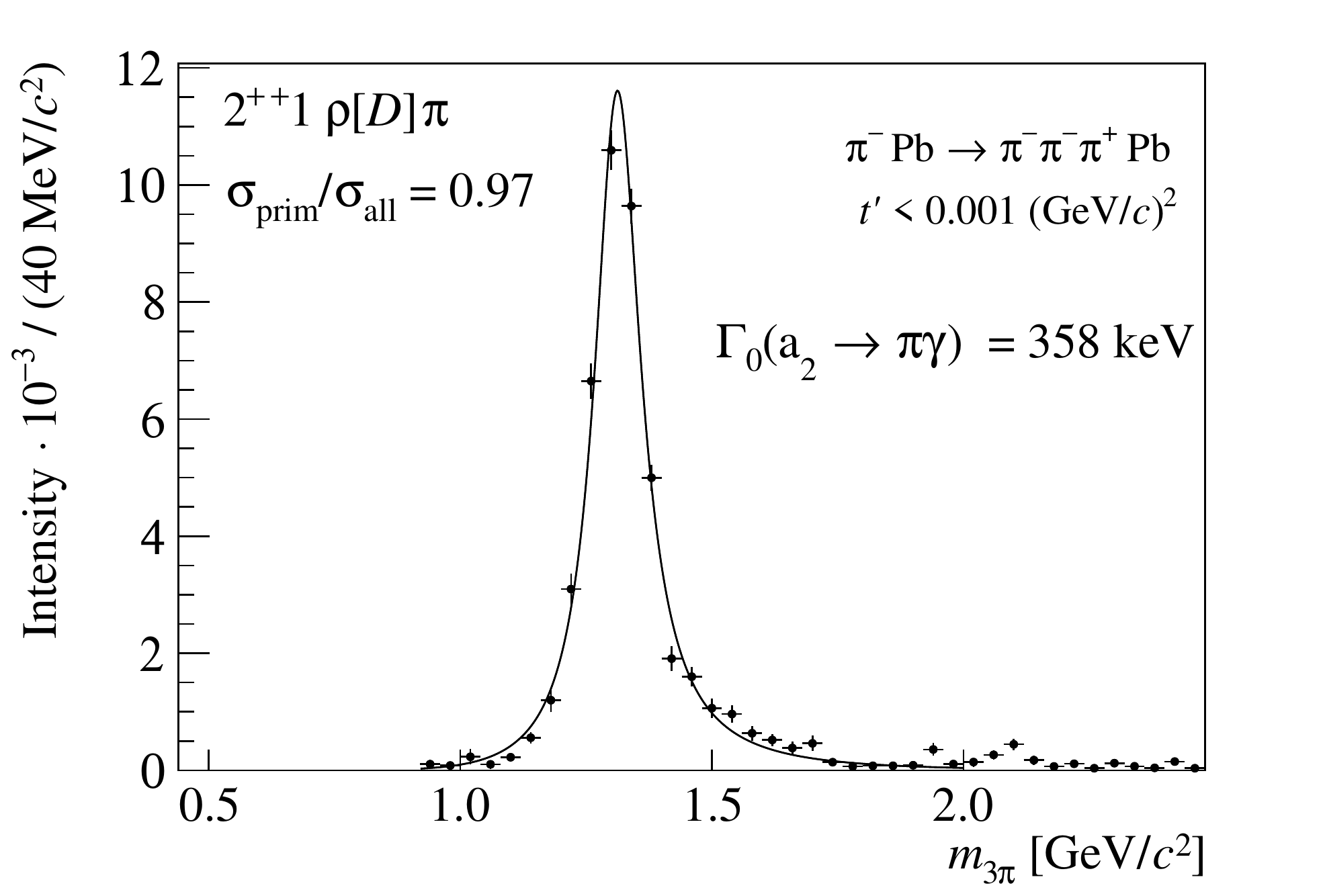}
\includegraphics[width=0.49\textwidth]{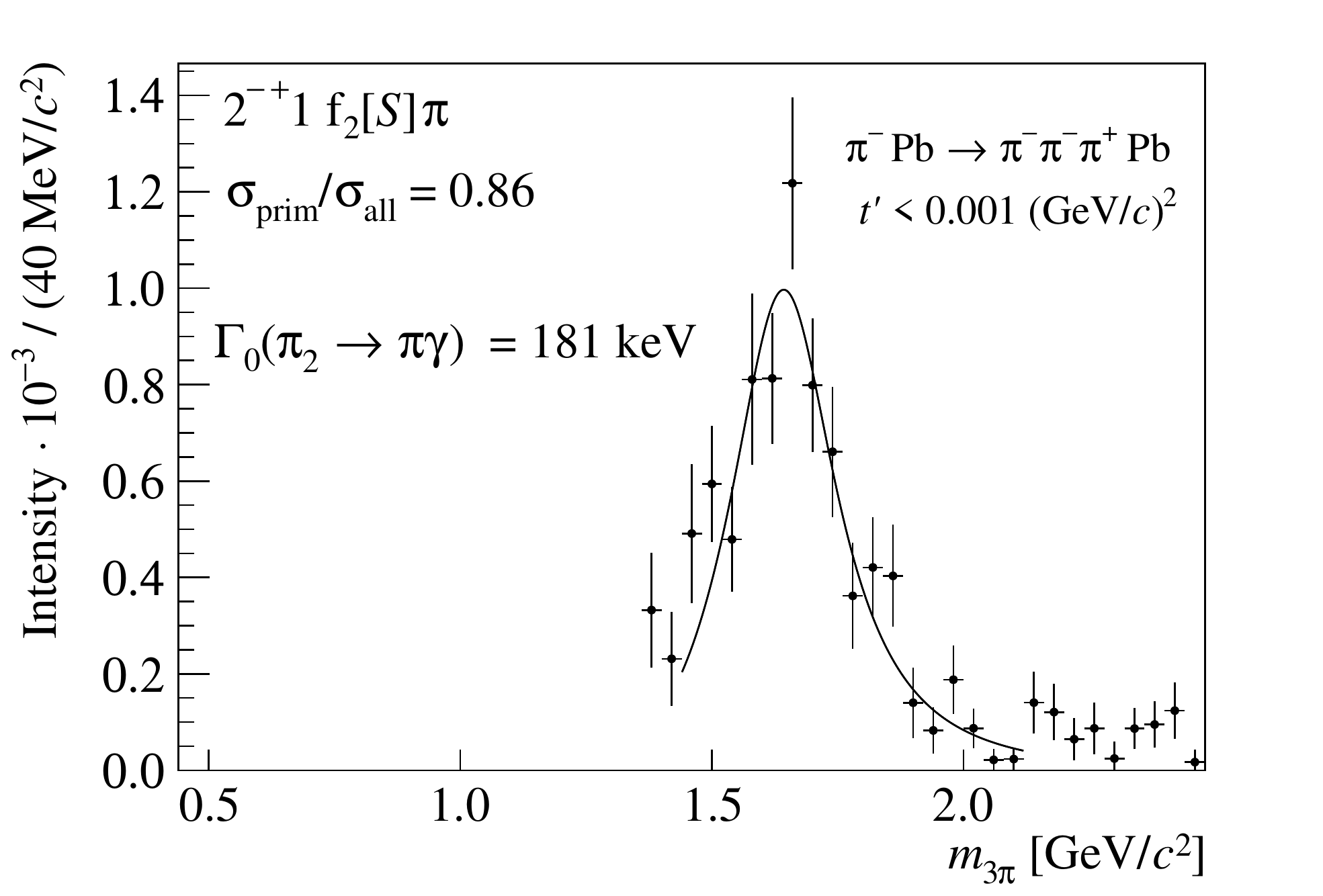}\\
 {\centerline{(a) \hspace{.47\textwidth} (b)}}
\caption{Total intensities in mass bins: $J^{PC}M=2^{++}1$ from $\rho[D]\pi$
  decay and $J^{PC}M=2^{-+}1$ from $f_2 [S]\pi$ decay. The error bars show the
  statistical uncertainties of the PWA. The Breit-Wigner fits used for the
  determination of the intensities of (a) the $a_2(1320)$ and (b) the
  $\pi_2(1670)$ are shown as well.}
\label{fig:extract:v130}
\ifthenelse{\equal{\EPJSTYLE}{yes}}
{
\end{figure*}
}{
\end{figure}
}
In the first step the acceptance-corrected intensities $N_{X,
  \textrm{prim}}/\epsilon_X$ used in eq.~\eqref{eq:intro:radwidth_from_data} are
determined.  The intensities in mass bins obtained from the PWA
(fig.~\ref{fig:extract:v130}) are fitted by the mass-dependent Breit-Wigner
terms from eq.~\eqref{eq:intro:xsec_primakoff_resonance}, while integrating over
$0\ \ttGeV <\tpGeV <0.001\ \ttGeV$. The mass-dependent widths $\Gamma_{\pi\gamma}(m)$,
$\Gamma_{\textrm{final}}(m)$ and $\Gamma_{\textrm{total}}(m)$ are evaluated
using the terms $f^{\textrm{dyn}}_{\pi\gamma}(m)$,
$f^{\textrm{dyn}}_{\pi\{\textrm{isobar}\}}(m)$, and $\sum_n \textrm{BR}_n
f^{\textrm{dyn}}_{n}(m)$ as explained in sect.~\ref{sec:extract:ingred}, $m_0$
and $\Gamma_0$ are introduced as fit parameters, and $\Gamma_0(X \rightarrow
\pi\gamma)$ is contained in the fit parameter for the overall normalisation of
the Breit-Wigner.  In the fitting procedure the bin width of
$40~\textrm{MeV}\!/c^2$ and the mass resolution are taken into account. The mass
resolution amounts to $\approx 16~\textrm{MeV}\!/c^2$ for the $a_2(1320)$ and
$\approx 18~\textrm{MeV}\!/c^2$ for the $\pi_2(1670)$.  More precisely, the mass
resolutions of both resonances are parameterised by a sum of three Gaussian
distributions, the parameters of which were obtained from the MC simulation
described in sect.~\ref{sec:prim_res:resol}.  The fit to the intensities shown
in fig.~\ref{fig:extract:v130} yields the following parameters.  For the
$a_2(1320)$ we obtain the mass $m_0 = (1319 \pm 1) ~\textrm{MeV}\!/c^2$ and the
width $\Gamma_0 = (105 \pm 4) ~\textrm{MeV}\!/c^2$ with a fit quality of
$\chi^2/\textrm{NDF} = 59.9/ 24$, and for the $\pi_2(1670)$ the mass $m_0 =
(1684 \pm 11) ~\textrm{MeV}\!/c^2$ and the width $\Gamma_0 = (277 \pm 38)
~\textrm{MeV}\!/c^2$ with a fit quality of $\chi^2/\textrm{NDF} = 20.0 / 14$.

From these fits, the following acceptance-corrected integrated intensities
$N_X/\epsilon_{X}$ are obtained that are used for the evaluation of
eq.~\eqref{eq:intro:radwidth_from_data}. The intensity for the $a_2(1320)$
($2^{++}1 \rho[D]\pi$) is integrated over the range $0.92\ \mmGeV <\mtpGeV
<2.00\ \mmGeV$ using
the fit function shown in fig.~\ref{fig:extract:v130}, which results in
$N_{a_2}/\epsilon_{a_2} = 44\,601 \pm 798$.  The uncertainty represents the
statistical uncertainty of the PWA fits in mass bins, which is propagated to the
parameters of the Breit-Wigner fit. After the correction for
$\sigma_{\text{prim}}/\sigma_{\text{all}} = 0.97$ we obtain the number of
Primakoff-produced events $N_{a_2, \text{prim}}/\epsilon_{a_2} = 43\,262\pm
774$.  The number of $\pi_2(1670)$ events is taken from the $2^{-+}1 f_2 [S]\pi$
intensity integrated over the range $1.44\ \mmGeV <\mtpGeV <2.12\ \mmGeV$, which results in
$N_{\pi_2}/\epsilon_{\pi_2} = 6\,977 \pm 435$. Applying
$\sigma_{\text{prim}}/\sigma_{\text{all}} = 0.86$, the number of
Primakoff-produced $\pi_2(1670)$ is $N_{\pi_2, \text{prim}}/\epsilon_{\pi_2} =
6\,000\pm 374$ in this decay channel.

\subsection{Normalisation constants}
\label{sec:extract:C}
In order to calculate the normalisation constant $C_X$, which is needed for the
evaluation of eq.~\eqref{eq:intro:radwidth_from_data}, we apply
eqs.~\eqref{eq:intro:xsec_primakoff_resonance} and
\eqref{eq:intro:xsec_radwidth} using $\Gamma(\pi\gamma) =
f^{\textrm{dyn}}_{\pi\gamma}(m) \cdot \Gamma_0(X \rightarrow \pi\gamma)$, with
$f^{\textrm{dyn}}_{\pi\gamma}(m)$ from eq.~\eqref{eq:extract:dynwidth_pigamma},
$\Gamma_{\textrm{final}}(m)$ as given in eq.~\eqref{eq:intro:gamma_final} with
$f^{\textrm{dyn}}_{\textrm{final}}(m) = f^{\textrm{dyn}}_{\pi \textrm{\{isobar\}
}}(m)$ from eq.~\eqref{eq:extract:dynwidth_final} and
$(\textrm{CG}\cdot\textrm{BR})$ divided out here, and
$\Gamma_{\textrm{total}}(m)$ as given in sect.~\ref{sec:extract:ingred}.  The
constant is obtained integrating over the same mass range $m \in [m_1, m_2]$ as
used for the extraction of $N_X/\epsilon_{X}$, and over $0\ \ttGeV <\tpGeV
<0.001\ \ttGeV$ which
reflects the $t'$ cut that is applied to the data:

\begin{align}
\ifthenelse{\equal{\EPJSTYLE}{yes}}
{
C_X= \int_{m_1}^{m_2} \int_{0}^{t'_{\textrm{max}}}& 16 \alpha Z^2 (2J+1) \left( \frac{m}{m^2 - m_{\pi}^2} \right)^3 \nonumber \\
&\cdot 
\frac{m_0^2 \, f^{\textrm{dyn}}_{\pi\gamma}(m) \, f^{\textrm{dyn}}_{\text{final}}(m)\Gamma_0(m_0)}{(m^2 - m_0^2)^2 + m_0^2 \Gamma^2_{\textrm{total}}(m)} \nonumber \\
&\cdot \frac{t'}{(t' + t_{\textrm{min}})^2} \ F^2_{\textrm{eff}}(t')  \ \textrm{d}t' \ \textrm{d}m \ .
}{
C_X= \int_{m_1}^{m_2} \int_{0}^{t'_{\textrm{max}}} 16 \alpha Z^2 (2J+1) \left( \frac{m}{m^2 - m_{\pi}^2} \right)^3 
\cdot 
\frac{m_0^2 \, f^{\textrm{dyn}}_{\pi\gamma}(m) \, f^{\textrm{dyn}}_{\text{final}}(m)\Gamma_0(m_0)}{(m^2 - m_0^2)^2 + m_0^2 \Gamma^2_{\textrm{total}}(m)} 
\ \frac{t'}{(t' + t_{\textrm{min}})^2} \  F^2_{\textrm{eff}}(t')  \ \textrm{d}t' \ \textrm{d}m \ .
}
\label{eq:extract:C}
\end{align}
For the numbers given in the following, the form factor $F^2_{\textrm{eff}}(t')$
and the Weizs\"{a}cker-Williams term are replaced by $|F^u_C(t',
t_{\textrm{min}})|^2$ from ref.~\cite{faeldt_2009_2013cor} as discussed before.
For $C_{a_2}$ we use $m_0=1320~\textrm{MeV}\!/c^{2}$,
$\Gamma_0(m_0)=107~\textrm{MeV}\!/c^{2}$, and obtain $C_{a_2} =
2\,236.23~\textrm{mb}/\textrm{GeV}$.  The constant $C_{\pi_2}$ is calculated
using $m_0=1672\,\textrm{MeV}/c^{2}$, $\Gamma_0(m_0) = 260\,\textrm{MeV}/c^{2}$,
and $\Gamma_{\textrm{total}}(m)$ is evaluated using the decay into $f_2 \pi$
only, which results in $C_{\pi_2} = 579.83~\textrm{mb}/\textrm{GeV}$.

\subsection{Luminosity determination using the beam kaon flux}
\label{sec:extract:lumi}
The determination of the absolute production cross section of the $a_2(1320)$
and the $\pi_2(1670)$ requires the knowledge of the luminosity, {\it i.e.\ }the
(well-known) thickness of the lead target and the incoming beam flux.  This flux
is not monitored precisely, and the absolute trigger and detector efficiencies
are only partly known, so that the absolute beam flux is not determined reliably
for the present data. Instead, the effective beam flux, which takes into account
spill structure and dead time, is determined with good precision by using $K^-
\rightarrow \pi^-\pi^-\pi^+$ decays observed in the target region.  These decays
originate from the kaon component in the negative hadron beam and are contained
in the data set preselected for the $3\pi$ production analysis, {\it i.e.\ }in
the same final state (see fig.~\ref{fig:mass_prim}).  As the systematics
concerning trigger and detector efficiencies are the same as for the $3\pi$
production from incoming pions, they cancel in the ratio of the two data sets.
 
\ifthenelse{\equal{\EPJSTYLE}{yes}}
{
\begin{figure*}
}{
\begin{figure}
}
\includegraphics[width=0.49\textwidth]{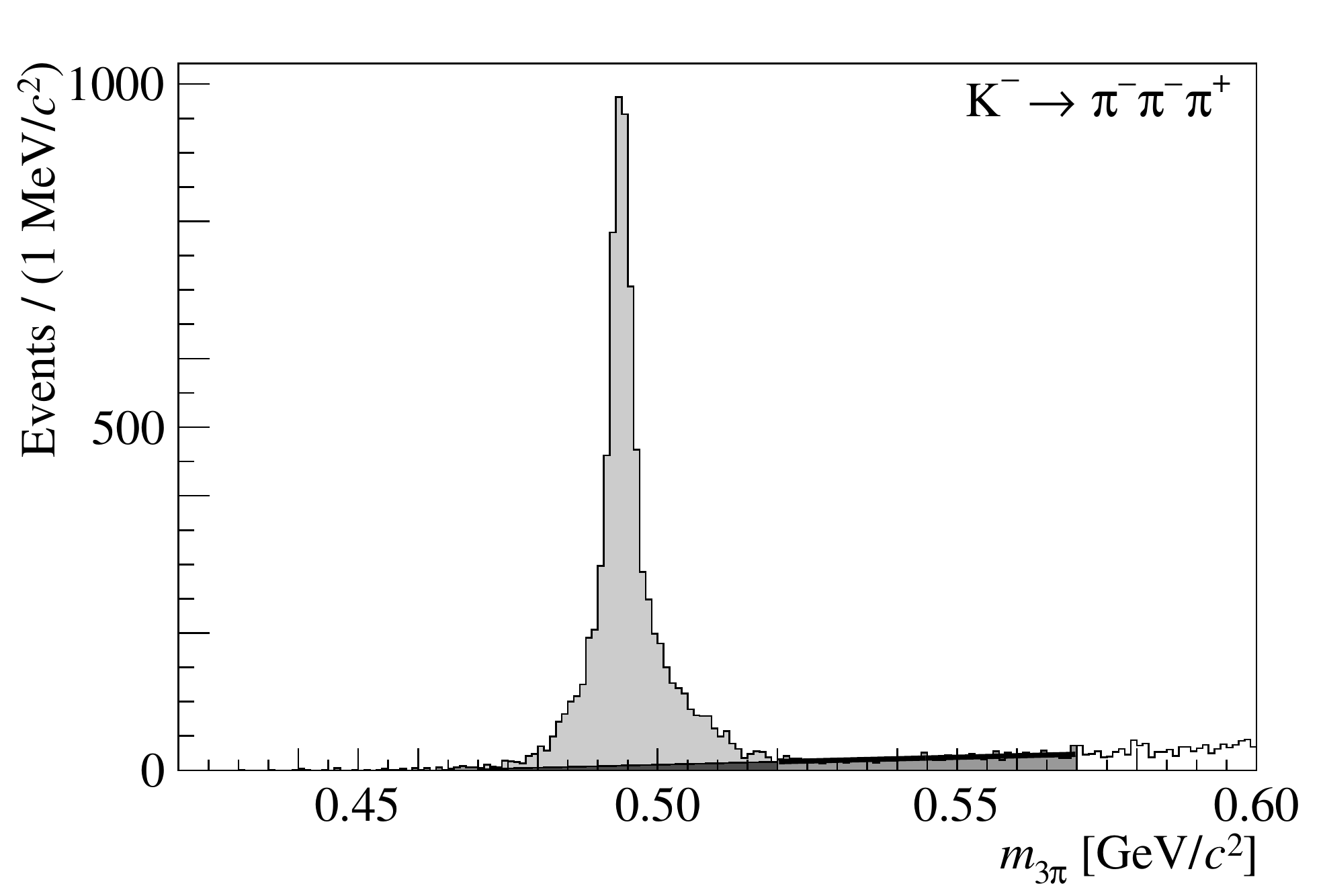}
\includegraphics[width=0.49\textwidth]{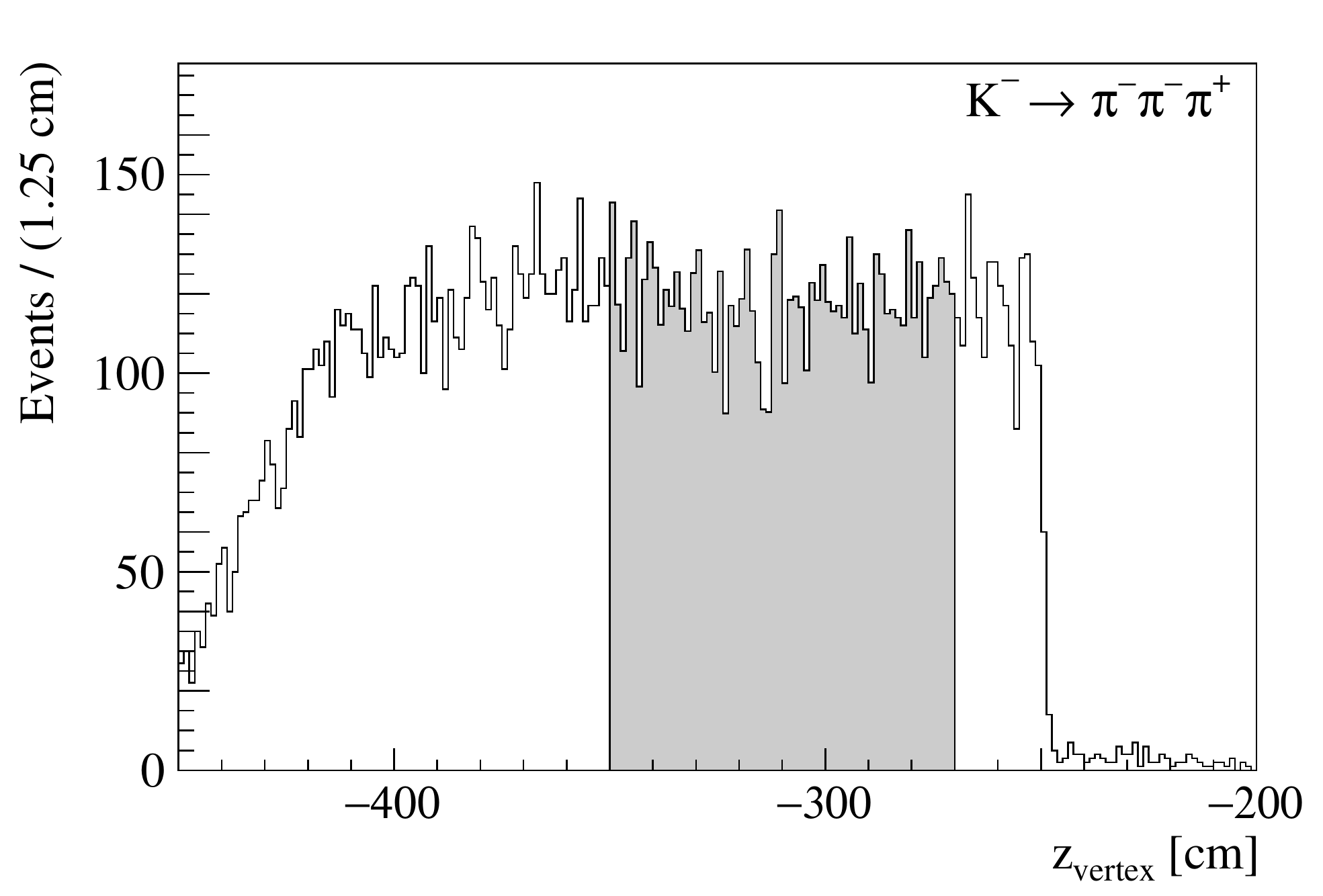}\\
 {\centerline{(a) \hspace{.47\textwidth} (c)}}\\
\includegraphics[width=0.49\textwidth]{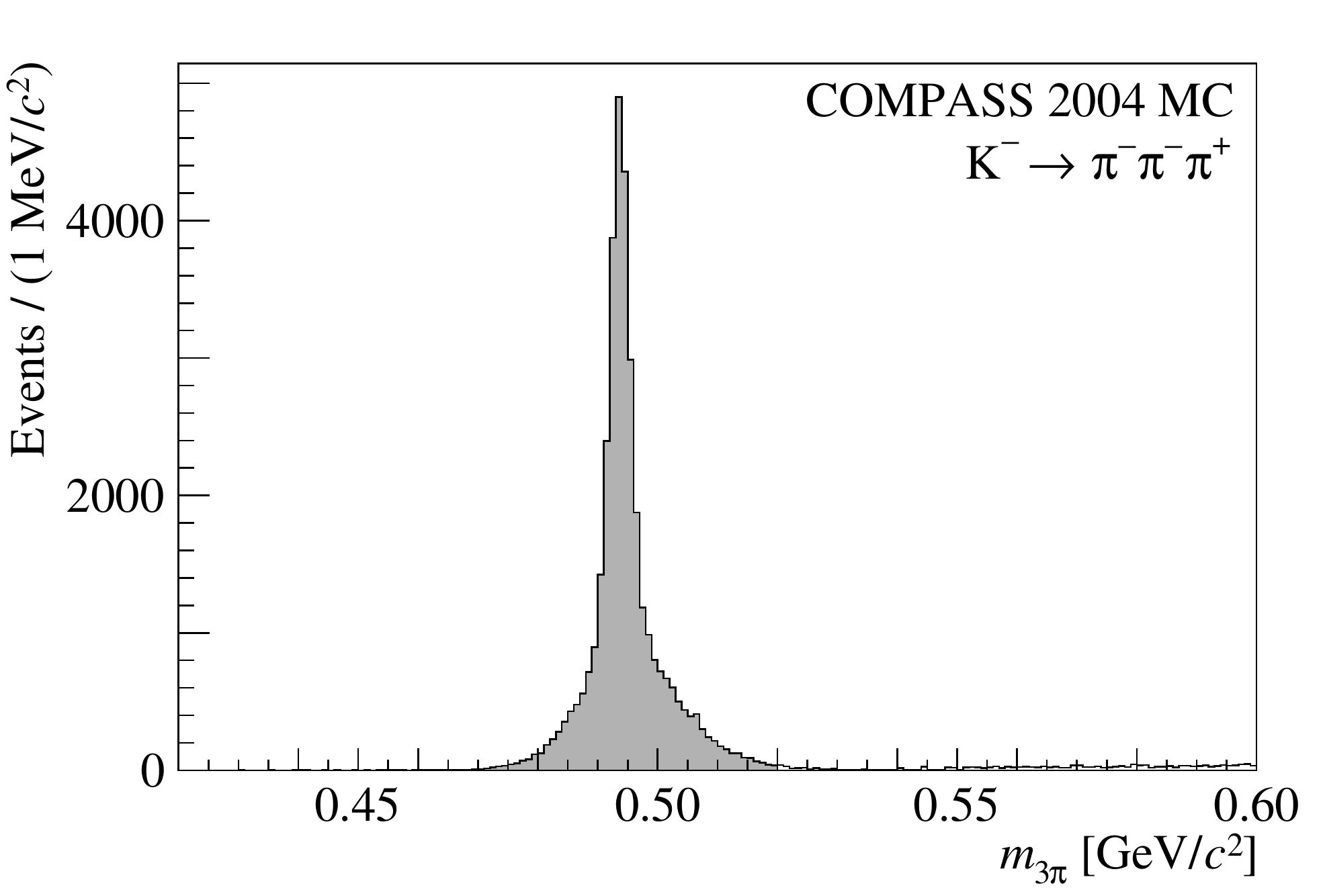}
\includegraphics[width=0.49\textwidth]{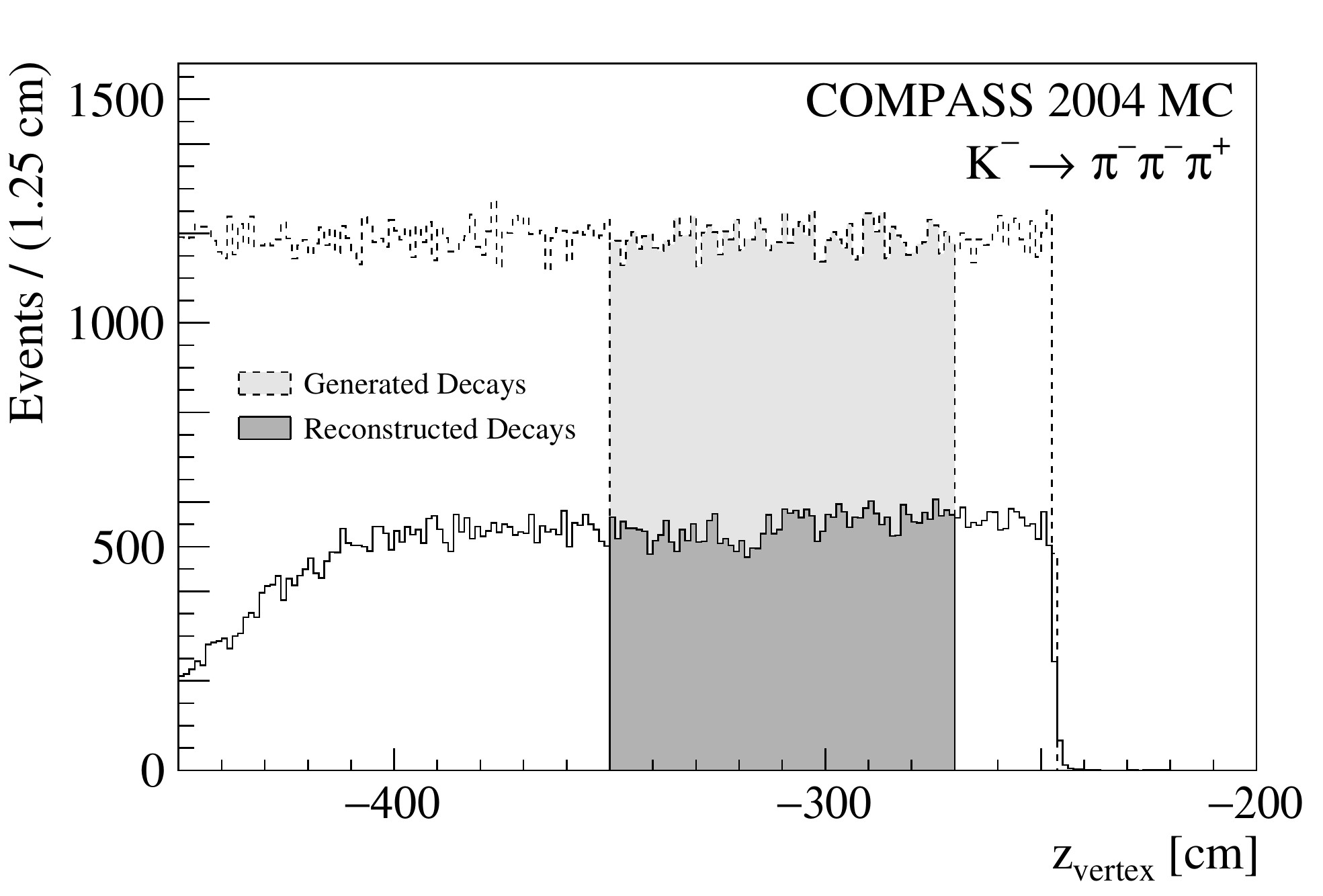}\\
 {\centerline{(b) \hspace{.47\textwidth} (d)}}
\caption{Luminosity determination from the in-flight decays of beam
  kaons. Reconstructed invariant mass spectrum (a) and decay vertex positions
  (c) for the real data, and for simulated kaon decays in the target region
  (b,d).}
\label{fig:kaon_decays}
\ifthenelse{\equal{\EPJSTYLE}{yes}}
{
\end{figure*}
}{
\end{figure}
}
For better statistical precision, the cut on the decay vertex position is
relaxed with respect to the usual cut for interactions in the
target. Figure~\ref{fig:kaon_decays}~(a) presents the resulting invariant $3\pi$
mass spectrum that shows a clean kaon signal at $\mtpGeV\approx 0.493~\mmGeV$.  The
number of kaons observed is obtained after subtraction of the small background
that is estimated by a linear fit to the mass spectrum near the peak and
extrapolated beneath the kaon signal.

The corresponding decay vertex distribution in fig.~\ref{fig:kaon_decays}~(c)
demonstrates that these vertices are reconstructed in free space along the beam
direction.  At the downstream (right) edge the distribution vanishes at the
position of the charged-particle multiplicity counter acting as the trigger
counter.  At the upstream (left) edge the positions of the beam telescope
detectors measuring an incoming beam track limit the fiducial decay volume.  The
contribution of the small background stemming from pion interactions in the lead
target is obtained from the vertex distribution in the neighbouring mass region
$0.52\ \mmGeV <\mtpGeV <0.57\ \mmGeV$ as indicated in fig.~\ref{fig:kaon_decays}~(a), scaled
according to the expected intensity below the peak of the mass spectrum, and
subtracted.  The resulting distribution is quite flat, as expected, but shows a
small drop in the region just upstream of the lead target. This can be explained
by multiple scattering of the three pion tracks in lead, which leads to a local
broadening of the decay vertex distribution.  The choice of the range of
positions of reconstructed decay vertices in [$-$350,~270]~cm (shaded area in
fig.~\ref{fig:kaon_decays}~(c)) assures uniform reconstruction efficiency.

These mass and decay vertex distributions were confirmed by a dedicated full MC
simulation of kaon decays in the respective region of the COMPASS spectrometer.
Figure~\ref{fig:kaon_decays}~(b) depicts the corresponding invariant mass
distribution of the reconstructed kaon decays. The contribution from pion
interactions in the target, which is present in the experimental data as
smoothly rising background, is absent here. The shape of the reconstructed kaon
mass spectrum is precisely reproduced, including the broad part at the base
which is traced back to kaons decaying upstream of the lead target.
Figure~\ref{fig:kaon_decays}~(d) presents the spatial distribution of simulated
and reconstructed kaon decay vertices, confirming the correct choice of the
fiducial decay volume.

The acceptance for kaon decay events is $\epsilon_{K} \approx 0.459$, calculated
from the ratio of number of reconstructed kaon decays (with all cuts applied) to
simulated decays in the same spectrometer region as used for the kaon flux
analysis of the experimental data.  Using the thickness of the lead target of
$3~\textrm{mm}$, we determine the integrated effective luminosity
\begin{equation}
\epsilon_{K} \cdot L = \epsilon_{K} \cdot \int \mathcal{L} \mathrm{d}t = 9.55 \cdot 10^{4} ~{\textrm{mb}}^{-1} \ . 
\label{eq:luminosity_value}
\end{equation}
The relative uncertainty on this number is estimated to be 6\%, with
contributions from the uncertainty of the kaon fraction in the negative hadron
beam of about 5\%, an uncertainty on the branching fraction of $K^{-}
\rightarrow \pi^{+}\pi^{-}\pi^{-}$ of less than 1\%, and the uncertainty on the
number of kaon decays in the analysed data set of less than 1\%. The statistical
uncertainty of the luminosity determination is added in quadrature to the linear
sum of the two other uncertainties.

\subsection{Results}
\label{sec:extract:result}
The radiative widths for both resonances are calculated using
eq.~\eqref{eq:intro:radwidth_from_data} with the corresponding cross sections
given by eq.~\eqref{eq:intro:xsec_radwidth}.  For the $a_2(1320)$, the radiative
width is calculated using the parameter $\epsilon_{\textrm{resol}} = 0.742$
determined by the Monte Carlo simulation shown in
sect.~\ref{sec:prim_res:resol}, $\textrm{BR}^{\textrm{PDG}}_{\rho\pi}=0.701$,
and the squared Clebsch-Gordan coefficient $\textrm{CG}=\frac{1}{2}$. The
obtained value for the radiative width is $\Gamma_0(a_2 \rightarrow \pi\gamma) =
358~\textrm{keV}$.  The radiative width of the $\pi_2(1670)$ is determined using
$\textrm{BR}^{\textrm{PDG}}_{f_2 \pi}=0.56$, $\textrm{CG}=\frac{2}{3}$,
$\epsilon_{\textrm{resol}}= 0.736$, which results in $\Gamma(\pi_2 \rightarrow
\pi\gamma) = 181~\text{keV}$ depending on the true $\textrm{BR}_{f_2 \pi}$, {\it
  i.e.\ }to be multiplied by $0.56/\textrm{BR}_{f_2 \pi}$.

The relative statistical and systematic uncertainties are summarised in
table~\ref{tab:extract:unc}.  The statistical uncertainties are obtained from
the uncertainty of the Breit-Wigner fits to the related total intensities from
fig.~\ref{fig:extract:v130}.  The systematic uncertainties have five
contributions that are added in quadrature. The uncertainties on the fraction of
diffractive background, determined by
$\sigma_{\text{prim}}/\sigma_{\text{all}}$, and the uncertainty from the kaon
normalisation were discussed above. 
\begin{table}[h]
\begin{center}
\caption{Summary of estimated statistical and systematic uncertainties for the
  measurement of the radiative widths of $a_2(1320)$ and $\pi_2(1670)$.}
\begin{tabular}{l|l|l}
 & $a_2(1320)$ & $\pi_2(1670)$ \\
 \hline\hline
Statistical & 1.8\% & 6.2\%  \\
\hline\hline 
Systematic &  &  \\
\cline{1-1}
Diffractive background & 1.2\% & 7.4\% \\
Kaon normalisation & 6.0~\% & 6.0~\%  \\
PWA models & 5.0\% & 7.7\%  \\
Parameterisation mass-dep. fit & 3.2\% & 3.1\%  \\
Radiative corrections & 8.0~\% & 8.0\% \\
\hline
\hline
Quadratic sum & 11.7\% & 15.0\%  %
\end{tabular}
\label{tab:extract:unc}
\end{center}
\end{table}
The systematic uncertainties related to the variation of the model used for the
PWA fits were derived comparing the total intensities obtained from models using
either different thresholds for the $1^{++}1$ amplitudes or an additional
$2^{++}0^{-} \rho[D]\pi$ amplitude with respect to the nominal fit model.  The
parameterisation of the mass-dependent widths covers again several aspects.  The
systematic uncertainty related to the fits determining the resonance parameters
via Breit-Wigner functions covers different parameterisations of the mass
dependence of the widths, as well as the inclusion or omission of the factor
$m_0/m$ and/or the $t_{\textrm{min}}$ dependent term.  The differences between
the parameterisation of the phase space by the traditional angular momentum
barrier factors and the phase space respecting the non-zero width of the isobars
were evaluated. In addition, for the $a_2(1320)$ we take into account also the
difference between our approach and a simplistic description of the $a_2(1320)$
shape given by the decay into $\rho\pi$ only ({\it i.e.\ }neglecting the
$\eta\pi$ decay), as implemented {\it e.g.\ }in ref.~\cite{selex_plb}. The fifth
contribution to the systematic uncertainties originates from radiative
corrections as discussed below. The employed PWA tools did not include possible
relativistic effects on the amplitude parameterisation as described in
ref.~\cite{relampl}.

\ifthenelse{\equal{\EPJSTYLE}{yes}}
{
\begin{table*}
}{
\begin{table}
}
\begin{center}
\caption{The values for the radiative widths of the $a_2(1320)$ and the $\pi_2(1670)$ from this analysis, compared to previous measurements and theoretical predictions.}
\begin{tabular}{l||l|l}
 & \multicolumn{1}{c|}{$a_2(1320)$} & \multicolumn{1}{c}{$\pi_2(1670)$} \\
 \hline\hline
This measurement & $(358 \pm 6 \pm 42)~\textrm{keV}$ & $(181 \pm 11 \pm 27) ~\textrm{keV} \cdot (0.56/\textrm{BR}_{f_2 \pi})$  \\
\hline\hline 
SELEX \cite{selex_plb} & $(284 \pm 25 \pm 25) ~\textrm{keV}$ &  \\
S.~Cihangir {\it et al.\ }\cite{e272_cihangir} & $(295 \pm 60) ~\textrm{keV}$ & \\
E.~N.~May {\it et al.\ }\cite{may_et_al} & $(0.46 \pm 0.11) ~\textrm{MeV}$ & \\
\hline
VMD model \cite{rosner_radwidths} & $ (375 \pm 50)~\textrm{keV}$ & \\
Relativ. Quark model \cite{aznauryan_oganesyan}  & $324~\textrm{keV}$ & \\
Cov. Osc. Quark model \cite{ishida_et_al} & $ 235 ~\textrm{keV}$ & \\
Cov. Osc. Quark model \cite{maeda_cov_osc_qm} & $ 237 ~\textrm{keV}$ & 2 values: $335 ~\textrm{keV}$ and $521 ~\textrm{keV}$
\end{tabular}
\label{tab:extract:comp}
\end{center}
\ifthenelse{\equal{\EPJSTYLE}{yes}}
{
\end{table*}
}{
\end{table}
}

There exists no full QED correction of the pion and resonance interaction with
the lead nucleus as $Z\alpha$ is ``not small''. From the size of the correction
of about $20\%$ and the omission of any further radiative corrections, as in
ref.~\cite{compass_3pichpt}, we conservatively estimate an $8\%$ contribution to
the systematic uncertainty \cite{comm_kaiser}.  As the only way to reduce this
uncertainty, we see a measurement on a medium-heavy nucleus, where the Primakoff
contribution is still sizeable but the discussed Coulumb correction has a minor
impact.

Our final results for the radiative widths of the $a_2(1320)$ and the
$\pi_2(1670)$ are listed in table~\ref{tab:extract:comp}.  Here and in the
following, the first uncertainty denotes the statistical and the second the
systematic one.  The value for the $a_2(1320)$ is $\Gamma_0(a_2(1320)
\rightarrow \pi\gamma) = (358 \pm 6 \pm 42) ~\textrm{keV}$, where $|F^u_C(t',
t_{\textrm{min}})|^2$ from ref.~\cite{faeldt_2009_2013cor} is used.  If
$F_{\mbox{\tiny eff}}^2(t')$ in eq.~\eqref{eq:intro:xsec_primakoff_resonance} is
approximated as $F_{\mbox{\tiny eff}}^2(t') = j_1^2(t')$ for the lead target,
{\it i.e.\ }the Coulomb correction is not applied, we obtain $\Gamma_0(a_2(1320)
\rightarrow \pi\gamma) = (312 \pm 6 ) ~\textrm{keV}$.  Most earlier measurements
reported lower values compatible with this value as given in
table~\ref{tab:extract:comp}.  The authors of ref.~\cite{selex_plb} report to
have taken into account the Coulomb corrections and they estimated that it had
``minor impact'' on their result (see table~\ref{tab:extract:comp}).  Our
calculation, however, shows that the effect is 24\% for our experiment and 15\%
for the conditions of the SELEX experiment. The result for the $\pi_2(1670)$ is
$\Gamma_0(\pi_2(1670) \rightarrow \pi\gamma) = (181 \pm 11 \pm 27) ~\textrm{keV}
\cdot (0.56/\textrm{BR}_{f_2 \pi})$.  In the case that $F_{\mbox{\tiny
    eff}}^2(t') = j_1^2(t')$ is used, we calculate $\Gamma_0(\pi_2(1670)
\rightarrow \pi\gamma) = (151 \pm 9 ) ~\textrm{keV} \cdot (0.56/\textrm{BR}_{f_2
  \pi})$.

\section{Conclusions}
We have measured the radiative widths of the $a_2(1320)$ and $\pi_2(1670)$
resonances produced in pion-nucleus interactions via the Primakoff mechanism
using a partial-wave analysis for a clean identification of the two states.  The
value for the $a_2(1320)$ is $\Gamma_0(a_2(1320) \rightarrow \pi\gamma) = (358
\pm 6 \pm 42) ~\textrm{keV}$.  Comparing our measurement with theoretical
predictions, we find our result consistent with the calculation from the VMD
model given in ref.~\cite{rosner_radwidths}, while predictions from quark models
are substantially lower. For the first time we present a value for the radiative
width of the $\pi_2(1670)$, $\Gamma_0(\pi_2(1670) \rightarrow \pi\gamma) = (181
\pm 11 \pm 27) ~\textrm{keV} \cdot (0.56/\textrm{BR}_{f_2 \pi})$. This is the
first observation of the radiative width of an E2 transition in meson
spectroscopy, which may provide constraints for further model calculations.

\subsection*{Acknowledgements}
We gratefully acknowledge the support of the CERN management and staff and the
skill and effort of the technicians of our collaborating institutes. Special
thanks go to V. Anosov and V. Pesaro for their technical support during the
installation and the running of this experiment. This work was made possible by
the financial support of our funding agencies. We would like to thank
Prof. Norbert Kaiser (TUM) for his helpful comments.

\begin{appendix}
\section{Appendix}
\label{sec:appendix}

\subsection{The extended maximum-likelihood fit}
\label{sec:app:likelihood}
The physics interpretation of the experimental data is developed in terms of
simple models based on eq.~\eqref{eq:pwa:xsec_mindep}. Within these models, the
transition amplitudes $T^{\epsilon}_{ir}$ have to be optimised individually for
each mass bin $\Delta m_{\textrm{fit}}$ such that the best possible agreement of
$\Delta \sigma_m(\tau, t')$ with the distribution of the experimental data in
the respective mass bin is achieved. This is realized by using the extended
likelihood method to maximise the following expression for every mass bin
$\Delta m$:

\begin{align}
\ln \mathcal{L} =& \sum^{N_{\textrm{events}}}_{n=1} \ln \Delta\sigma_{m_{n}}(\tau_{n}, t'_{n})
\ifthenelse{\equal{\EPJSTYLE}{yes}}
{
\nonumber \\ &-
}{
-
}
\underbrace{ \int \Delta\sigma_{m}(\tau, t') \; \eta(\tau, m, t') \text{d}\Phi(\tau) \, \text{d}m \, \text{d}t' }_{= N_{\textrm{events}} \textrm{ for the converged fit} } \nonumber\\ 
=&  \sum^{N_{\textrm{events}}}_{n=1} 
\ln \bigg[ \sum_{\epsilon, r} \sum_{ij} T^{\epsilon}_{ir} T^{\epsilon *}_{jr}  
\ifthenelse{\equal{\EPJSTYLE}{yes}}
{
\nonumber \\ &\quad \cdot
}{
}
\overline{f}_{i}^{\epsilon}(t'_{n}, m_{n})\,\overline{\psi}_{i}^{\epsilon}(\tau_{n}, m_{n}) \overline{f}_{j}^{\epsilon *}(t'_{n}, m_{n})\,\overline{\psi}_{j}^{\epsilon *}(\tau_{n}, m_{n})  \bigg] 
\ifthenelse{\equal{\EPJSTYLE}{yes}}
{
\nonumber  \\ &-
}{
-
}
\sum_{\epsilon, r} \sum_{ij} T^{\epsilon}_{ir} T^{\epsilon *}_{jr} I^{\epsilon}_{ij} \ .
\label{eq:pwa:likelihood_used}
\end{align}
The pre-calculated normalisation integrals are given by 

\begin{align}
I^{\epsilon}_{ij} =& \int  \overline{f}^{\epsilon}_{i}(t', m)\,\overline{\psi}^{\epsilon}_{i}(\tau, m)\; \overline{f}^{* \epsilon}_{j}(t', m)\,\overline{\psi}^{* \epsilon}_{j}(\tau, m)
\ifthenelse{\equal{\EPJSTYLE}{yes}}
{
\nonumber \\ & \quad \cdot
}{
\;
}
 \eta(\tau, m, t') \;  \text{d}\Phi(\tau)\,\text{d}m\,\text{d}t' \ .
\label{eq:pwa:int_acc}
\end{align}
For pairs of individual amplitudes $i$ and $j$, the integration is performed
over the phase space $\tau$, the three-pion mass $m$ inside $\Delta
m_{\textrm{fit}}$, and the $t'$ range used in this analysis.  The expression
$f^{\epsilon}_{i}(t', m)\,\psi^{\epsilon}_{i}(\tau, m)$ is evaluated using
phase-space Monte Carlo events.  Their number exceeds the experimental number of
events by typically a factor $5-10$, such that their statistical uncertainty can
be neglected.  The factor $\eta(\tau, m, t')$ takes into account the acceptance
of the spectrometer.  Note that the integral in eq.~\eqref{eq:pwa:int_acc} is
normalised such that if $\eta(\tau, m, t')=1$ then $I_{ij}=1$ for $i=j$, and
$I_{ij} =0$ for $i\neq j$. With this normalisation the fitted number $\sum_{r}
T^{\epsilon}_{ir} T^{\epsilon *}_{jr}$ refers to the number of events in a
partial wave $i$, cf. eq.~\eqref{eq:pwa:int_phase}.

The fitting procedure is carried out for each mass bin individually, with
typically 10-50 independent fit attempts with random starting values of the
parameters per mass bin until the best fit yields optimized sets of
$T^{\epsilon}_{ir}$ with their statistical uncertainties.

\subsection{The concept of partial coherence}
\label{sec:app:part_coh}
Incoherence effects may be observed in a PWA of experimental data, even if a
coherent production process takes place.  In the data presented in this paper,
these effects are related to the resolution effects discussed in
sect.~\ref{sec:prim_res:resol}.  In these cases, using $N_r>1$ in
eq.~\eqref{eq:pwa:xsec_mindep} often allows for too much freedom between the
production amplitudes $T_{ir}^{\epsilon}$ that appear $N_r$ times in the PWA
fit.  Instead, the observed incoherence can be taken into account by using
partial coherences. This allows limited coherence between selected sets of decay
amplitudes and thus reweighting of individual off-diagonal terms in the coherent
sum, as illustrated by:

\begin{align}
\sum_{ij} \; T_{i} \; T_{j}^{*} \; \psi_{i} \; \psi_{j}^{*} \quad \rightarrow \quad
\sum_{ij} \; \textrm{r}_{ij} \; T_{i} \; T_{j}^{*} \; \psi_{i} \; \psi_{j}^{*}  \ .
\label{eq:pwa:partcoh_simple}
\end{align}

Here, $\textrm{r}_{ij} \leq 1$ are real numbers that reflect the reduction of
coherence between the decay amplitudes $i$ and $j$, with $\textrm{r}_{ij} =
\textrm{r}_{ji}$.  They decrease the contribution of the interference of the
amplitudes $i$ and $j$, without introducing additional phases as $N_r>1$
does. These parameters are usually also fitted.  The intensities of the
individual decay amplitudes are preserved by fixing $\textrm{r}_{ii} \equiv 1 $.

\subsection{Parameterisation of the ChPT amplitude}
\label{sec:appendix:chpt}
The transition amplitude $\mathcal{A}$ of a process contributing to the reaction
$\pi^- \gamma \rightarrow \pi^- \pi^- \pi^+$ has the general form

\begin{align}
\mathcal{A} = \frac{2e}{m_{\pi}^2}( \Vec{\epsilon}\cdot\Vec{q_1} A_1 + \Vec{\epsilon}\cdot\Vec{q_2} A_2) \ .
\label{eq:chpt_amp_orig}
\end{align}
Here $\Vec{q_1}$ and $\Vec{q_2}$ are the three-momenta of the two outgoing
$\pi^-$ in the Gottfried-Jackson reference system, which are complemented with
$\Vec{q_3}$ and $\Vec{p_1}$, the three-momenta of the outgoing $\pi^+$ and the
incoming $\pi^-$, respectively, and $\hat{k} = (0, 0, -1)$, the unit vector of
the photon momentum $\Vec{k}$.  The reference system is determined in the rest
frame of the $3\pi$ system. Its $z$-axis is in the direction of the incoming
beam particle, the $y$-axis perpendicular to the production plane, {\it
  i.e.\ }given by $\Vec{p}_{\textrm{recoil}} \times \hat{z}$ with
$\Vec{p}_{\textrm{recoil}}$ being the three-momentum of the recoil particle, and
$\hat{x} = \hat{y} \times \hat{z}$.  Equation~\eqref{eq:chpt_amp_orig} uses the
elementary electric charge $e$, the transverse polarisation vector of the photon
$\Vec{\epsilon}$, and the amplitudes $A_1$ and $A_2$ that contain the dynamical
information. The calculation is performed using the radiation gauge, where
$\Vec{\epsilon} \cdot \hat{k} = 0 $.  When the cross section $\sigma$ is
calculated, the amplitude $\mathcal{A}$ is squared and the average of the
transverse polarisation states of the photon is evaluated, which leads to the
following square of the vector product:
\begin{align}
\sigma \propto [ (A_1 \Vec{q_1} + A_2 \Vec{q_2}) \times \hat{k} ]^2 \ .
\end{align}
The amplitudes $A_1$ and $A_2$ are expressed in terms of the dimensionless
Mandelstam variables $s = (p_1 + k)^2/m_{\pi}^2$, $s_1 = (q_1 +
q_3)^2/m_{\pi}^2$, $s_2 = (q_2 + q_3)^2/m_{\pi}^2$, $t_1 = (k-q_1)^2/m_{\pi}^2$
and $t_2 = (k-q_2)^2/m_{\pi}^2$. The amplitudes $A_1$ and $A_2$ transform into
each other by the relation $A_2(s, s_1, s_2, t_1, t_2) = A_1(s, s_2, s_1, t_2,
t_1)$.

For the implementation into the PWA the amplitude $\mathcal{A}$ is represented
in the reflectivity basis. The ChPT amplitude employed in our analysis is
implemented as
\begin{align}
\Psi_{\textrm{ChPT}}^{\epsilon = +1} &= \frac{1}{\sqrt{s - 1}} (A_1 \cdot q_1[x] + A_2 \cdot q_2[x]) \quad \textrm{and} \label{eq:chpt:refl_pos} \\
\Psi_{\textrm{ChPT}}^{\epsilon = -1} &= \frac{1}{\sqrt{s - 1}} (-A_1 \cdot q_1[y] -A_2 \cdot q_2[y]), 
\label{eq:chpt:refl}
\end{align} 
where $q_i[j]$ is the $j$-component of the momentum $\Vec{q_i}$ in the
Gottfried-Jackson reference system. The term $1/\sqrt{s - 1}$ originates from
the flux factor in the cross section.

The ChPT amplitude employed in the presented PWA takes into account the
tree-level calculation given in ref.~\cite{kaiser}, the calculation of loops and
the necessary counter-terms from ref.~\cite{kaiser2}, and also $\rho$-exchange
contributions.  The $\rho$-exchange contribution to $A_1$ of
eqs.~\eqref{eq:chpt:refl_pos} and \eqref{eq:chpt:refl}, which were provided by
ref.~\cite{comm_kaiser}, are explicitely written down as:
\begin{align}
A_1^{\rho}  &= 
	 \frac{g_{\rho\pi}^2 }{ b^2 (3 - s - t_1 - t_2)} 
\ifthenelse{\equal{\EPJSTYLE}{yes}}
{\nonumber \\ & \quad \cdot
}{
}		
		\bigg[ \frac{(2s +1 - 2 s_2 - s_1 + t_1) (s - 2 - s_1 + t_2)^2}{b + s - 2 - s_1 + t_2 - \Sigma_{\rho}(2 -s + s_1 - t_2)} \nonumber \\
	& \qquad + \frac{(2s + 1- 2s_1 - s_2 + t_2)(s-2-s_2 + t_1)^2 } { b + s - 2 -s_2 + t_1 -  \Sigma_{\rho} (2 - s + s_2 - t_1)} \bigg] \nonumber \\
	&+  \frac{g_{\rho\pi}^2}{ b^2 (t_1 - 1)} \bigg[ \frac{(2s - 2s_1 - s_2 +t_1 + 2t_2 -1) s_2^2}{b - s_2 - \Sigma_{\rho}(s_2)} \nonumber \\
	& \quad + \frac{(s + 1 - s_1 - 2 s_2 + t_1 + t_2) (s - 2 - s_1 + t_2)^2}{b+ s - 2 - s_1 + t_2 - \Sigma_{\rho} (2 -s + s_1 - t_2)} \bigg] \nonumber \\
	&+  \frac{g_{\rho\pi}^2 }{ b^2 } \bigg[ \frac{s_2(1 - s_2 - t_2)}{ b - s_2 - \Sigma_{\rho}(s_2)} + \frac{s_1 (s - 2 - 2 s_1 + t_2)}{ b - s_1 - \Sigma_{\rho}(s_1)} \nonumber \\
	& \quad + \frac{(s_2 + 1 - t_1) (s-2-s_2 + t_1)}{b + s - 2 -s_2 + t_1 -  \Sigma_{\rho} (2 - s + s_2 - t_1)} \bigg] \ .
\label{eq:chpt:amp_rho}
\end{align}
This notation uses the coupling constant $g_{\rho\pi} = 6.03$ and the squared
mass ratio $b = \big(m_{\rho}/m_{\pi}\big)^2 = 30.4367$.  The energy-dependent
self-energy $\Sigma_{\rho}$ of the $\rho$ is given by
\begin{align}
 \textrm{Im}\, \Sigma_{\rho}(x) 
= \frac{g_{\rho\pi}^2}{48 \pi \,\sqrt{x} } \Big(x - 4 \Big)^{3/2} \theta(x-4), 
\label{eq:Sigma_rho_imag} \\
\text{where}\quad  \textrm{Im}\, \Sigma_{\rho}(x=b) = m_{\rho} \Gamma_{\rho} \quad \text{with}\ \Gamma_{\rho}= 150~\text{MeV},
\label{eq:Sigma_rho_imag_b}
\end{align}
and
\begin{align}
\textrm{Re} \,\Sigma_{\rho}(x) = 
\ifthenelse{\equal{\EPJSTYLE}{yes}}
{&}{}
\frac{g_{\rho\pi}^2}{24 \pi^2} x  \Bigg[
   \frac{4}{b} - \frac{4}{x} 
\ifthenelse{\equal{\EPJSTYLE}{yes}}
{\nonumber \\
}{
}  
  &- \bigg(1-\frac{4}{x}\bigg) ^{3/2}  \ln \Big( 0.5 \big(\sqrt{|x|} + \sqrt{|x-4|} \big) \Big) \nonumber \\
&+ \left(1-\frac{4}{b}\right)^{3/2}  \ln \Big( 0.5 \big(\sqrt{b} + \sqrt{b-4} \big) \Big)
\Bigg] \ .
\label{eq:Sigma_rho_real}
\end{align}
Terms like $\Sigma_{\rho}(s_{1,2})$ in eq.~\eqref{eq:chpt:amp_rho} refer to
contributions from real $\rho$, which decay into $\pi^+\pi^-$ pairs so that
$s_{1,2}>4$, and are complex-valued. In case of a virtual $\rho$, where the
$\rho$-exchange mediates the $\pi\pi$-interaction, the self-energy reads like
$\Sigma_{\rho}(2 - s + s_{1,2} - t_{2,1})$, with $2 - s + s_{1,2} - t_{2,1} <
0$, and is purely real-valued.

\subsection{Set of decay amplitudes}
\label{sec:appendix:waveset}
Table~\ref{tab:waveset} shows the set of decay amplitudes used for the PWA
presented in this paper. The wave set includes established amplitudes with
$M^{\epsilon}=0^{+}$ attributed to diffractive dissociation.  The amplitudes
related to Primakoff production are provided as isobaric $M=1$ amplitudes or as
the ChPT amplitude. All Primakoff amplitudes are introduced with both $\epsilon=
\pm1$ which is necessary due to the limited resolution as explained in
sect.~\ref{sec:prim_res:resol}.  Also included in the fit are an amplitude
describing the decay of beam kaons into $\pi^-\pi^-\pi^+$, and an incoherent
background wave that is homogeneous in phase space.  Many amplitudes are
introduced with an upper or lower threshold which is chosen in order to
constrain them to the region where they are expected to contribute.

\begin{table}
\caption{The set of decay amplitudes used for the PWA fit for the extraction of the intensities of $a_2(1320)$ and $\pi_2(1670)$.}
	\begin{center}
          \begin{tabular}{c|c|c|c}
          $J^{PC}M^{\epsilon}$ & $L$ & \{Isobar\} $\pi$ & Thr. [GeV]\\
          \hline
          \hline
          $0^{-+}0^+$ & $S$ & $(\pi\pi)_s\pi$ & - \\
          $0^{-+}0^+$ & $S$ & $f_0(980)\pi$ & 1.12 \\
          $0^{-+}0^+$ & $P$ & $\rho\pi$ & - \\
          $0^{-+}0^+$ & $D$ & $f_2\pi$ & 1.24 \\
          $0^{-+}0^+$ & $S$ & $f_0(1500)\pi$ & 1.64 \\
          \hline
          $1^{++}0^+$ & $S$ & $\rho\pi$ & 0.48 \\
          $1^{++}0^+$ & $P$ & $f_2\pi$ & 1.40 \\
          $1^{++}0^+$ & $P$ & $f_0(980)\pi$ & 1.24 \\
          $1^{++}0^+$ & $P$ & $(\pi\pi)_s\pi$ & - \\
          $1^{++}0^+$ & $D$ & $\rho\pi$ & 1.04 \\
          $1^{++}1^+$ & $S$ & $\rho\pi$ & 0.76 \\
          $1^{++}1^+$ & $P$ & $(\pi\pi)_s\pi$ & 0.92 \\
          \hline
          $1^{-+}1^+$ & $P$ & $\rho\pi$ & 1.48 \\
          \hline
          $2^{++}1^+$ & $D$ & $\rho\pi$ & 0.92 \\ 
          \hline
         $2^{-+}0^+$ & $S$ & $f_2\pi$ & 1.24 \\
          $2^{-+}0^+$ & $P$ & $\rho\pi$ & 0.80 \\
          $2^{-+}0^+$ & $D$ & $(\pi\pi)_s\pi$ & 1.32 \\
          $2^{-+}0^+$ & $D$ & $f_2\pi$ & 1.52 \\
          $2^{-+}0^+$ & $F$ & $\rho\pi$ & 1.28 \\
          $2^{-+}1^+$ & $S$ & $f_2\pi$ & 1.38 \\
          $2^{-+}1^+$ & $P$ & $\rho\pi$ & 1.28 \\
          $2^{-+}1^+$ & $D$ & $(\pi\pi)_s\pi$ & 1.24 \\
          $2^{-+}1^+$ & $F$ & $\rho\pi$ & 1.52 \\
           \hline
          $3^{++}0^+$ & $P$ & $f_2\pi$ & 1.52 \\
          $3^{++}0^+$ & $D$ & $\rho\pi$ & 1.52 \\
          \hline 
          \multicolumn{3}{l|}{ChPT trees+loops+rho $\epsilon=+1$} &  $<1.56$\\
          \hline
          \hline
          $1^{++}1^-$ & $S$ & $\rho\pi$ & 0.76 \\ 
          $1^{++}1^-$ & $P$ & $(\pi\pi)_s\pi$ & 0.92 \\ 
          \hline
          $1^{-+}1^-$ & $P$ & $\rho\pi$ & 1.48 \\
          \hline
          $2^{++}1^-$ & $D$ & $\rho\pi$ & 0.92 \\
          \hline 
          $2^{-+}1^-$ & $S$ & $f_2\pi$ & 1.36 \\ 
          $2^{-+}1^-$ & $P$ & $\rho\pi$ & 1.28 \\ 
          $2^{-+}1^-$ & $D$ & $(\pi\pi)_s\pi$ & 1.24 \\ 
          $2^{-+}1^-$ & $F$ & $\rho\pi$ & 1.52 \\ 
          \hline 
          \multicolumn{3}{l|}{ChPT trees+loops+rho $\epsilon=-1$} & $<1.56$\\
          \hline
          \hline 
          \multicolumn{3}{l|}{Kaon decay} & $<0.56$\\
          \hline
          \hline
          \multicolumn{3}{l|}{Background} & - \\
          \hline
          \hline
        \end{tabular}
        \end{center}
        \label{tab:waveset}
\end{table}

\subsection{Parameterisation of the decay amplitudes for the PWA covering the mass region of the $\pi_2(1670)$ in bins of $t'$}
\label{sec:appendix:tbins_pi2}
For the PWA fit covering the mass region around the $\pi_2(1670)$, {\it
  i.e.\ }$1.50\ \mmGeV < m_{3\pi} < 1.80\ \mmGeV$, in small bins of $t'$
(see sect.~\ref{sec:prim_res:a2_pi2}), both mass dependences and phases have to
be taken into account. The resonances are parameterised by Breit-Wigners
functions, and the background by exponentials in $m_{3\pi}$. As the background
to the amplitudes can stem from tails of higher-mass resonances, it is added
coherently with a mass-dependent relative phase. Breit-Wigner functions and
background distributions are summed to single complex-valued terms, denoted in
the following by $B_{j}(m)$. Their parameters were obtained by fitting the
square of these $B_{j}(m)$ to the respective intensity of each amplitude.  These
terms are then multiplied by the normalised decay amplitudes, {\it i.e.\ }in
every $t'$ bin the PWA uses the following decay amplitudes

\begin{align}
\overline{\psi_{j}}(\tau, m) = \frac{\psi_{j}(\tau) \cdot B_{j}(m)} {\sqrt{\int \vert \psi_{j}(\tau) \vert \text{d}\Phi(\tau)}}
\label{eq:pwa:int_pi2_BWBKGmult}
\end{align}
with

\begin{align}
\ifthenelse{\equal{\EPJSTYLE}{yes}}
{
B_{j}(m) = &p_{4,j}\frac{\sqrt{m_j \Gamma_j}}{m_j^2 - m^2 - i m_j \Gamma_j} \nonumber \\
&+ p_{5, j} \exp(- \alpha_j m) \nonumber \\ 
&+ i p_{6, j} \exp(- \alpha_j m) \ .
}{
B_{j}(m) = p_{4,j}\frac{\sqrt{m_j \Gamma_j}}{m_j^2 - m^2 - i m_j \Gamma_j} + p_{5, j} \exp(- \alpha_j m) + i p_{6, j} \exp(- \alpha_j m) \ .
}
\label{eq:pwa:int_pi2_BWBKGform}
\end{align}
Here, $m_j$ is the nominal mass, $\Gamma_j$ is the nominal width of the
resonance, $\alpha_j$ describes the background, and $p_{4,j}$, $p_{5,j}$, and
$p_{6,j}$ are adjusted to the relative strengths of Breit-Wigner and background
contributions. This background is disregarded for the $J^{PC}=2^{-+}$ amplitudes
that contain the $\pi_2(1670)$, as satisfactory fits were obtained without this
background, thus avoiding artificial phase shifts in the resonance region. An
exception is the $2^{-+}0^{+} f_2[D]\pi$ amplitude that seems rather to contain
the higher-mass $\pi_2(1880)$, which is not further discussed here. All used
parameters are derived from fits to intensities in small mass bins, and given in
table~\ref{tab:int_pi2_BWpolynoms}.
\ifthenelse{\equal{\EPJSTYLE}{yes}}
{
\begin{table*}
}{
\begin{table}
}
\caption{Parameters used in eq.~\eqref{eq:pwa:int_pi2_BWBKGform} to describe the
  mass-dependence of decay amplitudes in the $\pi_2(1670)$ mass region
  considered in the analysis described in sect.~\ref{sec:prim_res:a2_pi2}.}
\begin{center}
\begin{tabular}{|l|c|c|c|c|c|c|}
\hline
amplitude(s)  & $m_j$ & $\Gamma_j$ & $\alpha_j$ & $p_{4, j}$ & $p_{5, j}$ & $p_{6, j}$ \\
$J^{PC}M^{\epsilon}\{\textrm{isobar}\}[L]\pi$ & \scriptsize{$[\textrm{GeV}/c^{2}]$} & \scriptsize{$[\textrm{GeV}/c^{2}]$} & \scriptsize{$[(\textrm{GeV}/c^{2})^{-1}]$} & & & \\
\hline  $0^{-+}0^{+}(\pi\pi)_S [S]\pi$ & 1.8188 & 0.23117 & 4.5104 & 43.180 & 12512.0 & $-43305.0$ \\
\hline  $0^{-+}0^{+}f_0(980)[S] \pi$ & 1.8094 & 0.23515 & $-0.36871$ & $28.042$ & $3.0863$ & $-0.57855$\\
\hline  $0^{-+}0^{+}\rho [P]\pi$ & 1.5150 & 0.20004 & 1.1676 & 47.787 & 43.133 & $-162.45$\\
\hline  $1^{++}0^{+}\rho [S]\pi$ & 1.3051 & 0.31616 & 7.3830 & 101.58 & 238700.0 & 23925.0\\
\hline  $1^{++}0^{+}\rho [D]\pi$ & 1.5943 & 0.26680 & 1.5255 & 53.629 & $-77.010$ & $-524.38$\\
\hline  $1^{++}0^{+}(\pi\pi)_S [P]\pi$ & 1.3763 & 0.74318 & 0.49296 & 89.914 & 76.893 & $-17.233$\\
\hline  $1^{++}0^{+}f_2(1270) [P]\pi$ & 1.6815 & 0.25355 & 1.3663 & 34.721 & $-33.388$ & $-232.13$\\
\hline  $2^{-+}0^{+}f_2(1270)[S] \pi$ & \multirow{4}*{1.672} & \multirow{4}*{0.260} & \multirow{4}*{--} & \multirow{4}*{--} & \multirow{4}*{--} & \multirow{4}*{--}\\
$2^{-+}0^{+}\rho [P/F]\pi$ &  &  &  &  &  & \\
$2^{-+}1^{\pm}f_2(1270)[S] \pi$ &  &  &  &  &  & \\
$2^{-+}1^{\pm}\rho [P/F]\pi$ & & & & & & \\
\hline  $2^{-+}0^{+}f_2(1270) [D]\pi$ & 1.7782 & 0.24392 & 3.1200 & 18.036 & $-1756.3$ & $-284.26$\\
\hline 
\end{tabular}
\end{center}
\label{tab:int_pi2_BWpolynoms}
\ifthenelse{\equal{\EPJSTYLE}{yes}}
{
\end{table*}
}{
\end{table}
}
The mass dependences of all waves of minor strength, not included in
table~\ref{tab:int_pi2_BWpolynoms} but used in the PWA fit, are described by a
polynomial behaviour

\begin{align}
\ifthenelse{\equal{\EPJSTYLE}{yes}}
{
&B_{j}(m) = \nonumber \\
&\sqrt{0.19746 - 0.32710 \,m + 0.18933 \, m^2 - 0.037275 \, m^3}
}{
B_{j}(m) = \sqrt{0.19746 - 0.32710 \,m + 0.18933 \, m^2 - 0.037275 \, m^3}
}
\label{eq:pwa:pol_pi2_BWbg_others}
\end{align}
instead of using eq.~\eqref{eq:pwa:int_pi2_BWBKGform}. 
\end{appendix}

\end{document}